\documentclass[usenatbib]{mn2e} 
\usepackage{epsfig}
\usepackage{subfigure}
\usepackage{amsmath}
\usepackage{amssymb}
\usepackage{lscape}
\usepackage{rotating}

\newcounter{savefig}

\newcounter{saveeqn}
\newcommand{\alpheqn}{\setcounter{saveeqn}{\value{equation}}%
  \stepcounter{saveeqn}\setcounter{equation}{0}%
  \renewcommand{\theequation}
      {\mbox{\arabic{saveeqn}\alph{equation}}}}
\newcommand{\reseteqn}{\setcounter{equation}{\value{saveeqn}}%
  \renewcommand{\theequation}{\arabic{equation}}}
\begin{document}
\title[Substructure in 2dFGRS Clusters]{Substructure Analysis
of Selected Low Richness 2dFGRS Clusters of Galaxies} 

\author[W. Burgett et al.]
{William~S.~Burgett$^{1}$\thanks{email: burgett@ifa.hawaii.edu},
%{William~S.~Burgett$^{1}$,
Michael~M.~Vick$^2$, David~S.~Davis$^{3,4}$, Matthew~Colless$^{5,6}$, 
\newauthor
Roberto~De~Propris$^6$,
Ivan~Baldry$^7$, Carlton~Baugh$^8$, Joss~Bland-Hawthorn$^5$, 
\newauthor
Terry~Bridges$^{21}$,
Russell~Cannon$^5$, Shaun~Cole$^8$, Chris~Collins$^9$, 
Warrick~Couch$^{10}$, 
\newauthor
Nicholas~Cross$^{7}$, Gavin~Dalton$^{11,12}$,
Simon~Driver$^6$, George~Efstathiou$^{13}$, 
\newauthor
Richard Ellis$^{14}$, Carlos~S.~Frenk$^8$, Karl
Glazebrook$^{7}$, 
Edward~Hawkins$^{15}$, 
\newauthor
Carole~Jackson$^{16}$, Ofer~Lahav$^{13}$, Ian~Lewis$^{11}$, Stuart~Lumsden$^{17}$,
Steve~Maddox$^{15}$, 
\newauthor
Darren~Madgwick$^{18}$, Peder~Norberg$^{19}$,
John~A.~Peacock$^{20}$, Will~Percival$^{20}$, 
\newauthor
Bruce~Peterson$^6$,
Will~Sutherland$^{20}$, and Keith~Taylor$^{14}$\\
$^1$Institute for Astronomy, University of Hawaii, 2680 Woodlawn Dr., Honolulu, HI 96822, USA;
~email: burgett@ifa.hawaii.edu\\
$^2$Department of Physics, University of Texas at Dallas, Richardson, TX 75083-0688, USA\\
$^3$Joint Center for Astrophysics, Dept. of Physics, UMBC, Baltimore, MD 21250 USA\\ 
$^4$Laboratory for High Energy Astrophysics, NASA GSFC, Code 662, Greenbelt MD 20771, USA\\ 
$^5$Anglo-Australian Observatory, P.O. Box 296, Epping, NSW 2121, Australia\\
$^6$Research School of Astronomy \& Astrophysics, The Australian National University,
Weston Creek, ACT 2611, Australia\\
$^{7}$Department of Physics \& Astronomy, Johns Hopkins University, Baltimore MD 21218-2686, USA\\
$^8$Department of Physics, University of Durham, South Road, Durham, DH1 3LE, UK\\
$^9$Astrophysics Research Institute, Liverpool John Moores University, Twelve Quays House,
Birkenhead, L14 1LD, UK\\
$^{10}$Department of Astrophysics, University of New South Wales, Sydney, NSW 2052, Australia\\
$^{11}$Department of Physics, University of Oxford, Keble Road, Oxford, OX1 3RH, UK\\
$^{12}$Rutherford Appleton Laboratory, Chilton, Didcot, OX11 0QX, UK\\
$^{13}$Institute of Astronomy, University of Cambridge, Madingley Road, Cambridge, CB3 0HA, UK\\
$^{14}$Department of Astronomy, California Institute of Technology, Pasadena, CA 91125, USA\\
$^{15}$School of Physics \& Astronomy, University of Nottingham, Nottingham, NG7 2RD, UK\\
$^{16}$CSIRO Australia Telescope National Facility, PO Box 76, Epping, NSW 1710, Australia\\
$^{17}$Department of Physics, University of Leeds, Woodhouse Lane, Leeds, LS2 9JT, UK\\
$^{18}$Lawrence Berkeley National Laboratory, 1 Cyclotron Rd., Berkeley, CA 94720, USA\\
$^{19}$ETHZ Institut fur Astronomie, HPF G3.1, ETH Honggerberg, CH-8093, Zurich, Switzerland\\
$^{20}$Institute for Astronomy, University of Edinburgh, Royal Observatory, Blackford Hill,
Edinburgh, EH9 3HJ, UK\\
$^{21}$Physics Department, Queen's University, Kingston, ON, K7L 3N6, Canada}

\maketitle
\begin{abstract}
Complementary one-, two-, and three-dimensional tests for detecting the presence of substructure
in clusters of galaxies are applied to recently obtained data from the 2dF Galaxy Redshift
Survey. The sample of 25 clusters used in this study includes 16 clusters not previously investigated 
for substructure.  
Substructure is detected at or greater than the 99\% CL level in at least one test for
21 of the 25 clusters studied here.
From the results, it appears that low richness clusters
commonly contain subclusters participating in mergers. About half of the clusters have two or more
components within $0.5 \, h^{-1} \, \mathrm{Mpc}$ of the cluster centroid, and 
at least three clusters (Abell 1139, Abell 1663, and Abell S333) exhibit velocity-position characteristics
consistent with the presence of possible cluster rotation, shear, or infall dynamics. 
The geometry of certain features is consistent with influence by the host supercluster
environments. In general, our results support the hypothesis that 
low richness clusters relax to structureless equilibrium states
on very long dynamical time scales (if at all).
\end{abstract}
\begin{keywords}
galaxies: clusters: general -- large-scale structure of Universe
\end{keywords}

%\clearpage
\section{Introduction}

It is well recognized that many clusters of galaxies exhibit substructure where
the operational definition of substructure adopted here will be departures from otherwise
smooth galaxy density and velocity equilibrium distributions. Representative
previous studies of samples of several clusters are given by, e.g., \citet{west_1}, 
\citet{rhee}, \citet{bird_1},
\citet{escalera}, \citet{west_2}, 
Pinkney, Roettiger, Burns, and Bird (1996),
\citet{girardi}, 
and \citet{solanes}.\footnote{Because of the large number of published substructure studies, it is not possible 
to cite all of them; the references herein are meant to be representative and directly relevant 
to the results presented in this paper.}
The presence of subclusters affects estimates of quantities such as 
the dynamical mass and mass-to-light ratio, the mean gravitational potential, and
the global ellipticity. Substructure may also occur in the intracluster gas distribution
traced by X-ray emission, and
cross-correlating galaxy-gas structure is an important part of 
characterizing cluster dynamics (see, e.g., \citealt{davis}, \citealt{bird_3},
or references contained in the review article by \citealt{rosati}). 
In addition to determining the dynamical state of
a cluster at a single epoch, the interaction of subclusters within the cluster
(including mergers) can significantly affect dynamical evolution.
Identifying and quantifying substructure may also reveal clues as to the initial 
conditions of the density perturbations from which clusters form and evolve.

The purpose of the present study is to apply a comprehensive, yet reasonably small, set
of tests for substructure to recently obtained data from the 2dF Galaxy Redshift Survey 
(2dFGRS)(\citealt{colless_1}, \citealt{colless_2}, and \citealt{maddox}).
With the large number of redshifts provided by the
Survey, it is possible
to explicitly correlate the substructure detected in the projected two-dimensional
surface distribution with that found from three-dimensional tests combining positional
and velocity information.  This is usually sufficient to distinguish true substructure
from projection effects including foreground and background contamination. 
Thus, the work presented in this paper is intended to provide survey results, identify
potential trends and common characteristics, and highlight features warranting further study; it is not
intended to necessarily provide a definitive last word.

Among its goals, the 2dFGRS is designed to allow investigation of
the properties of galaxy groups and clusters by providing a large, homogeneous sample in redshift
space. The source catalogue used as the Survey base is essentially a revised and
extended version of the APM catalogue with target galaxies of extinction-corrected magnitudes
$b_J \leq 19.45$ and redshifts with a measured RMS uncertainty of 
$85 \: \mathrm{km} \,\mathrm{s^{-1}}$~\citep{colless_2}.  The median redshift of the catalogue is $z = 0.11$.
A preliminary study of the Survey database
identified 431 Abell, 173 APM, and 343 EDCC clusters and provides precise redshifts, velocity
dispersions, and updated centroids~\citep{depropris}. 
Additionally, this new data affords an opportunity to
assess the completeness of, and contamination in, the older catalogues.
The 25 clusters of galaxies listed in Table~\ref{clusters} to be studied here are selected from
this new compilation, and include 16
clusters not previously investigated for substructure. 
The proper distances to each cluster at the epochs of emission and observation, $d_p(t_e)$ and  
$d_p(t_0)$, are calculated under the assumption of a flat universe with 
$\Omega_\Lambda = 0.7$ and $\Omega_m = 0.3$.  
All distances and linear scales are given in units of $h^{-1}$ Mpc
where $h\equiv H_0/100$. 

The last two columns of Table~\ref{clusters} contain the original
richness classification (where available) as well as a revised richness based on the current
2dFGRS catalogue of clusters. 
Based on the memberships in the present catalogue,
it can be seen that under the strict definition of the richness classes
proposed by Abell \citep{abell}, 9 of the 25 clusters in this study would be classified 
as ``large groups'' instead
of low richness clusters. Nevertheless, these structures are included in this study under the (loose)
heading of ``low richness clusters''. Also note that since the membership completeness here is estimated at $80-90\%$,
it is possible that 100\% completeness would alter these richness estimates. 
\begin{table*}
\begin{minipage}{115mm}
\caption{The sample of 2dFGRS clusters selected for substructure analysis. Where available,
and using the original definitions given by Abell,
the original ACO richness classification R is given 
(\citealt{abell} and 
\citealt{aco})
along with the richness class determined
from the 2dFGRS catalogue of clusters. The number in parentheses in the $\mathrm{R}_{2dF}$ column
is the number of galaxies between $m_3$ and $m_3 + 2$ in the current catalogue. The proper distances
$d_p(t_e)$ and $d_p(t_0)$ are given in $h^{-1} \, \mathrm{Mpc}$.}
\label{clusters}
\begin{center}
\begin{tabular}{lrrr@{,~}l@{,~}rr@{,~}l@{,~}rcccr}
\hline\\
\multicolumn{2}{c}{Cluster} & ~~N~~ & \multicolumn{3}{c}{RA(J2000)} & \multicolumn{3}{c}{~~~Dec(J2000)}
& $d_p(t_e)$ & $d_p(t_0)$ & $\mathrm{R}_{ACO}$ & $\mathrm{R}_{2dF}$\\
\hline\hline
Abell &   930 &  ~~91~~ & 10 & ~~7& ~1.32 &~~ $-5$ & 37 & 28.7 & 162 & 171 & 0  &  0~(46) \\
Abell &   957 &  ~~90~~ & 10 & 13 & 38.42 &~~ $-0$ & 55 & 32.6 & 128 & 134 & 1  &  0~(35) \\
Abell &  1139 & ~~106~~ & 10 & 58 & 10.98 &~~  1 & 36 & 16.4 & 113 & 118 & 0  &  $<0$~(20)\\
Abell &  1238 &  ~~86~~ & 11 & 22 & 54.36 &~~  1 & ~6 & 51.2 &  203 & 218 & 1  &  $<0$~(23)\\
Abell &  1620 &  ~~95~~ & 12 & 50 & ~3.89 &~~ $-1$ & 32 & 26.7 & 231 & 250 & 0  &  0~(49) \\
Abell &  1663 &  ~~94~~ & 13 & ~2 & 52.55 &~~ $-2$ & 31 & ~4.0 &  225 & 243 & 0  &  1~(72) \\
Abell &  1750 &  ~~78~~ & 13 & 31 & 11.00 &~~ $-1$ & 43 & 42.0 &  232 & 251 & 0  &  1~(51) \\
Abell &  2734 & ~~125~~ & ~0 & 11 & 21.60 &~~$-28$ & 51 & 16.6 &  172 & 183 & 1  &  1~(72) \\
Abell &  2814 &  ~~87~~ & ~0 & 42 & ~8.82 &~~$-28$ & 32 & ~8.8 &  284 & 315 & 1  &  2~(83) \\
Abell &  3027 &  ~~91~~ & ~2 & 30 & 49.41 &~~$-33$ & ~6 & 11.7 &  211 & 228 & 0  &  1~(51) \\
Abell &  3094 & ~~108~~ & ~3 & 11 & 25.00 &~~$-26$ & 55 & 52.2 &  188 & 201 & 2  &  $<0$~(23) \\
Abell &  3880 & ~~120~~ & 22 & 27 & 54.49 &~~$-30$ & 34 & 31.3 &  161 & 171 & 0  &  $<0$~(25) \\
Abell &  4012 &  ~~73~~ & 23 & 31 & 50.89 &~~$-34$ & ~3 & 16.6 &  152 & 160 & 0  &  0~(44)  \\
Abell &  4013 &  ~~85~~ & 23 & 30 & 22.62 &~~$-34$ & 56 & 48.3 &  154 & 162 & 1  &  0~(30)  \\
Abell &  4038 & ~~154~~ & 23 & 47 & 34.92 &~~$-28$ & ~7 & 29.3 &  ~87 & ~89  & 2  &  0~(39)  \\
Abell & S141 & ~~110~~ & ~1 & 13 & 47.09 &~~$-31$ & 44 & 53.0 &  ~57 & ~58  & 0  &  0~(33)  \\
Abell & S258 &  ~~87~~ & ~2 & 25 & 44.44 &~~$-29$ & 36 & 57.5 &  168 & 178 & 0  &  0~(36)  \\
Abell & S301 &  ~~95~~ & ~2 & 49 & 33.72 &~~$-31$ & 11 & 23.2 &  ~65 & ~66  & 0  &  $<0$~(21) \\
Abell & S333 &  ~~74~~ & ~3 & 15 & ~9.95 &~~$-29$ & 14 & 37.4 &  185 & 197 & 0  &  0~(47)  \\
Abell & S1043 & ~~111~~ & 22 & 36 & 27.96 &~~$-24$ & 20 & 30.8 &  105 & 109 & 0  &  $<0$~(22) \\
APM   &   268 &  ~~97~~ & ~2 & 29 & 55.78 &~~$-33$ & 10 & 37.2 &  212 & 228 &    &  0~(31)  \\
APM   &   917 &  ~~77~~ & 23 & 41 & 35.49 &~~$-29$ & 14 & 10.9 &  144 & 151 &    &    $<0$~(22) \\
APM   &   933 & ~~124~~ & 23 & 56 & 27.66 &~~$-34$ & 35 & 35.0 &  140 & 147 &    &    $<0$~(28) \\
EDCC  &   365 &  ~~80~~ & 23 & 55 & ~8.44 &~~$-32$ & 44 & 26.0 &  165 & 175 &    &    $<0$~(28) \\
EDCC  &   442 & ~~127~~ & ~0 & 25 & 31.36 &~~$-33$ & ~2 & 47.6 &  140 & 147 &    &    0~(36)  \\
\hline
\end{tabular}
\end{center}
\end{minipage}
\end{table*}

Among the previous substructure investigations involving relatively large samples of clusters is
the study by \citet{solanes}.
This work considered subclustering in 67 rich clusters contained
in the ESO Nearby Abell Cluster Survey (ENACS) catalogue.
Using the best data available at the time, the ENACS catalogue
is a reasonably homogeneous dataset with a completeness of $\sim 60-80\%$ (fraction of cluster members
with redshifts), less but similar to that of
the 25 2dFGRS clusters selected for this study. The completeness for the sample here is
estimated to be $\sim 80 - 90\%$ for each cluster, and the sample contains all clusters in
the 2dFGRS catalogue with $N \ge 70$.
Two important differences between the ENACS study and this work are (1) the different test suites
used to test for substructure, and (2) the average
number of cluster members in the ENACS study is significantly less than here:
the average membership
in the ENACS study was $N = 45$ with 48 of the 67 clusters having $N < 50$, whereas the average
membership number for this study is $N = 99$ with the minimum number being $N = 73$. This is
important due to the dependence of the reliability and sensitivity of the statistical tests as a function
of $N$. 

The paper is organized as follows.
Section 2 presents the method by which the ellipticity parameters, core radius, and central density
are estimated. The tests
used to detect the presence of substructure are described in Section 3 with a 
detailed analysis of the individual clusters presented in Section 4. A summary of the results
and a concluding discussion are given in Section 5. 
Finally, Appendix 1 contains visualization plots for each cluster.

\begin{table*}
\begin{minipage}{115mm}
\caption{Cluster Ellipticity and Position Angles.
Position angles are measured counterclockwise from north (positive y-axis).}
\label{ellipt_table}
\begin{center}
\begin{tabular}{lrrrrrrrrrrr}
\hline\\
&\multicolumn{11}{c}{R $=\sqrt{AB} \; h^{-1}$ Mpc}\\
%\cline{2-12}
%&\multicolumn{2}{c}{$\sqrt{AB}=0.50$}&~
&\multicolumn{2}{c}{0.50}&~
&\multicolumn{2}{c}{0.75}&~
&\multicolumn{2}{c}{1.00}&~
&\multicolumn{2}{c}{1.25}\\
\cline{2-3}\cline{5-6}\cline{8-9}\cline{11-12}
Cluster & $\epsilon$ & PA &
& $\epsilon$ & PA &
& $\epsilon$ & PA &
& $\epsilon$ & PA \\
\hline
Abell 930   & 0.29 &  75 && 0.41 &   4 && 0.43 &  16 && 0.45 &   9 \\
Abell 957   & 0.68 &  74 && 0.47 &  70 && 0.49 &  61 && 0.35 &  42 \\
Abell 1139  & 0.25 & 117 && 0.42 &  98 && 0.20 &  62 && 0.04 &  73 \\
Abell 1238  & 0.34 &  95 && 0.29 &  19 && 0.22 &  14 && 0.05 &   5 \\
Abell 1620  & 0.08 &   1 && 0.49 &  17 && 0.50 &  39 && 0.17 &  41 \\
Abell 1663  & 0.69 &  68 && 0.39 &  80 && 0.33 &  59 && 0.14 &  36 \\
Abell 1750  & 0.43 &   4 && 0.74 &  12 && 0.62 &  34 && 0.54 &  39 \\
Abell 2734  & 0.55 & 114 && 0.23 & 104 && 0.22 & 109 && 0.07 &  95 \\
Abell 2814  & 0.30 & 100 && 0.27 & 130 && 0.50 & 127 && 0.40 & 120 \\
Abell 3027  & 0.65 & 166 && 0.59 & 179 && 0.53 &  21 && 0.25 &  33 \\
Abell 3094  & 0.55 &   5 && 0.49 & 160 && 0.37 & 141 && 0.33 & 138 \\
Abell 3880  & 0.59 & 150 && 0.50 & 159 && 0.16 & 156 && 0.20 & 175 \\
Abell 4012  & 0.42 &  14 && 0.37 &   1 && 0.36 & 118 && 0.01 & 133 \\
Abell 4013  & 0.29 & 163 && 0.34 & 145 && 0.20 & 116 && 0.18 &  78 \\
Abell 4038  & 0.43 &  21 && 0.36 &  18 && 0.21 & 156 && 0.50 &  96 \\
Abell S141  & 0.67 &   0 && 0.66 & 168 && 0.63 & 169 && 0.58 & 168 \\
Abell S258  & 0.75 & 114 && 0.69 & 113 && 0.71 & 104 && 0.58 & 104 \\
Abell S301  & 0.36 & 161 && 0.45 &  32 && 0.61 &  48 && 0.54 &  48 \\
Abell S333  & 0.59 & 152 && 0.49 & 156 && 0.47 & 131 && 0.41 & 132 \\
Abell S1043 & 0.60 & 152 && 0.43 & 169 && 0.03 & 162 && 0.32 &  16 \\
APM 268     & 0.71 & 172 && 0.52 &   4 && 0.35 &   3 && 0.20 & 177 \\
APM 917     & 0.45 &  27 && 0.33 &  17 && 0.27 &  22 && 0.37 &  56 \\
APM 933     & 0.53 & 145 && 0.40 & 128 && 0.03 & 160 && 0.30 &  17 \\
EDCC 365    & 0.71 & 150 && 0.65 & 154 && 0.52 & 159 && 0.47 & 163 \\
EDCC 442    & 0.34 &  33 && 0.54 &  20 && 0.58 &  38 && 0.45 &  35 \\
\hline
\end{tabular}
\end{center}
\end{minipage}
\end{table*}

\section{Global Cluster Properties}

Prior to applying tests for substructure, it is necessary to characterize each cluster
in terms of global parameters such as a core radius and maximum central density, mean
velocity and velocity dispersion, and global ellipticity and position angle.
Here, ``global'' refers to the calculation of properties over an angular area that contains most or all
cluster members. First, the cluster selection and foreground/background rejection methods are briefly
summarized. 

Essentially, this first catalogue of 2dFGRS clusters is a preliminary catalogue based on comparing
the 2dFGRS results with previously identified clusters of galaxies
sourced from the catalogues of Abell \citep{abell} and the revised/supplemented Abell
\citep{aco}, APM \citep{dalton}, and EDCC \citep{lumsden}.
The 2dFGRS catalogue was searched to identify clusters with centroids given in
the above catalogues within $1^\circ$ of the centre of an observed survey tile. If the
centroid of a catalogued cluster was found in a 2dFGRS tile, the 2dFGRS redshift catalogue was then searched for
objects within a specified radius of the centroid. This process isolates a cone in redshift space
containing candidate cluster members as well as foreground and background galaxies. The list of candidate
members was then refined by inspection of the Palomar Observatory Sky Survey (POSS) plates.
Finally, foreground and background contamination was eliminated (or significantly reduced)
through subsequent detailed analysis of the
redshift cone diagrams and, when necessary, redshift histograms.
Further details concerning cluster selection can be found in \citet{depropris}.

\subsection{Cluster Centroid and Ellipticity}

Each galaxy position in $(RA, Dec)$ is converted to
cartesian $(x,y)$ coordinates with relative separations calculated using 
the proper distance at photon time of emission, $d_p(t_e)$.
The cluster centroid is determined using an iterative search algorithm. A trial centre is computed
from the arithmetic mean of all galaxies in the cluster, and
is then recalculated using only the galaxies within an approximately
0.5 $h^{-1}$ Mpc radius of the initial centre.  This process is repeated until
it converges and the final centre determined. The reason for restricting the centroid computation
in this manner is to avoid bias from large substructures well removed from the central
regions.

For a quantitative measurement of cluster shape, it is convenient to use
the dispersion ellipse of the bivariate normal frequency function of position vectors 
(see \citealt{trumpler}) and first used to study cluster ellipticity by \citet{carter}.
The dispersion ellipse is defined as the contour at which
the density is 0.61 times the maximum density of a set of points distributed normally with
respect to two correlated variables although the assumption of normally distributed coordinates
is not necessary to achieve accurate shape parameters. Using the moments 
\alpheqn
\begin{equation}
\mu_{10} = \bar{x}
\end{equation}
\begin{equation}
\mu_{01} = \bar{y}
\end{equation}
\begin{equation}
\mu_{20} = \frac{1}{N}\sum_{i=1}^N x^2_i - \bar{x}^2
\end{equation}
\begin{equation}
\mu_{02} = \frac{1}{N}\sum_{i=1}^N y^2_i - \bar{y}^2
\end{equation}
\begin{equation}
\mu_{11} = \frac{1}{N}\sum_{i=1}^N x_iy_i - \bar{x}\bar{y} \; \; \mbox{,}
\end{equation}
\reseteqn
the semi-principal axes of the ellipse $\Gamma_A$ and $\Gamma_B$ are the solutions of
\begin{equation}
\left| \begin{array}{cc}
\mu_{20}-\Gamma^2 & \mu_{11} \\
\mu_{11}     & \mu_{02}-\Gamma^2 \; \; \mbox{.}\\
\end{array} \right|
= 0
\end{equation}
The position angle of the major axis with respect to north is then
\begin{equation}
\theta = \cot^{-1}\left( -\frac{\mu_{02} - \Gamma_A^2}{\mu_{11}}\right) + \frac{\pi}{2}
\end{equation}
with $\Gamma_A$ ($ > \Gamma_B$) being the semi-major axis, and the ellipticity is
\begin{equation}
\epsilon = 1 - \frac{\Gamma_B}{\Gamma_A} \; \; \mbox{.}
\end{equation}

To achieve the best results, an iterative approach is adopted here similar to that used in
previous studies such as \citet{carter} or \citet{burgett}.
An initial circle of a given radius $R$ is defined
about a trial centre and a new centre and dispersion ellipse are then calculated using only
the galaxies contained within the circle. The semi-major and semi-minor axes $A$ and $B$ of the new
ellipse are constrained to satisfy $R = \sqrt{A\,B}$. The process is repeated
until convergence is achieved (usually requiring less than four iterations). It is
possible for the iterative solution to become trapped in a local extremum, 
but this can be avoided by tracking each iteration. In applying the algorithm to the 18
richest clusters in the Dressler 1980 catalogue, it was noted that the results
are sensitive to the presence of substructure \citep{burgett}. The ellipticities and position
angles for the 25 clusters are shown in Table \ref{ellipt_table} 
for four different average distances
from the cluster centroids; the dispersion ellipses for mean radii of 0.5 and 1.0 
$\mathrm{h}^{-1}$ Mpc are superimposed on the plots of galaxy positions 
in Appendix A for each cluster
analyzed in Section 4.

Since a model density
distribution can be significantly affected
by ellipticity, this must be accounted for when constructing Monte Carlo catalogues of random clusters for
comparison against actual data. Also note that
it is sometimes nontrivial to distinguish true ellipticity from subcluster bias, and, in fact,
the algorithm used to compute ellipticity can be used to also probe for substructure. 

\begin{table*}
\begin{minipage}{115mm}
\caption{Cluster Core Radii and Central Densities. The errors quoted are for the particular fit.}
\label{core}
\begin{center}
\begin{tabular}{lrcccr}
\hline\\
& & $A_c$ & $B_c$ & $R_c \equiv \sqrt{A_c B_c}$ & $\sigma_0$~~~~~~~~~ \\
\multicolumn{2}{c}{Cluster} & ($h^{-1}$ Mpc) & ($h^{-1}$ Mpc) & ($h^{-1}$ Mpc)
& {gal}/($h^{-1}$ Mpc)$^2$ \\
\hline\hline
Abell &   930  & ~~0.62~~ & ~~0.35~~ & ~$0.47 \pm 0.01$ &  ~$61 \pm 2$~~~~~ \\
Abell &   957  & ~~0.25~~ & ~~0.10~~ & ~$0.16 \pm 0.01$ & ~$318 \pm 1$~~~~~ \\
Abell &  1139  & ~~0.33~~ & ~~0.21~~ & ~$0.26 \pm 0.01$ & ~$138 \pm 4$~~~~~ \\
Abell &  1238  & ~~0.50~~ & ~~0.35~~ & ~$0.41 \pm 0.02$ &  ~$68 \pm 3$~~~~~ \\
Abell &  1620  & ~~0.57~~ & ~~0.40~~ & ~$0.47 \pm 0.02$ &  ~$59 \pm 3$~~~~~ \\
Abell &  1663  & ~~0.65~~ & ~~0.32~~ & ~$0.46 \pm 0.03$ &  ~$57 \pm 4$~~~~~ \\
Abell &  1750  & ~~0.61~~ & ~~0.25~~ & ~$0.39 \pm 0.03$ &  ~$41 \pm 4$~~~~~ \\
Abell &  2734  & ~~0.40~~ & ~~0.24~~ & ~$0.31 \pm 0.01$ & ~$121 \pm 5$~~~~~ \\
Abell &  2814  & ~~0.43~~ & ~~0.30~~ & ~$0.36 \pm 0.01$ &  ~$81 \pm 4$~~~~~ \\
Abell &  3027  & ~~0.49~~ & ~~0.17~~ & ~$0.29 \pm 0.01$ &  ~$90 \pm 3$~~~~~ \\
Abell &  3094  & ~~0.53~~ & ~~0.29~~ & ~$0.39 \pm 0.01$ &  ~$88 \pm 2$~~~~~ \\
Abell &  3880  & ~~0.34~~ & ~~0.18~~ & ~$0.25 \pm 0.01$ & ~$161 \pm 13$~~~~~ \\
Abell &  4012  & ~~0.46~~ & ~~0.17~~ & ~$0.28 \pm 0.01$ &  ~$71 \pm 3$~~~~~ \\
Abell &  4013  & ~~0.19~~ & ~~0.14~~ & ~$0.17 \pm 0.01$ & ~$224 \pm 2$~~~~~ \\
Abell &  4038  & ~~0.23~~ & ~~0.15~~ & ~$0.19 \pm 0.01$ & ~$261 \pm 15$~~~~~ \\
Abell & S141  & ~~0.44~~ & ~~0.15~~ & ~$0.26 \pm 0.02$ &  ~$158 \pm 18$~~~~~ \\
Abell & S258  & ~~0.51~~ & ~~0.15~~ & ~$0.28 \pm 0.04$ &  ~$108 \pm 1$~~~~~ \\
Abell & S301  & ~~0.34~~ & ~~0.17~~ & ~$0.24 \pm 0.01$ &  ~$133 \pm 9$~~~~~ \\
Abell & S333  & ~~0.52~~ & ~~0.24~~ & ~$0.35 \pm 0.01$ &   ~$62 \pm 2$~~~~~ \\
Abell & S1043  & ~~0.44~~ & ~~0.22~~ & ~$0.31 \pm 0.01$ &  ~$107 \pm 3$~~~~~ \\
APM   &   268  & ~~0.44~~ & ~~0.20~~ & ~$0.30 \pm 0.01$ &  ~$81 \pm 3$~~~~~ \\
APM   &   917  & ~~0.17~~ & ~~0.12~~ & ~$0.14 \pm 0.01$ & ~$270 \pm 1$~~~~~ \\
APM   &   933  & ~~0.49~~ & ~~0.27~~ & ~$0.36 \pm 0.01$ &  ~$91 \pm 2$~~~~~ \\
EDCC  &   365  & ~~0.56~~ & ~~0.25~~ & ~$0.37 \pm 0.01$ &  ~$70 \pm 2$~~~~~ \\
EDCC  &   442  & ~~0.47~~ & ~~0.24~~ & ~$0.33 \pm 0.01$ & ~$113 \pm 4$~~~~~ \\
\hline
\end{tabular}
\end{center}
\end{minipage}
\end{table*}

\subsection{Density Profiles and Core Fitting}

The assumed models for the density and velocity distributions are those 
for a cluster in dynamical and thermal equilibrium, viz., the empirical King
approximation to an isothermal sphere and a Gaussian velocity distribution.
It should be remembered that while the projected number density distributions of many
clusters are fit well by King profiles, the core radius and central density are
fitting variables that are convenient, but arbitrary, parameters for characterising 
the compactness of cluster central regions. In particular, the core radius has
no intrinsic dynamical significance. Thus, the results presented in Table~\ref{core} should
be regarded as comparative geometrical measures across the sample without dynamical implications.
Similar considerations apply to the usage of the term `core' throughout this paper.

For a circularly symmetric cluster projection, the two-dimensional surface 
King profile \citep{king} takes the form
\begin{equation}
\sigma(r) = \frac{\sigma_0}{\displaystyle 1 + \left( \frac{r}{R_c} \right)^2} \; \; \mbox{,}
\label{kingcirc}
\end{equation}
where $\sigma(r)$ is the galaxy density at a distance r from the centre of the cluster,
$\sigma_0$ is the maximum galaxy density of the cluster, and $R_c$ is the core radius of
cluster. 
By extension, the two-dimensional King profile for an elliptically symmetric surface density distribution is 
\begin{equation}
\sigma(x,y) = \frac{\sigma_0}{\displaystyle 1 + \left( \frac{x}{A} \right)^2 
 + \left( \frac{y}{B} \right)^2}
\label{kingellip}
\end{equation}
where $A$ and $B$ are the semi-major and semi-minor axes, respectively, and where it is conventional
to define an equivalent mean core radius, $R_c = \sqrt{A\,B}$. 

As a starting point, values of $R_c$ and $\sigma_0$ for the circular profile of Eqn. (\ref{kingcirc})
are calculated using a least squares method. First, the expected galaxy
count $n(r)$ at a given distance from the centre is found by integrating Eqn. (\ref{kingcirc}),
\begin{equation}
n(r) = \pi R^2_c \sigma_0 \ln{\left[ (R^2_c + r^2)/R^2_c \right]} \; \; \mbox{,}
\label{1s1}
\end{equation}
\noindent
and inverting for the radius $r(n)$ of the circle containing $n$ galaxies, 
\begin{equation}
r(n) = R_c \sqrt{\exp{\left( n / \pi R^2_c \sigma_0 \right)} - 1} \; \; \mbox{.}
\label{ls2}
\end{equation}
\noindent
Each galaxy then becomes a data point in the merit function 
\begin{equation}
\chi^2 (R_c,\sigma_0) = \sum^N_{i=1} \left( \frac{r_i - r(i)}{\sigma_i} \right)^2 \; \; \mbox{,}
\label{ls5}
\end{equation}
\noindent
where $r_i$ is the actual distance of the $i$th galaxy from the centre of the
cluster and $r(i)$ is given by Eqn. (\ref{ls2}).  The minimisation of the 
merit function requires the derivatives 
with respect to parameters $R_c$ and $\sigma_0$ to be set equal to zero,
\alpheqn
\begin{equation}
\frac{\partial \chi^2}{\partial R_c} = 2 \sum^N_{i=1} \left(
\frac{r_i -  r(i)}{\sigma^2_i} \right) \left( \frac{\partial r}{\partial R_c}
\right) = 0 \quad\mbox{,}
\label{ls6}
\end{equation}
and
\begin{equation}
\frac{\partial \chi^2}{\partial \sigma_0} = 2 \sum^N_{i=1} \left(
\frac{r_i -  r(i)}{\sigma^2_i} \right) \left( \frac{\partial r}{\partial \sigma_0}
\right) = 0 \quad\mbox{,}
\label{ls7}
\end{equation}
\reseteqn
where
\alpheqn
\begin{equation}
\frac{\partial r}{\partial R_c} = \frac
{(\pi \sigma_0 R^2_c - n) \exp{(n/ \pi R^2_c \sigma_0)} - \pi R^2_c \sigma_0}
{\pi \sigma_0 R^2_c \sqrt{\exp{(n/ \pi R^2_c \sigma_0)} - 1}} \quad\mbox{,}
\label{ls8}
\end{equation}
and
\begin{equation}
\frac{\partial r}{\partial \sigma_0} = \frac
{ - n \exp{(n/ \pi R^2_c \sigma_0)} - \pi R^2_c \sigma_0}
{2 \pi \sigma_0 R^2_c \sqrt{\exp{(n/ \pi R^2_c \sigma_0)} - 1}} \quad\mbox{.}
\label{ls9}
\end{equation}
\noindent
\reseteqn
The Levenberg-Marquardt algorithm is then used to find the best fit for $R_c$ and 
$\sigma_0$~\citep{press}. 

\begin{figure*}
\begin{minipage}{115mm}
\centering
\epsfig{figure=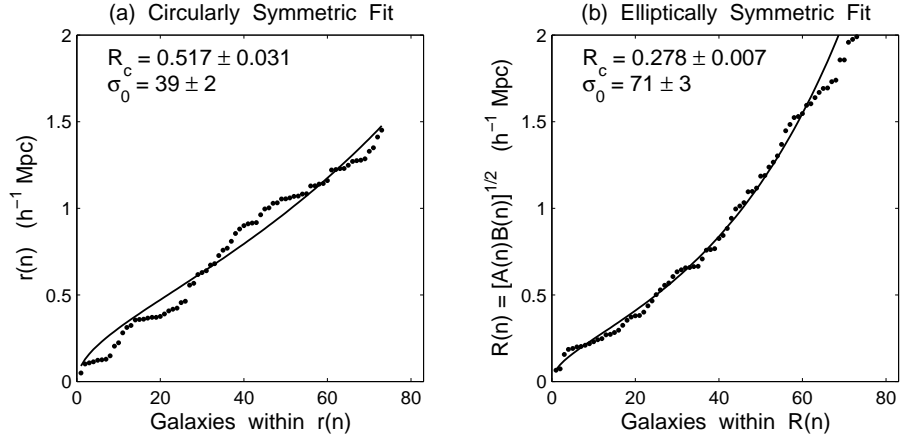,height=5.975cm}
\caption{Comparison of the circular and elliptical King fits for Abell 4012.
The solid line represents the expected count $n$ of galaxies
for the fitted density model within either a circle of radius $r(n)$ (circular
King profile) or a circle of geometric mean
radius $R(n) \equiv [A(n) \, B(n)]^{1/2}$ for an ellipse with semimajor and
semiminor axes $A(n)\mbox{,} \, B(n)$ (elliptical King profile).
The points in each plot represent a galaxy at a specified distance from
the cluster centroid. Points above the fit line are outside the test circle,
but should have been inside at that distance; points underneath the fit line
are within the test circle, but yield a higher density than the model density
at that distance.}
\label{a4012_fits}
\end{minipage}
\end{figure*}

The circular fit above requires modification when the cluster core is highly elongated or multimodal.
While it is straightforward to extend the fitting method above for the four free parameters in the
elliptical King profile of Eqn.~(\ref{kingellip}), it is possible to employ a simpler approach.
If an ellipse with ellipticity $\epsilon$
is constructed such that $\sqrt{AB} = 1 h^{-1}$ Mpc $\equiv \sqrt{A_{R=1}B_{R=1}} \equiv R_{=1}$,
then the coordinates of the perimeter satisfy
\begin{equation}
\frac{x^2}{A_{R=1}^2} + \frac{y^2}{B_{R=1}^2} = 1 \; \; \mbox{.}
\end{equation}
Given the coordinates $(x_i,y_i)$ of the $i$th galaxy in the cluster, there exists a 
concentric ellipse which includes the coordinate $(x_i,y_i)$ defined by
\begin{equation}
\frac{x_i^2}{A_i^2} + \frac{y_i^2}{B_i^2} = 1
\end{equation}
This ellipse is defined by $\epsilon_i = \epsilon$ and $R_i = \sqrt{A_i B_i}$  
with $R_i$ satisfying
\begin{equation}
R_i = R_{=1}\sqrt{\frac{x_i^2}{A_{R=1}^2} + \frac{y_i^2}{B_{R=1}^2}} \mbox{.}
\end{equation}

The circular core fitting technique described above can now be utilized in the following manner:
\begin{enumerate}
\item
Calculate values for the semi-major axis $A_{R=1}$ and the
semi-minor axis $B_{R=1}$ using values for the ellipticity $\epsilon$
from Table \ref{ellipt_table} (this requires a certain amount of judgement to
select the best values),
\item
Transform the $(x_i,y_i)$ coordinates for each galaxy to $(x_i^\prime,y_i^\prime)$ for the coordinate system
rotated to have the $x^\prime$-axis along the semi-major axis of the ellipse,
\item
Calculate for all galaxies the quantity
\begin{equation}
R_i  = R_{=1}\sqrt{(x_i^\prime)^2/A_{R=1}^2 + (y_i^\prime)^2/B_{R=1}^2}\quad\mbox{,}
\end{equation}
\item
Sort the values of $R_i$ in ascending order.  This new order of galaxies,
denoted by $n$, along with their values for $R_i(n)$, gives the number
of galaxies found on or within the ellipse where $R$ = $R_i(n)$,
\item
Use the above values for $R_i(n)$ in place of $r(n)$ used by the circular
fitting algorithm to obtain new values for $R_c$ and $\sigma_0$ 
from the least-squares fit,
\item 
Recompute $A_c$ and $B_c$ using the original $\epsilon = 1 - B/A$ and $R_c = \sqrt{A_c\,B_c}$.  
\end{enumerate}
\indent
The results for elliptical surface density profile fits are shown in Table \ref{core} for
the 25 clusters. 
Comparisons of the results for the circular and elliptical fits 
shows that the fit to a circular profile provides a fair representation of the number
distribution even when the actual shape is noticeably elongated. On average, compared to the elliptical
profile, the circular fit also is better over the extent of the entire cluster. This can be due to
the cluster ellipticity
generally decreasing as a function of distance from the centroid as well as the circular fit essentially 
averaging over clumpy distributions that might bias the elliptical fit.
However, the elliptical fit reproduces the density
profile in the central regions significantly better than the circular fit for clusters such as, e.g., Abell 4012 
(Figure \ref{a4012_fits}). 

For some clusters containing multi-component core regions such as Abell 930, the circular
fit is slightly better than the elliptical fit in the central regions. This is presumably due to 
both components containing approximately the same number of galaxies so that the computed
centroid is not at the centre of either component. Further detailed comparisons and fitting plots for all clusters
can be found in ~\cite{vick}.

\section{Tests for Cluster Substructure}

The dangers of inferring the presence of substructure based on visual inspection of contour or
gray scale plots alone is well known, yet this remains an appropriate component of a comprehensive
test suite in order to better guide and interpret a statistical analysis. 
Two early studies utilizing contour plots to detect possible
substructure in clusters of galaxies were conducted by \citet{geller} and \citet{burgett}.
In creating contour plots of clusters drawn from the \citet{dressler_1} catalogue, 
the Geller and Beers study applied a ``boxcar smoothing'' technique that was a best-available
technique at that time. The improvements gained
using an adaptive kernel approach to contour plotting 
are shown for many of these same clusters in \citet{kriessler}. 
In addition to number density contours, the Burgett study presented luminosity-weighted contour plots
of clusters in an attempt to correlate light with mass.
As substructure studies have evolved with experience and numerical sophistication,
it is now conventionally accepted
that a synergistic combination of quantitative tests is necessary to detect different
types of structure as well as to mitigate the results of false
positives occasionally given by almost all estimators.

The test suite adopted for this study
consists of a variety of one-, two-, and three-dimensional tests.  These include
visualization plots such as contour and nearest neighbor plots, velocity distribution statistics,
and the $\alpha$ test \citep{west_1}, $\beta$ test \citep{west_3}, and $\kappa$ test 
\citep{colless_3}. 
For the latter three statistical tests, and for all clusters, a sample of 10,000 simulations was used to normalize
the test and to determine the statistical significance.
Use of the two point angular correlation function and the Fourier elongation test
were considered but rejected on the grounds that they
would not provide any new information compared to the chosen test suite. 

\begin{figure}
\centering
\epsfig{figure=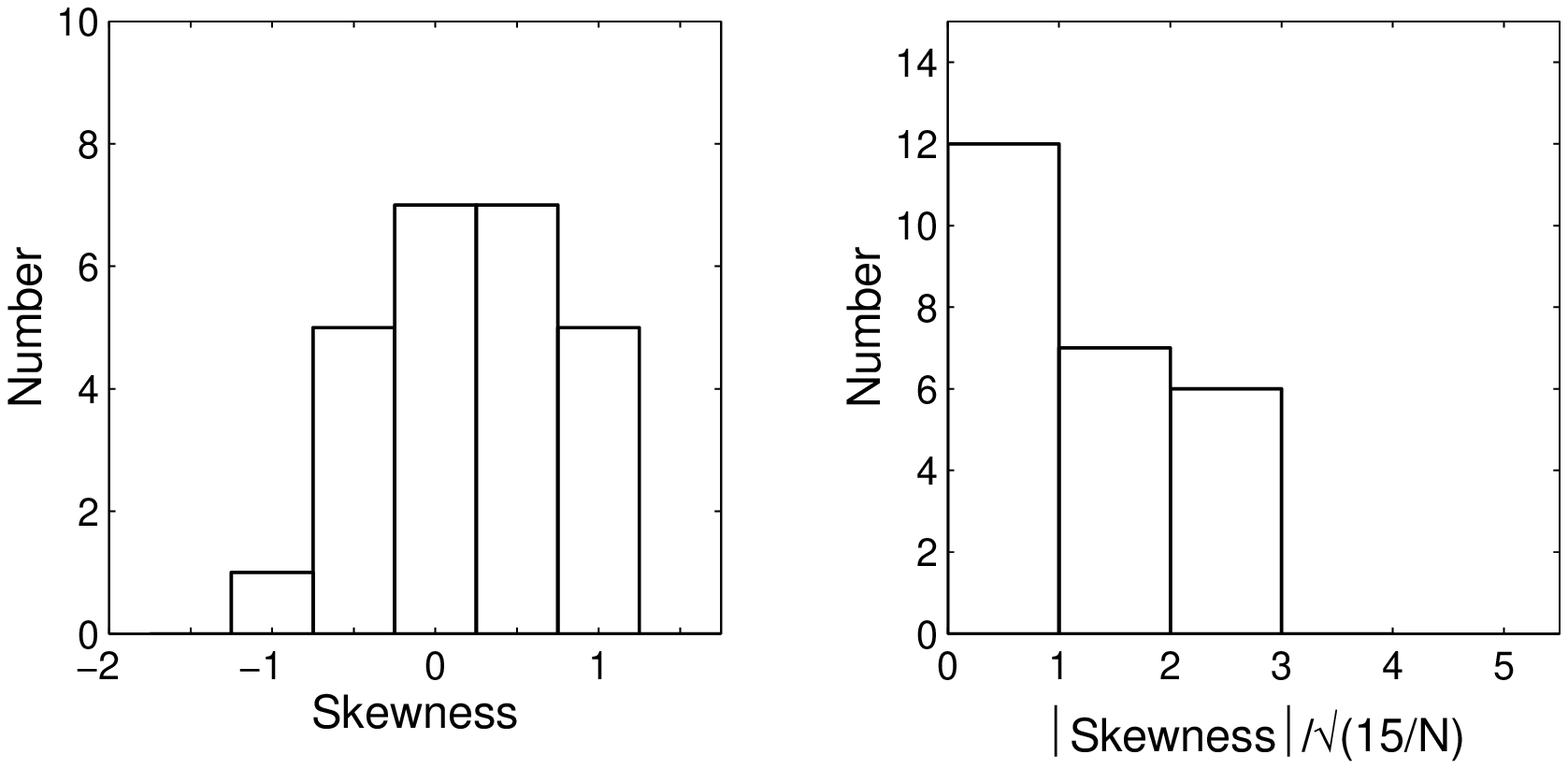, width=0.48\textwidth}
\caption{The skewness values for the 25 clusters with the standard deviation of
the computed skewness value for a single cluster being approximately $\sqrt{15/N}$.
From the figures it is evident that there is an excess of high skewness values over
that expected if they were normally distributed.}
\label{skew2}
\end{figure}
\begin{figure}
\centering
\epsfig{figure=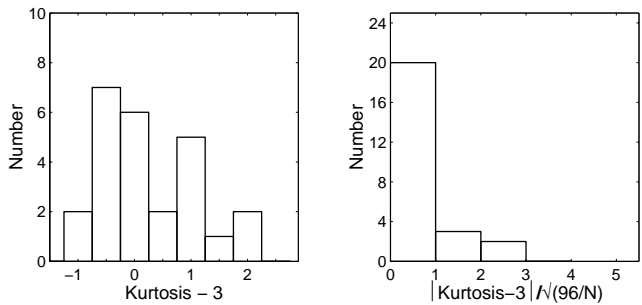, width=0.48\textwidth}
\caption{The kurtosis values for the 25 clusters with the standard deviation of
the computed kurtosis value for a single cluster being approximately $\sqrt{96/N}$.
From the figures it is evident that the velocity distributions
for the majority cannot be distinguished from Gaussian distributions based solely on kurtosis values.}
\label{kurt2}
\end{figure}

\begin{table*}
\begin{minipage}{115mm}
\caption{Statistical moments of the cluster velocity distributions, equivalent standard
deviations for the skewnness and kurtosis, and the confidence level (CL) derived from the
$\chi^2_{vel}$ result for consistency with a Gaussian distribution.  The entry for Abell S333*
presents the velocity statistics for Abell S333 without the 8 galaxies
in the group having $cz < 18,500$ km/s, and the entry for Abell 4013* presents the velocity
statistics without the 14 galaxies having $ cz > 17,500$ km/s.}
\label{velstats}
\begin{center}
\begin{tabular}{rlrrcrcc}
\hline\\
\multicolumn{2}{c}{Cluster} & $\sigma_{vel}$~~ &
~Skewness~ & ~~\# of ~~ & ~Kurtosis~ & ~~\# of ~~ & CL to reject\\
\hfill & \hfill & (km/s) & \hfill & $\sigma_S$ & \hfill & $\sigma_K$ & Gaussian (\%)\\
 \hline\hline
 Abell & ~~930  &  856~~ &  0.30~~ & ~~~0.7~ & $-1.00$~~~ & ~~1.0~ & $>$99.9\\
 Abell & ~~957  &  708~~ &  0.31~~ & ~~~0.8~ & $-0.45$~~~ & ~~0.4~ & 49\\
 Abell & ~1139  &  484~~ & $-0.21$~~ & ~~~0.6~ &  0.18~~~ & ~~0.2~ & 76\\
 Abell & ~1238  &  551~~ &  0.30~~ & ~~~0.7~ &  0.04~~~ & $<0.1$~ & 14\\
 Abell & ~1620  & 1001~~ &  0.77~~ & ~~~1.9~ &  0.65~~~ & ~~0.6~ & 97\\
 Abell & ~1663  &  881~~ & $-0.60$~~ & ~~~1.5~ &  0.86~~~ & ~~0.8~ & 36\\
 Abell & ~1750  &  897~~ &  0.43~~ & ~~~1.0~ & $-0.87$~~~ & ~~0.8~ & 98\\
 Abell & ~2734  &  984~~ &  0.71~~ & ~~~2.1~ &  1.38~~~ & ~~1.6~ & 85\\
 Abell & ~2814  &  894~~ & $-0.14$~~ & ~~~0.3~ &  0.09~~~ & ~~0.1~ & 68\\
 Abell & ~3027  &  838~~ & $-0.57$~~ & ~~~1.4~ & $-0.76$~~~ & ~~0.7~ & ~~99.9\\
 Abell & ~3094  &  728~~ &  0.42~~ & ~~~1.1~ & $-0.33$~~~ & ~~0.4~ & 93\\
 Abell & ~3880  &  784~~ & $-0.33$~~ & ~~~0.9~ & $-0.02$~~~ & $<0.1$~ & 36\\
 Abell & ~4012  &  471~~ & $-0.08$~~ & ~~~0.2~ &  0.37~~~ & ~~0.3~ & 96\\
 Abell & ~4013  &  854~~ &  0.93~~ & ~~~2.2~ & $-0.12$~~~ & ~~0.1~ & $>$99.9\\
 Abell & ~4013* &  455~~ &  0.07~~ & ~~~0.2~ & $0.15$~~~ & ~~0.1~ & 97\\
 Abell & ~4038  &  835~~ & $-0.04$~~ & ~~~0.1~ & $-0.31$~~~ & ~~0.4~ & 96\\
 Abell & ~S141  &  403~~ &  0.82~~ & ~~~2.2~ &  0.58~~~ & ~~0.6~ & 98\\
 Abell & ~S258  &  557~~ &  0.07~~ & ~~~0.2~ & $-0.52$~~~ & ~~0.5~ & 70\\
 Abell & ~S301  &  679~~ &  $-0.01$~~ & $<0.1$~ &  1.92~~~ & ~~1.9~ & ~~99.9\\
 Abell & ~S333  &  933~~ & $-1.13$~~ & ~~~2.5~ &  1.04~~~ & ~~0.9~ & ~~99.9\\
 Abell & S333*  &  573~~ &  0.01~~ & $<0.1$~ & $-0.93$~~~ & ~~0.8~ & 93\\
 Abell & S1043  & 1271~~ &  0.86~~ & ~~~2.3~ & $-0.53$~~~ & ~~0.6~ & $>$99.9\\
 APM   &  ~~268 &  769~~ & $-0.65$~~ & ~~~1.7~ & $-0.47$~~~ & ~~0.5~ & 98\\
 APM   &  ~~917 &  471~~ &  0.18~~ & ~~~0.4~ &  0.03~~~ & $<0.1$~ & 73\\
 APM   &  ~~933 & 1033~~ &  0.77~~ & ~~~2.2~ &  0.81~~~ & ~~0.9~ & 97\\
 EDCC  &  ~~365 &  531~~ &  0.72~~ & ~~~1.7~ &  1.00~~~ & ~~0.9~ & 83\\
 EDCC  &  ~~442 &  726~~ &  0.00~~ & $<0.1$~ &  1.72~~~ & ~~2.1~ & 51\\
 \hline
 \end{tabular}
 \end{center}
\end{minipage}
 \end{table*}

\subsection{Velocity Statistics as a Substructure Indicator}

The velocities for the galaxies in a relaxed cluster in thermal equilibrium are expected
to be distributed normally with respect to any one cartesian coordinate.  
However, it should be recognized that taken by itself, an apparently non-Gaussian velocity
distribution may reflect either a cluster not in equilibrium or a cluster
containing one or more dynamically-bound subgroups embedded in an otherwise smooth, approximately
equilibrium, distribution. 
Since all semi-invariants of higher
order than two vanish identically for a Gaussian (normal) distribution, deviations from normality 
can be estimated by calculating the skewness and kurtosis 
(quantities proportional to the third and fourth order semi-invariants). In the following, all
cluster redshifts have been transformed to velocities via the standard relativistic Doppler
shift formula.

The skewness is given by 
\begin{equation}
S = \frac{1}{\sigma^3} \left[ \frac{1}{N} \sum^N_{i=1} (v_i - \overline{v})^3 \right]
\label{skew}
\end{equation}
with $\overline{v}$ and $\sigma$ the mean velocity and standard deviation
determined from the observed line-of-sight velocities $v_i$ of the $N$ cluster
members.  A positive (negative) value of $S$ implies the distribution is
skewed toward values greater (less) than the mean.
\noindent
The kurtosis coefficient is defined as
\begin{equation}
K = \frac{1}{\sigma^4} \left[ \frac{1}{N} \sum^N_{i=1}
(v_i - \overline{v})^4 \right] - 3.
\label{kurtosis}
\end{equation}
Because the kurtosis of a normal distribution is identically equal to 3, kurtosis values 
are conventionally
presented with this subtracted from the result.
Note that positive (negative) values of $K$ indicate distributions more strongly peaked 
(flatter) compared to normal distributions.
Figures \ref{skew2} and \ref{kurt2} show the distribution of the skewness and
kurtosis values for the 25 clusters as well as estimates of their statistical significance
in terms of the number of standard deviations away from the Gaussian values.
Examples of previous studies of cluster structure using computed skewness and kurtosis 
to characterize the velocity distribution include \citet{bird_2}, \citet{ashman},
\citet{bird_1}, \citet{west_2}, \citet{pinkney}, and
\citet{solanes}.

In practice, for a cluster containing $N \sim 50-100$ members,  
the skewness and kurtosis can provide suggestive but 
not necessarily definitive estimates as to whether a given distribution is significantly
non-Gaussian. In order to obtain a better quantitative check for normality, it is convenient
to use a $\chi^2$ statistic to estimate the probability that a test distribution
can be distinguished from a Gaussian with the null hypothesis that the
test distribution is, in fact, normal. 
To be reliable, the $\chi^2$ algorithm requires
the binned test distribution to have a minimum number of datapoints in each bin to guarantee
a reasonable probability for occupancy. This can be accomplished either by shifting/changing
the number of fixed-width bins, or by using variable-width bins with (chosen) occupation
probabilities above the minimum acceptable level. In testing the velocity distributions 
here, the adaptive method was implemented with a bin width varied to yield a
minimum $5\%$ occupation probability for each bin.

Table \ref{velstats} shows the velocity dispersion
$\sigma_{vel}$, skewness $S$, the number of standard deviations $\sigma_S$ away from the skewness 
value expected for a normal distribution, kurtosis $K$, the number of standard deviations $\sigma_K$ 
away from the kurtosis value expected for a normal distribution, and the confidence level CL derived  
from the $\chi^2_{vel}$ test to
reject the hypothesis of a normal distribution. 
Inspection of the results
shows that 13 of the 25 clusters have catalogue distributions inconsistent with Gaussian distributions at
$\geq 95\%$ confidence level (CL). Comparison of the $\chi_{vel}^2$ statistic with the
skewness and kurtosis values shows no consistent correlation as to whether a distribution
is (or is not) normal. In general, the skewness
and kurtosis are not necessarily reliable indicators of departures from normality
for these cluster membership sizes.
However, it should be noted that even the $\chi^2$ test is not $100\%$ reliable,
and the results depend somewhat on the degrees of freedom used in each case. 

\subsection{Contour, Segmentation, and Nearest Neighbor Visualization Plots}

In spite of the dangers of giving too much weight to visual appearances, contour
and other visualization plots remain useful when attempting to correlate the results of
purely statistical tests to spatial structure or when comparing galaxy structure
to gas structure revealed through X-ray emission. Thus, in the analysis of the
clusters considered in this study, four different types of two-dimensional visualization plots are
included in addition to the scatter plot of the galaxy positions.

The number density contour plots for the clusters are constructed with galaxy positions 
binned in $0.3 \times 0.3 \; (h^{-1} \, \mathrm{Mpc})^2$ square cells and then smoothing the contours 
with a bicubic interpolation algorithm. 
In order to provide a sufficient number of contour lines to give a visual sense of cluster structure,
it was found that the sample could be broken into two general categories depending
on whether, for a given cluster, any $0.3 \times 0.3 \; (h^{-1} \, \mathrm{Mpc})^2$ cell exceeded a count of 8
or more galaxies.
For these clusters, contour levels were drawn at 20 gal/$(h^{-1} \, \mathrm{Mpc})^2$
intervals, while for clusters with no cell exceeding this threshold, contour lines were drawn at
10 gal/$(h^{-1} \, \mathrm{Mpc})^2$ intervals. In the first category (20 gal/$(h^{-1} \, \mathrm{Mpc})^2$)
are Abell clusters 930, 957, 1139, 2734, 1663, 2734, 2814, 3094, 3880, 4012, 4013, 4038, S141, S258, S301, S1043, 
APM clusters 268, 917 and 933, 
and EDCC 442.
In the second category (10 gal/$(h^{-1} \, \mathrm{Mpc})^2$) are Abell clusters 1238, 1620, 1750, 3027,
and S333 and EDCC365.

As a means of comparing the cluster luminosity distribution to the number density
distribution, we have also constructed luminosity-weighted contour plots. As for the number 
density contour plots, the initial cell size is  
$0.3 \times 0.3 \; (h^{-1} \, \mathrm{Mpc})^2$. 
The $k$-corrected absolute
magnitude for each galaxy, $M_J$, is calculated using the standard Taylor expansion to first
order in $z$ (see, e.g., \citealt{peebles}, pp. 328-330),
\begin{eqnarray}
M_{J} = b_J - k - 42.386+ 5\log_{10}h - 5\log_{10}z \nonumber \\ 
  - 1.086(1-q_0)z + \, \ldots \; \; \mbox{,} 
\end{eqnarray}
where $q_0 = -0.55$ is the deceleration parameter value for $\Omega_m = 0.3$
and $\Omega_\Lambda = 0.7$. 
The $k$-corrections are computed from the emprical fits presented in~\cite{depropris_2}.
Each contour line
represents an increase of $\approx 5 \times10^{10} \, L_{\bigodot}/(h^{-1} \, \mathrm{Mpc})^{2}$
(except for A2814 and A1750 which were set to $\approx 1 \times10^{11} \, L_{\bigodot}/(h^{-1} \, \mathrm{Mpc})^{2}$), 
and, as for the number density
contours, the luminosity density contours are smoothed using bicubic interpolation. 

The segmentation plots included in this study are meant primarily as a visual aid that provides
a bridge between the density contour plots and the 
nearest neighbor plots described below. The basic idea is to connect ``segments'' of
occupied cells to give some idea as to the location and extent of both connected and island regions
in the cluster.
Another appropriate terminology might be ``tree''structure, but as that is used to denote
a specific analysis technique, the term ``segmentation'' is used instead.
The plot is constructed in much
the same manner as are the contour plots: cell counts of galaxies are tabulated for a specified
grid (again, $0.3 \times 0.3 \; (h^{-1} \, \mathrm{Mpc})^2$ square cells), 
the total number of cells is increased by subdividing the grid further in order to
increase the resolution, and new cell counts
are computed using the bicubic spline interpolation
scheme as both an estimator and a smoothing mask.
The segmentation algorithm is then simply the imposition of an additional step that thresholds
the interpolated cell counts: the count is reset to zero for values below the threshold, and
a user-defined gray scale is applied for counts above the threshold.
In practice, we found the following choice of thresholds most useful: \\
\indent{1. For central core densities $\sigma_0 < 100 \; \mathrm{gal}/(h^{-1} \; \mathrm{Mpc})^2$, the cutoff
is 1.0 galaxy per bin (or, equivalently, $11.1 \;  \mathrm{gal}/(h^{-1} \; \mathrm{Mpc})^2$),}\\
\indent{2. For central core densities $100 < \sigma_0 < 200 \; \mathrm{gal}/(h^{-1} \; \mathrm{Mpc})^2$, the cutoff
is 1.5 galaxies per bin (or, equivalently, $16.7 \;  \mathrm{gal}/(h^{-1} \; \mathrm{Mpc})^2$),}\\
\indent{3. For central core densities $\sigma_0 > 200 \; \mathrm{gal}/(h^{-1} \; \mathrm{Mpc})^2$, the cutoff
is 2.0 galaxies per bin (or, equivalently, $22.2 \;  \mathrm{gal}/(h^{-1} \; \mathrm{Mpc})^2$).}\\
The resulting plot exhibits the connectivity
of the cluster galaxy distribution as a function of cell count threshold.

A final visualization technique useful for identifying clumps that may represent true subclusters
is the plot of nearest neighbors. The maximum distance used in order to define two galaxies as
neighbors and the number of neighbors required to define a neighborhood of galaxies is,
of course, arbitrary. In the plots presented in Appendix A, the selected distance is
$0.25 \, \mathrm{h}^{-1}$ Mpc and the minimum number of galaxies is set to four, i.e., each point
on the plot is the position of a
galaxy with four or more galaxies closer to it than $0.25 \, \mathrm{h}^{-1}$ Mpc. 
Where appropriate, different choices are utilized to highlight certain features, and are presented
in the detailed analyses of individual clusters.

\subsection{The $\alpha$ Test}

The 3-dimensional $\alpha$ test used to detect substructure in clusters of galaxies was
introduced by \citet{west_1} and slightly revised by
\citet{pinkney}. The test measures the shift
in the arithmetic mean centroid of a cluster with the galaxy locations weighted by 
kinematics in specified neighborhoods about each galaxy.  The centroid used 
is simply the arithmetic mean of all galaxy positions in the cluster.
The location of each galaxy is weighted by the inverse of
the velocity dispersion of nearby galaxies. Thus, a neighborhood of galaxies with
similar velocities (i.e., low dispersion) has a higher weight and results
in a centroid shift in their direction.  If the member velocities are distributed randomly
among the member positions, the centoid shift is statistically insignificant.

Thus, the steps leading to construction of the $\alpha$ statistic are as follows: \hspace*{\fill}
\begin{enumerate} 
\item Convert (RA, Dec) sky coordinates to cartesian coordinates (x, y), and 
calculate the centroid of the two-dimensional galaxy distribution,
\begin{equation}
x_c= \frac{1}{N} \sum_{i=1}^N x_i,\mbox{ and\ } 
y_c = \frac{1}{N} \sum_{i=1}^N y_i.
\label{a1}
\end{equation}
\item Assign a weight $w_i=1/\sigma_i$ to each galaxy,
where $\sigma_i$ is the line of sight velocity dispersion for galaxy $i$ and
its $N_{nn}$ nearest neighbors 
where an arbitrary choice is made to set $N_{nn} = \sqrt{N}$,  
\item For each
galaxy $i$ and its $N_{nn}$ nearest neighbors, calculate the weighted centroid,
\begin{equation}
x^{\prime}_c = \frac{\sum_{i=1}^{N_{nn}+1} x_i w_i}{\sum_{i=1}^{N_{nn}+1} w_i}
\mbox{ and  }  
y^{\prime}_c = \frac{\sum_{i=1}^{N_{nn}+1} y_i w_i}{\sum_{i=1}^{N_{nn}+1} w_i}.
\label{a2}
\end{equation}
\item The difference in centroid of these nearest neighbor velocity groups
and the unweighted centroid is calculated,
\begin{equation}
\alpha_i = \sqrt{\left( x_c - x^{\prime}_c \right)^2 + \left( y_c - y^{\prime}_c \right)^2}
\label{a3}
\end{equation}
\item Finally, the $\alpha$ statistic quantifies the cluster substructure as an 
average of the $\alpha_i$ values,
\begin{equation}
\alpha = \frac{1}{N} \sum_{i=1}^N \alpha_i.
\label{a4}
\end{equation}
\end{enumerate}
The test is normalized by comparison with the value $\alpha$ takes for
Monte Carlo distributions created by randomly shuffling the velocity among
galaxy positions fixed by the data.
The $\alpha$ statistic for the cluster is compared to the mean $\alpha$
statistic from Monte Carlo simulations with the difference expressed in standard
deviations used to estimate the probability for the presence of substructure.
A subcluster will influence the centroid of its velocity group
provided it deviates from the global distribution both spatially and in velocity
dispersion. Detailed results of the $\alpha$-statistic computed for the clusters
themselves as well as for the Monte Carlo simulations are contained in 
\citet{vick}.

\subsection{The $\beta$ Test}

The $\beta$ test is used to detect deviations from mirror symmetry about the cluster centre
where mirror symmetry is defined as a 2-dimensional coordinate inversion $(x,y) \rightarrow (-x,-y)$
\citep{west_3}.
After determining the cluster centroid, the mean distance to each 
galaxy's nearest neighbors is calculated.  Then, the location of the inverted position of each galaxy 
is computed, and the mean distance from that location to the nearest neighbors 
is calculated.  Actual mean distances are compared to the mirror image distances.  Large 
differences indicate asymmetry and possible substructure. This test is particularly useful in
quantifying visual impressions of asymmetry conveyed by the visualization plots, and can often be
correlated with results from the ellipticity algorithm.

Here, the asymmetry parameter for galaxy $i$ is defined as 
\begin{equation}
\beta_{i} = \log{\left(\frac{d_0}{d_i}\right)}
\label{b1}
\end{equation}
\noindent{where $d_0$ is the mean distance of the galaxy's $N_{nn}$ nearest neighbors
($N_{nn}=\sqrt{N}$ 
rounded to the nearest integer),
$d_i$ is the mean distance of the mirror image point of each galaxy to the $N_{nn}$
nearest neighbors, and $N_{Total}$ is the number of total galaxies in the cluster.
The $\beta$ statistic for a cluster is the mean $\beta$ value for all the
galaxies times one thousand or
\begin{equation}
\beta = 1000\left\langle \beta_i \right\rangle.
\label{b2}
\end{equation}
The significance of the $\beta$ statistic of the data is estimated by comparison against a
distribution obtained from Monte Carlo simulation. This is accomplished by holding constant the
distance from the centre of each galaxy in the data, but uniformly varying the polar angle
from 0 to 2$\pi$. The galaxy coordinates are then relocated at $x_i = r_i\cos{\phi}$ and
$y_i = r_i\sin{\phi}$.  The $\beta$ statistic is recalculated for each reshuffled galaxy cluster.  
The mean and standard deviation computed from the Monte Carlo simulations are then compared to the value
from the original data. A Monte Carlo probability of random occurrence is used to estimate
the significance of any deviation from mirror symmetry. A significant deviation may indicate
an asymmetrical distribution caused by one or more subclusters. 
Detailed results for both the $\beta$-statistic computed for the clusters
as well as for the Monte Carlo simulations are contained in \citet{vick}.

\subsection{The $\kappa$ Test}

The $\kappa$ test is a three-dimensional substructure test based on nearest neighbor velocity
relationships first introduced in a substructure study of A1656~\citep{colless_3}.
As discussed in that study, the $\kappa$ test is similar in intent to the Dressler-Shectman 
$\Delta$ test \citep{dressler_2} in that it probes deviations in localized velocity variations compared to
the global cluster velocity distribution. It is especially useful for detecting subgroup dynamics
in a cluster core region including the case of a projected subgroup centroid overlapping the projected
cluster centroid.

The velocity distribution of a user-selected number of nearest neighbors $n$ is compared to
the velocity distribution of the entire cluster with the test statistic $\kappa_n$ constructed as
\begin{equation}
\kappa_n = \sum^{N}_{i=1} -\mathrm{log}[P_{KS}(D > D_{obs})] \; \; \mbox{,}
\label{kappa_stat}
\end{equation}
where the sum is over the $N$ galaxies in the cluster. Thus, $P_{KS}(D>D_{obs})$ is the probability
that the K-S statistic $D$ is greater than the observed value $D_{obs}$ which is simply a
standard K-S two sample test with a significance that can be computed 
in a straightforward manner as given in, e.g., \citet{press}.
The $\kappa$ statistic is then the (negative) log-likelihood that there is no localized deviation in
the velocity distribution on the scale of sets of $n$ nearest neighbors.
The larger the value
of $\kappa$, the greater the likelihood that the local velocity distribution is different
from the overall cluster velocity distribution. The significance of $\kappa_{n}$ can also be
computed from Monte Carlo simulations by randomly shuffling the velocities of the individual galaxies
while holding their positions constant. 

For this study, the default number of nearest neighbors is set
to $\sqrt{N}$, but the results are relatively insensitive to the precise choice when it is $\sim \sqrt{N}$; 
see \citet{vick} for a detailed
comparison between the $\kappa$ and $\Delta$ tests as well as the sensitivity of the 
$\kappa$ test to the selected number of nearest neighbors.
It is convenient to display the results utilizing ``bubble skyplots'' where at each galaxy position, a
circle is drawn of radius proportional to the log-likelihood of the local and overall distributions
\textit{not} being the same under the K-S two-sample test. In this paper, the radii of the $\kappa$
bubbles was set to $\kappa_n/30$ for visualization purposes.

\section{Detailed Analysis of Individual Clusters}

This section provides a detailed analysis of the individual clusters from
consideration of the results of the statistical tests combined with visualization
plots. It will be seen that while detecting
substructure is straightforward, unambiguous interpretation remains problematic.
Further correlation with X-ray data is expected to remove some ambiguities,
especially as relates to cluster membership. The projected (2-dimensional) galaxy
separations are inferred from the proper distance at photon time of emission, $d_p(t_e)$.

\subsection{Abell 930}

From Figures \ref{a930}(a)-(e), inspection of Abell 930 surface 
plots shows a low density, moderately elliptical cluster containing 
a (projected) bimodal central region aligned east to west with the two components 
having a projected separation of approximately $0.50 \, h^{-1}$ Mpc. Outside the 
core there is a relatively high density region (greater than 40 galaxies per ($h^{-1}$ Mpc)$^{2}$)
located approximately $1.0 \; h^{-1}$ Mpc due north of the cluster centre.
The segmentation and nearest neighbor plots, Figures \ref{a930}(d)-(e), again
highlight the two-component core region (labels A and B) and concentration of galaxies to the 
north (label C).
As discussed below, these three components appear to be 3-dimensional structures as well.

In spite of the relatively high concentration of galaxies to the north,
the $\beta$ test yields no indication of global asymmetry ($P_\beta$ = 30\%).
This is due to the positioning of the clustre center between the two components in the
central regions; positioning the center at the centre of one of these components would result
in a significant asymmetry signal. This re-positioning was not done for this cluster as for others
in this sample because of the near equality in number density and total number of galaxies
in each of the components, and the lack of an X-ray map that might indicate which of the
components possesses more mass. 

\begin{figure}
\centering
\epsfig{figure=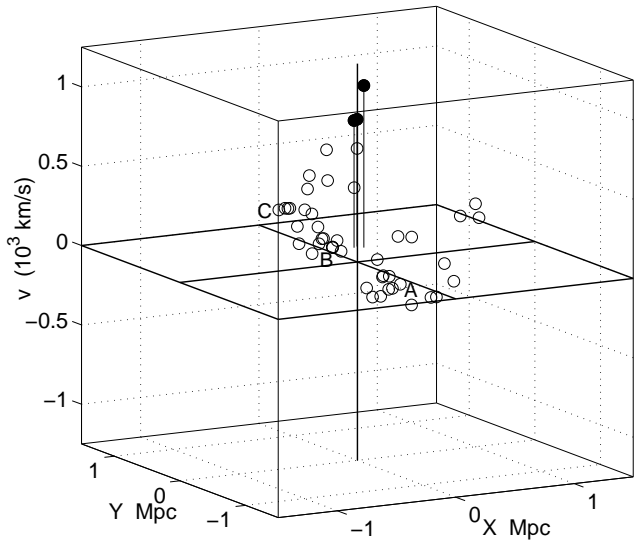,width=0.48\textwidth}
\caption{Local velocities of the nearest neighbors in Figure~\ref{a930}(e) for Abell 930.}
\label{a930_nnlv}
\end{figure}

The ellipticity computation yields an $\epsilon$ = 0.29 
-- 0.45 over the entire test range from radius R = 0.50 -- 1.25 $h^{-1}$ Mpc (Table \ref{ellipt_table}).
As expected, the mean radius of $R = 0.50 \; h^{-1}$ Mpc results in a position angle
aligned approximately east to west due to the bimodal core ($PA = 75^\circ$).
Average dispersion ellipse radii from $R$ = 0.75 - 1.25
$h^{-1}$ Mpc are aligned from north to south with an
average position angle of $PA \approx 10^\circ$.
We use an average ellipticity and position angle
$\epsilon$ = 0.4, $PA = 70^\circ$ to calculate a core radius and central density of 
$R_c = 0.47 \pm 0.01$ and $\sigma_0 = 61 \pm 2$.

Inspection of the velocity distribution in Figure \ref{a930}(i) shows a bimodal 
profile with the two peaks separated by approximately 1500 km/s.  
Comparison of the members of the two spatial core components to 
the two velocity modes yields no relationship, but  
due to this bimodality, the $\chi^2_{vel}$ velocity test rejects the hypothesis
of normality with $>99.9\%$ CL (Table \ref{velstats}). 

The 3D plots and tests also indicate the presence of
substructure (Figures \ref{a930}(f)-(h)and Table 
\ref{sumresults}).  In particular, the peculiar velocity plot of Figure \ref{a930}(g) 
shows the two-component central region in the 2D plots separating into the main body of the cluster with peculiar 
velocities less than zero (due to the averaging of the two peaks to arrive at the cluster mean velocity) 
while approximately 15 galaxies have peculiar velocities of $< 1000 \, \mathrm{km/s}$ (this is the eastern component seen
in the 2D plots). 
It is easier to distinguish the western (A), eastern (B), and northern (C) components
in the central regions as distinct subgroupings by plotting the local velocities of only the nearest
neighbors (see Figure~\ref{a930_nnlv}).
It should also be noted that the possibility cannot be excluded that each of these two larger groupings
may contain finer-detailed structure not resolved here, or that there is additional structure to the
north.

Both the $\alpha$ and $\kappa$ tests indicate some level of substructure
with probabilities of random occurrence of 3.0\% and 5.8\%, respectively.
The $\kappa$ bubble skyplot presented in Figure \ref{a930}(f) 
shows two areas of possible subclustering:  a group of seven galaxies show 
relatively large radii bubbles (indicating high $\kappa$ statistics) 
from $0.5 - 1.0 \, h^{-1}$ Mpc north of the centre, and a second group of above average-sized bubbles includes 
galaxies in the western component of the core region. Among the seven galaxies to the north 
with relatively high $\kappa$ values, three of the seven are well separated from a compact grouping of
four galaxies indicated by the arrows in Figures \ref{a930}(a), (e), and (f) 
(two have nearly coincident angular positions and are indistinguishable in the 2D plots).
Due to their apparent proximity coupled with the results of the $\kappa$ test,
and the fact that these four galaxies possess
velocities consistent with cluster membership, we hypothesize that they comprise a dynamically bound
subgroup belonging to the cluster. 

In summary, there appears to be significant substructure present in this cluster.
The substructure of the central region makes Abell 930 a potential site for
merger dynamics, the velocity structure is also potentially significant, and there are
probably at least one or two bound subclusters well removed from the central regions.  

\subsection{Abell 957}

Abell 957 possesses a dense core with a core radius $R_c = 0.16 \pm 0.01 \, h^{-1}$ Mpc
and a central density $\sigma_0 = 318 \pm 1$ (see Figures \ref{a957}(a),(b),(d)). 
In both two and three dimensions, the cluster central regions are distinctly bimodal, with a second high 
density concentration of galaxies approximately 0.5 
$h^{-1}$ Mpc east of the core also possessing a large fraction of the total luminosity 
due to two relatively bright galaxies whose velocities are consistent with cluster membership
(see Figure \ref{a957}(c)). Due to a selection effect in the catalogue, a very bright D galaxy 
residing at the centre
of the peak number density that is a well-established cluster member (based on redshift as well
as sky location) was not included in the present sample. However, we have added it to the sample,
and have also re-positioned the cluster center to coincide with the location of peak number density.
As can be seen by comparing Figures~\ref{a957}(b) and (c), the peak luminosity coincides with the
peak galaxy density at the position of the D galaxy. 

With a probability for random occurrence of $P_\beta = 0.5\%$,
the $\beta$ test detects the asymmetry
due to the single structure east of the core.  
The bimodal characteristic gives the 
cluster a highly elongated shape (see Table \ref{ellipt_table} 
and Figure \ref{a957}(a)) leading to $\epsilon = 0.68 - 0.35$ 
from $R = 0.50 - 1.25 h^{-1}$ Mpc, and where the semimajor 
axis is consistently aligned with the two central components in 
this range with a $PA = 74^\circ - 42^\circ$.  
Even at larger distances from the centre, with the outer regions being relatively diffuse,
the dispersion ellipses are strongly influenced by the dense central region.
However,  from Figure \ref{a957}(d), it can be seen that the 
core itself possesses some ellipticity with a position angle pointed directly toward the
eastern subgroup, possibly indicating tidal interactions between
these two central components.  

Considered in its entirety, the velocity distribution is consistent with a normal distribution.
However, note from Figure \ref{a957}(a) that the 0.5 $h^{-1}$ Mpc dispersion ellipse region
contains several galaxies with velocities that are slower (open circles) and faster (plus signs)
by more than $1.3\sigma$ compared to the cluster mean velocity.
While the $\kappa$ test for the entire cluster shows no significant signal with
$P_{\kappa} = 14.4\%$, 
four galaxies in the core show a 5\% 
or less probability that the average of their nearest 
neighbor velocities belong to the clusters velocity 
distribution (see Figures \ref{a957}(g)--(h)).
In particular, some of these galaxies may form part of a small subcluster with projected (surface)
centroid nearly coincident with the core centroid (see Figure \ref{a957}(h)), 
and it is the relatively slow local
velocities giving rise to the large bubbles in Figure \ref{a957}(f). This may be evidence
for a line-of-sight merger.

Although the $\kappa$ test does not indicate anything significant in the eastern subgroup,
it is not unreasonable to speculate that this group may still be moving toward or away
from the core in a direction perpendicular to the line of sight.
It should also be noted that Rosat All Sky Survey (RASS) data does exist for this
cluster, and confirms that the cluster's central potential well is very close to the
computed centroid of the core. The region appears dynamically active, and also hints
that the eastern clump is, in fact, a bound subgroup. However, 
the RASS observations provide only 454s of exposure time. This cluster
is also at the edge (44' off-axis) of a PSPC observation with a total of 1.95 ksec.
A much longer exposure time with better resolution is required to
infer further dynamical properties. 

In previous studies of this cluster, the bimodal core has been identified by \citet{beers_2} and
\citet{kriessler}. The analysis presented in \citet{beers_2} concludes that the two components
are not grabvitationally bound, but uncertainty in the radial velocity difference does
allow for bound solutions. With the larger sample of redshifts available here, this question
will be investigated further in a following paper. 
Using redshift data for 34 galaxies, \citet{solanes} found the velocity distribution 
to be consistent with
a normal distribution with statistics similar to those presenterd here for 90 galaxies. The  Solanes et al.
study also performed the $\Delta$ test with a result of $P_\Delta = 3.4\%$ compared to results found here
of $P_\Delta = 17\%$.
 
\subsection{Abell 1139}

The analysis of A1139 presented here is based on 106 galaxies.
The fitted optical core
radius of Abell 1139 is $0.26 \, h^{-1} \, \mathrm{Mpc}$ with a central density of 138 which is
fairly typical of the poor clusters in this sample. Figure~\ref{a1139}(b) shows the core and the more
asymmetric envelope. The azimuthally averaged fits to the cluster
shape are not sensitive to the relatively low density concentration of galaxies seen to
the west of the core at ($0.75$, $-0.2$) Mpc (see Figure~\ref{a1139}(d)).
At large radii the ellipticity is low, $\sim 0.2$, and the position
angle is $\sim \, 65^\circ$.  The inner part of this cluster is elongated
and the ellipticity within 0.75 Mpc is $\approx \, 0.4$ with a position angle
of $\approx \, 100^\circ$ reflecting the core elongation seen in the number
density map. It should be noted that there are three additional relatively bright galaxies 
located near the centre of this cluster having redshifts consistent
with cluster membership that are not included in the 2dFGRS catalogue.
Their inclusion shifts the centre of luminosity toward the number density centroid
compared to the sample of 106 galaxies used here, but does not alter any of the results
concerning substructure.

The two-dimensional $\beta$ test indicates no significant asymmetry ($P_\beta = 15\%$).
However, the three-dimensional $\alpha$ and $\kappa$ tests indicate the presence
of strong substructure, each with a probability of $< 0.1\%$ that the
values are drawn from a relaxed distribution. The bubble plot in Figure~\ref{a1139}(f)
shows the
two disjoint areas
that deviate from the cluster mean lying to
the west at (1.0, 0.0) Mpc and south (0.0, $-1.0$) Mpc of the main cluster
component. These two systems can be seen in the number density plot
(Figure~\ref{a1139}(b)) as low-level extensions from the main cluster. These two
structures are also seen in the nearest neighbor plots (Figure~\ref{a1139}(e)).
The segmentation plot (Figure~\ref{a1139}(d)) shows the core and western subclump
clearly while only a hint of the southern subclump is evident, but both
substructures can been seen in the local velocity plot (Figure~\ref{a1139}(h)).

Intriguingly, an inspection of the local velocity plot indicates
that there is a velocity gradient across the cluster.  This shear seen
in the velocity structure may indicate that this cluster has a
significant rotation. Alternatively, the cluster may be in the late
stages of a double merger with two subclumps spiraling in to merge
with the main system. 

In contrast to the $\alpha$ and $\kappa$ tests, the velocity distribution 
in Figure~\ref{a1139}(i) does not show any statistical evidence for substructure
(See Table~\ref{velstats}). 
In addition, the velocity dispersion of 
484 km/s is somewhat on the low side for the poor clusters in this sample. 
However, the three-dimensional tests and the cluster morphology indicate
that this cluster is dynamically interesting. This along with the possible
presence of a velocity gradient in the system indicates that additional 
data from X-ray observations would be useful in clarifying the 
dynamical state. 

Previous investigations of this cluster have found little or no evidence for
substructure, and have not identified the velocity gradient found here.
In particular, \citet{kriessler} and \citet{krywult} concluded that the cluster
is unimodal with no substructure while \citet{west_1} found marginal evidence
for substructure.

\subsection{Abell 1238}

With a central density of $\sigma_0 = 68 \pm 3 \, \mathrm{gal}/(h^{-1} \mathrm{Mpc})^{2}$, Abell 1238
is a relatively low density cluster with a somewhat irregularly shaped central region of
core radius $R_c = 0.41 \pm 0.02 \, h^{-1} \, \mathrm{Mpc}$~(see Figures \ref{a1238}(a), (b), and (d)).
In general, the cluster has a low ellipticity in the central regions ($\epsilon = 0.34$ at $0.5 \, h^{-1}$
Mpc) that steadily decreases to insignificance at the largest mean radius of $1.25 \, h^{-1} \, \mathrm{Mpc}$.
Inspection of the luminosity-weighted
contours, Figure \ref{a1238}(c), reveals the brightest galaxy in this catalogue to the east of the centre with 
$cz \approx 50$ km/s different
from the cluster mean (21,199 km/s compared to $c\overline{z} = 21,148$ km/s).

In general, this cluster shows no statistical evidence for substructure at greater than the 95\% CL.
These results are consistent with those found by \citet{rhee}.
The projected surface distribution is relatively asymmetric ($P_\beta = 50\%$), and the velocity
distribution is indistinguishable from a normal distribution. 
The large bubbles belonging to the galaxies southwest of the core in Figure \ref{a1238}(f) most likely
do not indicate significant cluster dynamics. One possible subgrouping is a small
clump of 7--8 galaxies to the east (labeled B in Figure \ref{a1238}(e)) 
and slightly south at a distance of about $1.1 \, h^{-1} \, \mathrm{Mpc}$
from the centre (see Figures \ref{a1238}(a) and (d)); this grouping also contains the brightest cluster
member mentioned previously, and is also tightly grouped in the local velocity-position plot of Figure \ref{a1238}(h).
A second clump appears due south of the centre labeled A in Figure \ref{a1238}(e). Both of
these potential subgroups show some marginally significant substructure as given by the $\kappa$ test, but
a long exposure X-ray observation is probably required to ascertain whether these are bound subgroups
and the extent to which their presence affects the overall cluster dynamics. 

\subsection{Abell 1620}

At a distance of $d_p(t_e) = 231 \, h^{-1} \, \mathrm{Mpc}$, Abell 1620 is one of the 
most distant clusters in this study.
Inspection of Abell 1620 2D plots (Figures
\ref{a1620}(a)--(e)) reveals an irregularly-shaped,
low-density central region with four distinct groupings
outside the central region labeled A, C, D, and E with a fifth possible
subcluster just northwest of the core labeled B.
Groupings A and C possess maximum number
densities equivalent to that of the central region,
$\sim 60 \, \mathrm{gal}/(h^{-1} \, \mathrm{Mpc})^2$.  Groupings D and E are less
prominent with densities of $\sim 30-40 \, \mathrm{gal}/(h^{-1} \, \mathrm{Mpc})^2$.

The central regionof the cluster show almost no ellipticity at a mean radius  
of 0.5 $h^{-1}$ Mpc, but are moderately elliptical with $\epsilon = 0.49$ at a mean
radius of $0.75 \, h^{-1} \, \mathrm{Mpc}$ as grouping B influences the computation.
At mean radii of 1.0 $h^{-1}$ Mpc, the ellipticity
is still around $\epsilon$ = 0.50, but there is a position angle change due to the influence of C and E. 
Overall, the position angles
maintain a fairly narrow range at all mean radii, ranging from
$PA$ = 17$^\circ$ to 41$^\circ$, in
approximate alignment with E, the core, and C with the exception of the $0.5 \, h^{-1} \, \mathrm{Mpc}$.
The core parameters are computed using the average ellipticity and position angle
to yield a mean core radius of $R_c = 0.47 \pm 0.02$
and a central density of $\sigma_0 = 59 \pm 3$ gal/($h^{-1}$ Mpc)$^{2}$
indicating the relatively diffuse nature of the central region.

The velocity distribution is skewed toward higher velocities (Figure
\ref{a1620}(i)) resulting in a skewness value of 
$S = 0.77 = 1.9\sigma_S$. The distribution is also somewhat broad compared to a Gaussian
with $K = 0.65 = 0.6\sigma_K$. The $\chi_{vel}^2$ test rejects the
hypothesis of normality at the 97\% CL.

The $\beta$ test for asymmetry indicates a very high probability
of substructure with a $P_\beta \approx 0.1\%$.  While the subgroups
are somewhat evenly distributed around the centre, the density of
the subgroups is significantly different, causing the computed high asymmetry value.

The 3D $\kappa$ test indicates a high probability of substructure
with a $P_\kappa = 0.9\%$.  As seen in Figure \ref{a1620}(f),
the level of $\kappa$ is relatively high in the core, and subgroups
C, D, and E. The absence of significant $\kappa$ in A can be explained by
the similarity of A's local velocity distribution relative to the global cluster
velocity distribution: the
mean velocity of A is  24,486 km/s compared to 24,430 km/s for the
cluster, and the related dispersions are 1103 km/s and 1001 km/s.
Also note the presence of significant $\kappa$ in the (projected) central
region that may indicate the presence of an additional subcluster or substructure.

\subsection{Abell 1663}

Surface plots of Abell 1663 (Figures \ref{a1663}(a)-(e)) reveal a large,
low density, bimodal central region.  
The centroid calculation used here is biased by the (projected) bimodal structure
region so that the centre is re-positioned to be coincident with the peak number density.
The northeast component contains
the most luminosity for this particular cluster sample.

Three galaxies are questionable cluster members.  The second brightest
galaxy in the cluster is also the slowest galaxy in the cluster
(2.9$\sigma$ from the mean) and on the periphery of the
cluster (1.4 $h^{-1}$ Mpc from the centre at coordinates
($-0.72$, $-1.18$)). The other two galaxies
are not overly bright ($b_J$ = 18.6 and 18.7), but are very close
to the bright galaxy in both space (within 0.19 $h^{-1}$ Mpc) and
velocity (within 350 km/s). This proximity suggests a bound group of 
three galaxies that, based upon the spatial, velocity, and luminosity data, 
is probably not associated with this cluster.
As their influence on the statistical tests is negligible, it does not matter
if these three galaxies are included or removed from the cluster sample.

Compared to the other clusters in the study, Abell 1663 is moderately
elliptical with a position angle biased by the presence of the subcluster to
the northeast (see Table \ref{ellipt_table}).
The core parameter values of $R_c = 0.46 \pm 0.03$ and $\sigma_0 = 57 \pm 4$ are computed
using the ellipticity parameters for the 1.0 $h^{-1}$ Mpc ellipse that improves
the fit in this region compared to the circular fit while degrading the fit
in the outer regions.

The velocity distribution is skewed toward lower velocities resulting in a skewness
statistic of $S = -0.60 = 1.5\sigma_S$ with a kurtosis of $K = 0.86 = 0.8\sigma_K$
(also see Figure~\ref{a1663}(i)). 
However, the $\chi^2_{vel}$ test indicates this distribution is indistinguishable from a normal
distribution with only a 36\% CL to reject the normality hypothesis. 

With the re-positioning of the cluster centroid,
the $\beta$ test then yields a marginally significant result of $P_{\beta} = 4\%$. 
However, the strong indication from the $\kappa$ test of 
$P_\kappa = 0.1\%$ together with inspection of the bubbleplot in Figure~\ref{a1663}(f)
seems to indicate additional substructure beyond the apparent bimodality.
Two subclusters are evident in the 3D
plots (Figures \ref{a1663}(f)-(h)).  One appears slightly east of the
centre and the other approximately 0.8 $h^{-1}$ Mpc west of the centre.
The subcluster nearest the centre is in very close proximity to but not
actually part of the bimodal central region.  With
the faster subcluster east of the centre and the slower subcluster west
of the centre, the velocity gradient is a possible signature for cluster rotation or infall
dynamics, but further analysis is required for an accurate determination. The $\kappa$
bubble plot (Figure \ref{a1663}(f)) also shows another possible subcluster
approximately 1.0 $h^{-1}$ Mpc due south of the centre.

\subsection{Abell 1750}

With a mean redshift of $z = 0.086$, Abell 1750 is one of the more distant
clusters in this study ($d_p(t_e) = 232 \, h^{-1} \, \textrm{Mpc}$), and
the present sample comprises 78 measured redshifts.
It has been investigated for substructure in both the galaxy
distribution and intracluster X-ray emission by \citet{forman}, \citet{quintana},
\citet{ramirez}, \citet{beers_2}, \citet{west_2}, \citet{bliton},
\citet{jones}, and \citet{donnelly}. Overall, our results are consistent
with these previous studies although we do find some differences described below.

The cluster shows two prominent galaxy concentrations in both the galaxy
and the X-ray distributions. In the following analysis, the cluster center is 
set to be coincident with the peak X-ray emission (labeled as Core in Figure~\ref{a1750})
although the number density in the C subgroup is comparable (and the visual comparison
between the two is affected by binning).
Fitting any smooth profile over the extent of this cluster is problematic due to the
clumpy nature of the galaxy distribution. For the Core region, the elliptical King fit
yields the smallest value in this sample for the
central density of only $41 \pm 4 \; \textrm{gal}/(h^{-1}\, \textrm{Mpc})^2$.
At mean radii greater than $\gtrsim 0.50 \, h^{-1} \; \textrm{Mpc}$,
the ellipticity calculation is heavily biased by galaxies to the north and northeast
of the core region (see Figures~\ref{a1750}(a)-(d)). 

\begin{figure}
\centering
\epsfig{figure=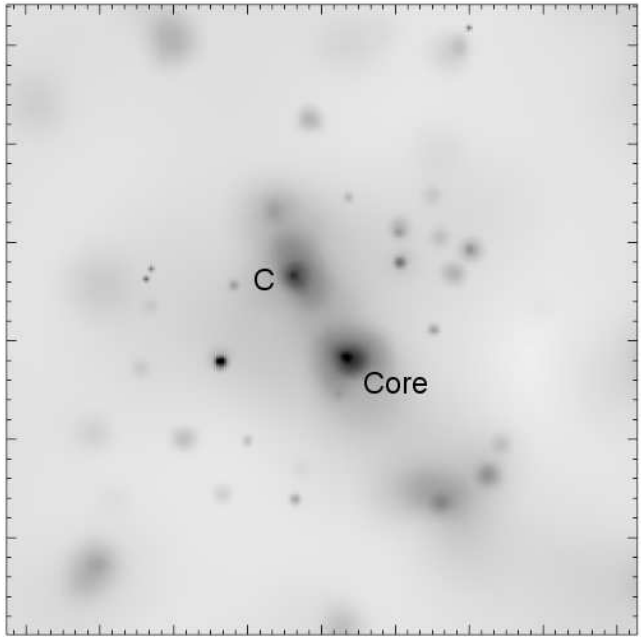,width=0.48\textwidth}
\caption{ROSAT PSPC image of Abell 1750 showing the two major components are associated
with X-ray emission indicative of bound clumps. The brightest cluster member is located
near the peak of the C subcluster emission.}
\label{a1750_xray}
\end{figure}

\begin{figure}
\centering
\epsfig{figure=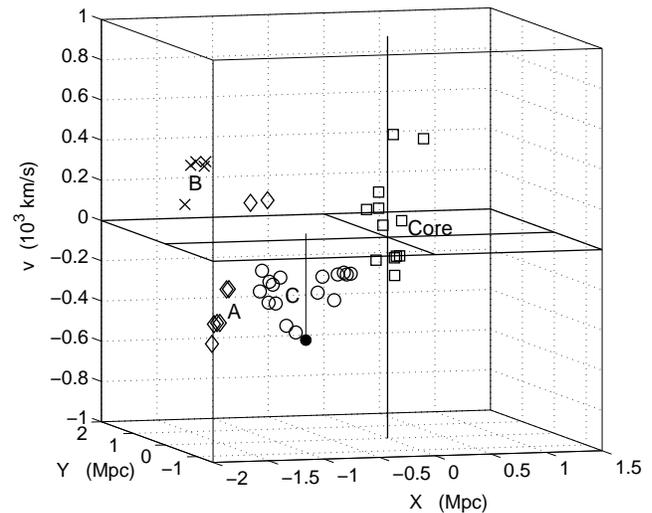,width=0.48\textwidth}
\caption{Local velocities of the nearest neighbors in Figure~\ref{a1750}(e) for Abell 1750.
In this plot, the
Core galaxies are marked by squares, the A galaxies by diamonds, the B galaxies by crosses,
and the C galaxies by circles.}
\label{a1750_nnlv}
\end{figure}

The clumpiness in the projected surface density is especially evident in Figures~\ref{a1750}(d)-(e) 
where at least 3 potential subclusters are indicated by labels A, B, and C with the C grouping possibly
possessing additional structure.
The Core and
C grouping correspond to the SW and NE subclump of
\citet{donnelly}. The other clumps identified here have no
corresponding structure in the X-ray maps presented by Donnelly et al.
or in the ROSAT PSPC X-ray map of Figure~\ref{a1750_xray}.
The peak luminosity for this cluster sample is found in subcluster C 
due to the presence of a single bright galaxy approximately 1.0 Mpc
northeast of the centre. The X-ray emission to the southeast of the core region
has no counterpart in the galaxy distribution in the current catalogue. 

In addition, the velocity histogram in Figure~\ref{a1750}(i) shows an additional
velocity component corresponding to $cz \sim 27,000$ km/s that is not present in the
Donnelly et al. data.  The hypothesis that the velocity distribution
for Abell 1750 can be drawn from a normal distribution can be rejected
at greater than the 98\% CL (Table~\ref{velstats}). 

Quantifying the highly asymmetric visual and clumpy appearance of the distribution of galaxies on the sky,
the $\beta$ test indicates that substructure is present at the $< 0.1\%$.
The higher order $\kappa$ test (Figure \ref{a1750}f) also shows distinct signs of substructure
with relatively clear evidence for the Core, A, B, and C subclusters ($P_\kappa \, < \, 3\%$). This test reveals that
that both the Core and C regions seen in the number density map may be composed of 2-3 units. There is some
evidence for this in the Xray map of Figure~\ref{a1750_xray} for C, but the Core shows no obvious substructure
in this figure. 
Further support for subclustering is revealed by the plot
of local velocities of the nearest neighbors shown in Figure~\ref{a1750_nnlv} 
where again the Core galaxies may possibly break into smaller clumps separated in local velocity
by $\sim 180 \, \mathrm{km/s}$, and with clear
identifications for the A, B, and C subclusters. 

Overall, the results above are consistent
with previous X-ray results that Abell 1750 is at least a triple
system \citep{jones}.  The velocity data presented in
\citet{beers_2} reveal that this system is even more complex with
at least four components of which the subclusters labeled as Core and D here appear
to be gravitationally bound and infalling.
Further dynamical analysis similar to that presented
in \citet{beers_1} and \citet{beers_2} applied to the subclusters identified here
will be presented in a future paper in order to confirm the conclusion in \citet{beers_2}
that the Core and C subclusters are gravitationally bound, and to ascertain whether
any of the other subclusters are also bound components.

\subsection{Abell 2734}

Abell 2734 has a well defined, elongated core (see Figures \ref{a2734}(a) - (e)).
With $\epsilon$=0.55, $PA$= 114$^\circ$ with low to insignificant ellipticity at
larger radii (Table \ref{ellipt_table}).
Comparison of the number density contours with the luminosity density contours,
Figures~\ref{a2734}(b) and (c), shows that the centre of luminosity is slightly offset
from the number density centroid due to the position of the brightest galaxy in the catalogue of this cluster
($b_J = 15.60$) at relative coordinates ($x = -0.14$ and $y = -0.23$).
Further comparison of the position angle of the 0.5 $h^{-1} \, \mathrm{Mpc}$ ellipse
with the luminosity density contours in Figure~\ref{a2734}(c) reveals that the line
connecting the centre of luminosity to the cluster centroid is approximately
parallel to the position angle of the dispersion ellipse.
The core parameters are $R_c= 0.31 \pm 0.01 \; h^{-1}\, \mathrm{Mpc}$ with a maximum
density $\sigma_0 = 121 \pm 5 \; \mathrm{gals}/(h^{-1} \, \mathrm{Mpc})^{2}$,
and where the elliptical fit of the core parameters provides a noticeably better
representation of the inner regions compared to the circularly symmetric fit.

Inspection of the redshift distribution shows the presence
of three galaxies with redshifts close to $cz = 22,000 km/s$. Each of those three are
possible background galaxies, but only one has a magnitude that makes it an obvious
non-member candidate. These three galaxies have an insignificant influence on the computation
of the velocity statistics that yields
relatively high values for the skewness and kurtosis
of $S = 0.71 \; \mathrm{and} \; K = 1.38$. These values are $2.1\sigma \; \mathrm{and} \; 1.6\sigma$,
respectively, away from the expected value for a normal distribution (see Table~\ref{velstats}).
However, the $\chi^2_{vel}$ test yields an insignificant $85\%$ CL that the distribution can
be rejected as normal.

Of the four statistical tests, the 2D $\beta$ test yields the strongest indication of
structure/substructure with the computed asymmetry having a probability of random occurrence
of only $P_{\beta} = 0.4\%$. Closer examination reveals this asymmetry is caused by what appears
to be a chance distribution -- the mirror image of a relatively isolated galaxy
happens to fall within a small group of galaxies.
The member with the highest $\beta$
statistic is located near coordinate ($0.5$, $-1.0$).  Its mirror image at coordinate ($-0.5$, $1.0$)
is located close to a group that also exhibits substructure in the $\kappa$ test (see below).
Other mirror images with high $\beta$ statistics do not fall in groups with any appreciable
3D structure, suggesting these groups may be the result of projection.  If so, the positive
$\beta$ test result could be a false positive.

With $P_\alpha = 44.8\%$ and $P_\kappa = 35.3\%$, the 3D tests show
no significant indication of dynamical substructure.
However, the $\kappa$ bubble skyplot in Figure \ref{a2734}(f) shows
a small subcluster in the northeast at approximate coordinates ($-0.4$, $1.1$), and
possibly some structure in the core. The structure in the core could be a merger
signature, but more detailed examination is required to be certain.

Prior substructure studies of this cluster found similar results. \citet{solanes}, 
using 45 members with redshift data, found no substructure ($P_S$ = 28.3\%,
$P_K$ = 41.5\%, $P_\Delta$ = 14.0\%).  Kolokotronis et al. (2001) found weak or
no substructure activity comparing optical and X-ray surface brightness parameters
($\epsilon$, $PA$, and centroid shift, $sc$).  \citet{biviano}, using
77 members with redshift data, found no substructure with the $\Delta$ test
($P_\Delta$ = 10.0\%). The $\Delta$ test result for our sample is $P_\Delta = 38\%$
(see \citealt{vick}).

\subsection{Abell 2814}

At a distance of $d_p(t_e) = 284 \, h^{-1} \, \mathrm{Mpc}$, Abell 2814 is the most distant cluster in this study. 
Figures~\ref{a2814}(a)-(b) reveal an irregularly-shaped cluster possessing a relatively well-defined core region
with isolated lower density concentrations of galaxies well removed from the central regions to the 
east and southeast labelled as groupings A and B in Figures~\ref{a2814}(d)-(e). 
A few, possibly isolated, galaxies to the north and south are also present but do
not appear to form subgroupings. Comparison of Figures~\ref{a2814}(b) and (c) show that the luminosity density
distribution is very similar to the number density distribution with most of the light emanating
from the central regions. 
\begin{figure*}
\begin{minipage}{115mm}
\centering
\epsfig{figure=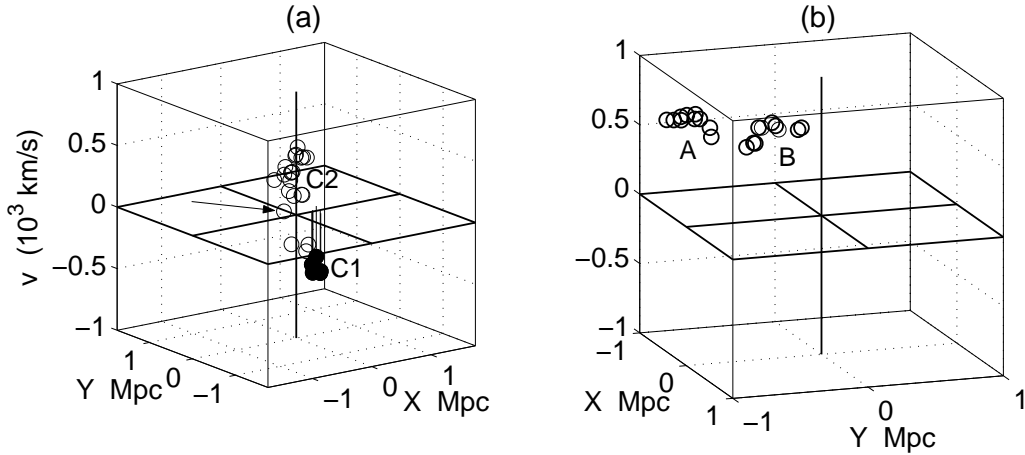,width=1.2\textwidth}
\caption{Local velocities of the nearest neighbors for Abell 2814 in Figure~\ref{a2814}(e). For visualization purposes,
only the galaxies in the (projected 2D) core are shown in (a) with the brightest cluster galaxy indicated by the arrow, 
while only the galaxies in groupings A and B are shown in (b).}
\label{a2814_nnlv}
\end{minipage}
\end{figure*}

The ellipticity and position angle of the $0.5 \, h^{-1} \, \mathrm{Mpc}$ 
mean radius ellipse are $\epsilon = 0.30$ and $PA = 100^\circ$ with the position
angle connecting the centres of the regions marked C1 and C2. Due to the galaxies to
the north and south becoming included in the ellipticity computation, the position angle 
shifts by $30^\circ$ for the 
$0.75 \, h^{-1} \, \mathrm{Mpc}$ ellipse ($\epsilon = 0.27$ and $PA = 130^\circ$)
At larger mean radii, the galaxies to the east and southeast again increase the
ellipticity while the position angle does not change appreciably
(see Table~\ref{ellipt_table}). In fitting the core parameters, 
the circular fit does well
with the outer regions while using the ellipse parameters for the $0.5 \, h^{-1} \, \mathrm{Mpc}$
ellipse provides a noticeably better fit to the inner regions, yielding 
values of $R_c = 0.36 \pm 0.01 \, h^{-1} \, \mathrm{Mpc}$ and 
$\sigma_0 = 81 \pm 4 \, \mathrm{gal}/(h^{-1} \, \mathrm{Mpc})^2$.  

The velocity distribution shown in Figure~\ref{a2814}(i) appears relatively normal, and
the $\chi^2_{vel}$ test 
gives an insignificant 68\% CL to reject the normality hypothesis. Similarly, the
skew and kurtosis show relatively small deviations from normal values (see Table \ref{velstats}).

Due to the asymmetry introduced by the concentrations to the east and southeast, the $\beta$ test  
gives a $P_\beta = 8\%$ chance of random occurrence.   
The three-dimensional $\alpha$ and $\kappa$ tests yield  $P_\alpha =1.3\%$ and $P_\kappa = 6.5\%$
probabilities of random 
occurrence, respectively.  However, the results from these latter two tests are influenced 
not by the potential substructure in the eastern and southeastern concentrations as much as by a potential
subcluster of at least eight galaxies just
west of the cluster centre labelled C1 (see Figures~\ref{a2814}(f) and (h)).  
As can be seen from the plot of nearest neighbor local velocities in Figure~\ref{a2814_nnlv}(a), the
groups of bubbles in Figure~\ref{a2814}(f) labeled as C1 and C2 appear as a distinct clumps with a few
remaining galaxies in the projected (2D) core not appearing to clump together. The brightest cluster galaxy
resides near the centre of the position-local velocity space. 
These isolated galaxies may reside outside the core region in 3D space. Whether C1 is a
subcluster in or near the core itself or is in the foreground remains to be determined,
but the structure does indicate probable infall/merger dynamics. 
From Figure~\ref{a2814_nnlv}(b), the groupings A and B are also probable subclusters.

\subsection{Abell 3027/APM 268}

Abell 3027 and APM 268 are essentially the same cluster with the primary differences
in the 2dFGRS catalogue being some galaxies to the east and northeast included in A3027 are not included
in APM268, and some galaxies to the southwest that are included in APM268 but not included in A3027
(see Figures~\ref{a3027}(a) and \ref{apm268}(a)). The central regions are similar but
not identical, leading to small but noticeable differences in the number and
luminosity density contour plots, Figures~\ref{a3027}(b), (c) and \ref{apm268}(b),
(c). Analysed together, these two clusters present an interesting
test case for checking the sensitivity of the various algorithms to small differences in
cluster membership. 

The surface plots for A3027 indicate the central region may consist of two components of approximately equal
galaxy density, but this is not as evident in the plots for APM268. 
With the centre of luminosity residing at the computed origin in both cases due to
the brightest cluster member ($b_J = 15.75$), no center re-positioning was done
(see Figures~\ref{a3027}(c) and \ref{apm268}(c)). An X-ray map of this cluster
would be useful in determining the central region distribution.
The elongated (or bimodal) central structure biases the ellipticity computation to yield high ellipticities with
position angles oriented north to south for mean radii out to $\sim 0.75$ Mpc at which point the galaxies
in the outer region become included and reduce the calculated ellipticity values (see Table~\ref{ellipt_table}).
Using the ellipticity and position
angle of the 0.75 Mpc mean radius, the core parameters are computed to be
$R_c = 0.29 \pm 0.01$ and $\sigma_0 = 90 \pm 3$ for A3027 and
$R_c = 0.30 \pm 0.01$ and $\sigma_0 = 81 \pm 3$ for APM268.

Four groupings of galaxies outside the core are common to both
clusters, and are marked as A, B, C and E in Figures~\ref{a3027}(d) and
\ref{apm268}(d). A concentration to the northeast in Abell 3027 is marked as F in
Figure~\ref{a3027}(d), and a concentration to the southwest is marked as D in Figure~\ref{apm268}(d).
Grouping C contains the second brightest galaxy in this sample ($b_J = 16.08$) so that the luminosity 
density is only slightly less than in the core.

Inspection of the velocity histograms in Figures \ref{a3027}(i) and
\ref{apm268}(i) hint at superpositions of distributions that considered 
as a single entity is clearly not normal. The $\chi^2_{vel}$ test 
rejects the hypothesis of a single normal distribution at greater than the
99.9\% CL and $98\%$ CL for APM268. The galaxies in the north contribute most of the slower galaxies (marked
by the open circles in Figures~\ref{a3027}(a) and \ref{apm268}(a)),
and the distribution  centered at $\sim 23,000 \, \mathrm{km/s}$ 
resulting from eliminating galaxies with velocities less
than $\sim 22,000 \, \mathrm{km/s}$ is decidely more gaussian-shaped. 

The $\alpha$ and $\beta$ tests yield no significant substructure for A3027 but do for APM268 
(Table \ref{sumresults}). With $P_{\kappa} = 0.2\%$ and $2.2\%$, the $\kappa$ test
does show strong indications of substructure. The slower galaxies to the north
have relatively large bubbles, but do not appear to be part of any organized structure,
and may be in the foreground, and grouping A does not appear to be a 3-dimensional structure. 
Grouping B, just outside the core region, does not show significant $\kappa$, but does show
some separation from the core in Figures~\ref{a3027}(f) and \ref{apm268}(f).
However, groupings C and E show significant $\kappa$
and are possible subclusters.  
Groupings D and F do not show a large amount of $\kappa$,
but appear in the local velocity plots of Figures~\ref{a3027}(h) and \ref{apm268}(h)
(along with C and E), and may be subclusters with dominant galaxy motions tranverse to the line-of-sight,
or subclusters with similar velocity distributions to the core region. 
Overall, A3027/APM268 shows substructure in the core and outer regions that should be
further investigated through a detailed X-ray study.

\begin{figure*}
\begin{minipage}{115mm}
\centering
\epsfig{figure=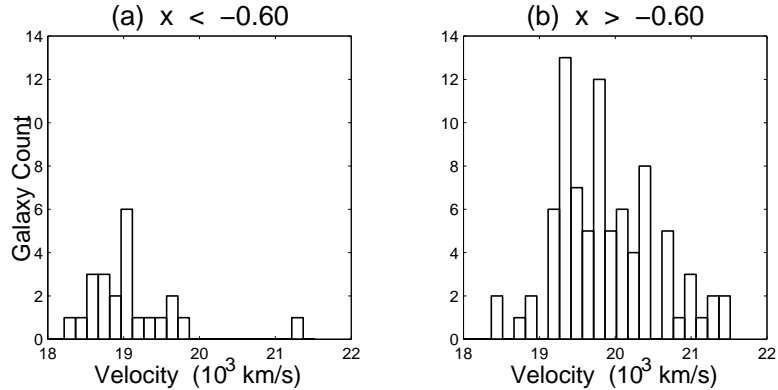,width=0.9\textwidth}
\caption{Abell 3094 Velocity Distribution of (a) the East Group
($x < -0.60 \, h^{-1}$ Mpc)  and (b) the Other Cluster Members
($x > -0.60 \, h^{-1}$ Mpc ).}
\label{a3094_2velhist}
\end{minipage}
\end{figure*}

\begin{figure*}
\begin{minipage}{115mm}
\centering
\epsfig{figure=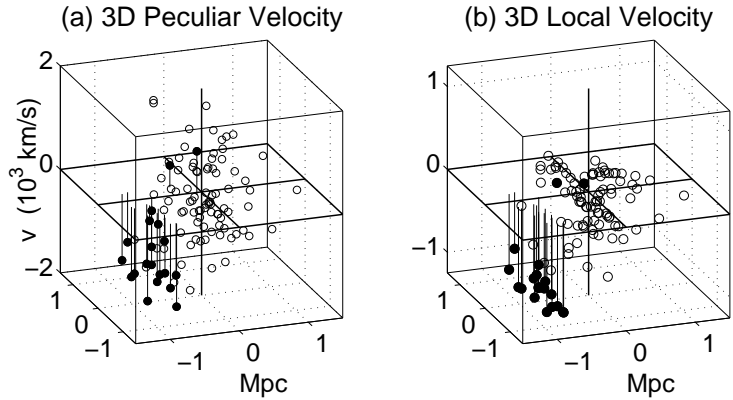,width=0.85\textwidth}
\caption{Abell 3094 (a) 3D peculiar velocity, and (b)
3D local velocity plots with the outlier removed.}
\label{a3094_2vel3D}
\end{minipage}
\end{figure*}

\subsection{Abell 3094}

The 2D plots in Figures \ref{a3094}
(a)-(e) show a cluster with an extended low density
central region with two projected surface density peaks
oriented north to south.
Concentrations of galaxies with densities greater than 
$40 \, \mathrm{gal}/(h^{-1} \, \mathrm{Mpc})^{2}$ appear 0.75 - 1.0 $h^{-1}$
Mpc to the east and north-northwest of the cluster centre.
The luminosity contour plot has a similar shape as the high
density regions in the isodensity contour plot, but without
a bimodal central region (at the resolution shown here).

Due to the apparent bimodal central region, ellipticity at the
smaller mean radii is noticeable
with $\epsilon = 0.55 \, \mathrm{and} \, 0.49$ at mean radii
of 0.5 and 0.75 $h^{-1} \, \mathrm{Mpc}$, respectively.
At mean radii of 1.0 to 1.25
$h^{-1}$ Mpc, the groupings to the east and north-northwest
bias the ellipticity computation resulting in values of
$\epsilon = 0.37 - 0.33$ at postion angles of $PA \approx 140^\circ$.
We use the $\epsilon$ and $PA$
for the mean radius of 1.0 $h^{-1}$ Mpc to compute a mean
core radius $R_c = 0.39 \pm 0.01 \, h^{-1}$ Mpc and a
central density $\sigma_0 = 88 \pm 2$ (Table \ref{core}).

The velocity distribution shown in Figure \ref{a3094}(i) is skewed
toward higher velocities with $S = 0.42$
($1.15 \sigma$ from normal) while the kurtosis is close to normal.
The relatively high value of skewness is probably the primary cause
that the $\chi^2_{vel}$ test rejects of the normal
hypothesis at a 93\% CL (Table \ref{velstats}).

The $\beta$ test indicates cluster asymmetry with a marginally significant
$P_\beta$ = 10\%, and 
the $\kappa$ test indicates a very high probability of
substructure with $P_\kappa < 0.1\%$. Most of the significant
$\kappa$ in the core is associated with galaxies in the
northern projected component.
However, while the local velocity plot of
the nearest neighbors in Figure~\ref{a3094_2vel3D}(b) does indicate
some spatial structure in the core,
it is not entirely clear whether it should be classified as bimodal.
Further inspection of the $\kappa$ bubble skyplot in Figure \ref{a3094}(f)
shows that most members with a location of $x < -0.60 h^{-1}$
Mpc have a relatively high $\kappa$ statistic.  Analysis
shows that this group, with the exception of one galaxy,
has a significantly slower velocity distribution
than the other members of the cluster with $x > -0.60 \, h^{-1}$
Mpc (see Figure \ref{a3094_2velhist}).
%\begin{figure*}
%\begin{minipage}{115mm}
%\centering
%\epsfig{figure=fig_17.eps,width=0.9\textwidth}
%\caption{Abell 3094 Velocity Distribution of (a) the East Group
%($x < -0.50 \, h^{-1}$ Mpc)  and (b) the Other Cluster Members
%($x > -0.50 \, h^{-1}$ Mpc ).}
%\label{a3094_2velhist}
%\end{minipage}
%\end{figure*}

The mean velocity of the eastern grouping (not including the outlier)
is 18,989 km/s or 898 km/s slower than cluster members to
the west.  The velocity dispersion of the eastern grouping is
$\sigma_{vel}$ = 401 km/s, significantly smaller than the
$\sigma_{vel} = 662 \, \mathrm{km/s}$ dispersion of the galaxies to the west 
(and $\sigma_{vel}$ = 728 km/s for the entire cluster).
For ease of visualization, a dashed line
is drawn on Figures \ref{a3094}(a)-(f) showing the approximate border
of the eastern grouping.
Ten of the 23
members in the east group show a $> 5\%$ probability that
the member and its nearest neighbors are not from the same
distribution as the entire cluster.  When the outlier is
removed, the number of members with a $> 5\%$ probability
increases from 10 to 18.  A possibility for shearing or infall motion is also
more evident in the 3D plots with the outlier removed as
shown in Figures \ref{a3094_2vel3D}(a) and (b). 
%\begin{figure*}
%\begin{minipage}{115mm}
%\centering
%\epsfig{figure=fig_18.eps,width=0.65\textwidth,angle=-90}
%\caption{Abell 3094 (a) 3D peculiar velocity, and (b)
%3D local velocity plots with the outlier removed.}
%\label{a3094_2vel3D}
%\end{minipage}
%\end{figure*}

In the substructure study of ENACS clusters, \citet{solanes}
used 46 redshifts to compute a $\Delta$ test statistic 
of $P_\Delta$ = 0.4\%, similar to that given in \citet{biviano}.
These results are consistent with the value of $P_\Delta = 0.4\%$ found by \cite{vick}
and the $\kappa$ test result reported here
for the sample of 108 2dFGRS redshifts. 
Other tests in the ENACS study were less conclusive.

In summary, Abell 3094 shows a high probability of substructure with
the $\kappa$ test, and some indication with the
$\chi^2_{vel}$ and $\beta$ tests (see Table \ref{sumresults}).  A grouping of 22
galaxies located east of $x = -0.60 \, h^{-1}$ Mpc  may represent
a foreground group or a large subcluster with infall or possibly shearing
motion.

\subsection{Abell 3880}

Inspection of Figures~\ref{a3880}(a)-(b) shows Abell 3880 as an elliptical cluster 
with a well-defined, dense core. Comparison of Figures~\ref{a3880}(b)-(c) shows that 
the luminosity density of the core region closely traces the number density.
From Figure~\ref{a3880}(b), there are at least three potential subclusters(A, B, and C) with two 
having density greater than 40 galaxies per ($h^{-1}$ Mpc)$^{2}$, each of which is just
over 1.0 $h^{-1}$ Mpc from the cluster centroid (see Figure~\ref{a3880}(d)). 
Two of these groupings, A and B, show a relatively high level of luminosity
(Figure \ref{a3880}(c)). This combination of high luminosity with
slow peculiar velocity hints at possible foreground contamination; inspection of the
velocity-magnitude plot reveals at least four galaxies as candidate, but not obvious, non-members. 

Closer inspection of the three groupings shows each to contain five
galaxies in close projected (2D) spatial proximity. In particular, 
grouping B contains the brightest galaxy in this sample ($b_J = 14.69$) with a velocity
a little more than $1.3\sigma$ away from the cluster mean (one of the non-member
candidates mentioned above). Of the remaing four galaxies in grouping B, two have
velocities very similar to the brightest galaxy, $\approx 1050 \; \mathrm{km/s}$ 
less than the cluster mean, while the other two have velocities $\approx 365$ km/s
greater than the cluster mean. Thus, it is possible that grouping B consists of a
foreground group of three galaxies with projected surface positions close to two cluster members.
Grouping C is similar to B in that of the five galaxies, one is relatively bright ($b_J = 15.45$)
with a velocity more than 1200 km/s less than the cluster mean, two others have similar velocities,
while the remaining two have velocities more than 300 km/s greater than the cluster mean.
Finally, grouping A possesses relatively low luminosity, and consists of 
three galaxies having velocities close to the cluster mean and two galaxies having
velocities 705 and 917 km/s faster than the mean.

Corresponding to the elongated contours tracing the central regions, the ellipticity 
is relatively high for the $0.50\; \mathrm{and} \; 0.75 \; h^{-1}$ Mpc ellipses
with $\epsilon = 0.59 \; \mathrm{and} \; 0.50$ at position angles of $PA = 150^\circ \;
\mathrm{and} \; 159^\circ$. As 
the mean radius is increased to 1.0 $h^{-1}$ Mpc, the ellipticity 
decreases to a relatively low value ($\epsilon = 0.16, PA = 156^\circ$) due to 
the influence of the three high density areas outside the core. 
To reduce the influence of the groupings in the determination of the core
parameters, the ellipticity and position angle are averaged for the $0.75 \; \mathrm{and} \;
1.0 \; h^{-1}$ Mpc mean radii to yield best fit parameters of $R_c = 0.25 \pm 0.01 \;
h^{-1}$ Mpc and $\sigma_0 = 161 \pm 13 \; \mathrm{gal}/(h^{-1} \; \mathrm{Mpc})^2$ (a relatively
high value for the central density). The resulting elliptical fit to the density profile is
noticeably better than the circular fit in the inner regions of the cluster.

Other than the three groupings described above, there is little obvious substructure present in this
cluster. The statistical $\alpha$, $\kappa$, and $\chi^2_{vel}$ yield results for
substructure below the 90 \% CL. However, it is worth noting that the nearest neighbor plot and the $\kappa$
bubble skyplot indicate that a fourth grouping of at least five galaxies, marked D in Figure~\ref{a3880}(e),
is also a possible subgroup (and might be detectable in a long exposure X-ray observation).
\citet{stein} initially considered but rejected A3880 for investigation due to there being less
than 25 galaxies within $0.5 \, h^{-1} \, \mathrm{Mpc}$ of the centre in the catalogue used for that
study; in the study here, there are $\sim 45-50$ galaxies within a $0.5 \, h^{-1} \, \mathrm{Mpc}$
circle or dispersion ellipse. However, Stein quotes a mean cluster velocity and dispersion for the
core galaxies consistent with what is found here.

\subsection{Abell 4012}

Abell 4012 has a low-density core with one obvious subgrouping of galaxies
approximately $1.0 \, h^{-1}$ Mpc
due north of the centre with a density exceeding 40 gal/($h^{-1}$ Mpc)$^2$ 
(see Figures \ref{a4012}(a), (b)). 
These five galaxies in the northern grouping are marked as A in Figure~\ref{a4012} 
and are discussed further below
(two galaxies have nearly coincident angular positions and are indistinguishable in the plots). 

At small mean radii, the cluster has a moderate level of ellipticity
that decreases to insignificance at the larger mean radii (Table 
\ref{ellipt_table}).  At a mean radius of 0.5 $h^{-1}$ Mpc,
$\epsilon = 0.42$ with $PA = 14^\circ$ where the position angle points                                   
directly toward Abell 4013.
By a mean radius of 1.25 $h^{-1}$ Mpc, the ellipticity is  
negligible ($\epsilon$ = 0.01, $PA$ = 133$^\circ$).   
Compared to a circular fit, the elliptical fit is noticeably better with 
a mean core radius of  $R_c = 0.28 \pm 0.01 h^{-1}$ Mpc 
and $\sigma_0 = 71 \pm 3$ galaxies/$(h^{-1} \mbox{Mpc})^{2}$.

With a value of $\sigma_{vel} = 471$ km/s, the velocity dispersion is one of the smallest found in this study
(Table \ref{velstats}).  Inspection 
of the velocity distribution in Figure \ref{a4012}(i) shows what appears to be a somewhat 
normal distribution. The skewness and kurtosis values of $S = -0.08$ and $K = 0.37$ are each
less than $0.5\sigma$ from what is expected for a normal distribution (see Table \ref{velstats}).
However, the $\chi^2_{vel}$ test shows a 96\% confidence level to reject the normal distribution 
hypothesis.  

Inspection of Figures \ref{a4012}(a)-(e) shows the only clear sign of subtructure 
as the previously mentioned small clump of galaxies to the north of centre. 
However, consisting of only five galaxies, this grouping is not enough to
cause the $\beta$-statistic to indicate significant asymmetry ($P_\beta = 26\%$). 

The $\alpha$ test indicates substructure 
($P_{\alpha} = 1$\% probability of random occurrence) but the
$\kappa$ test does not ($P_{\kappa} = 21$\%).
In order to better isolate any apparent dynamical clumpiness in this cluster, the nearest
neighbor positions in Figure~\ref{a4012}(e) are displayed in a local velocity plot in
Figure~\ref{a4012_nnlv}. This plot clearly shows the grouping A to be a probable bound subcluster.
If this interpretation is correct, the local gravitational potential minimum should be detectable in a 
long exposure X-ray observation.
The other hints of clumpiness in Figure~\ref{a4012}(e) are not as clearly manifested in Figure~\ref{a4012_nnlv},
but cannot be completely excluded as small subclusters or merger remnants.

\begin{figure}
\centering
\epsfig{figure=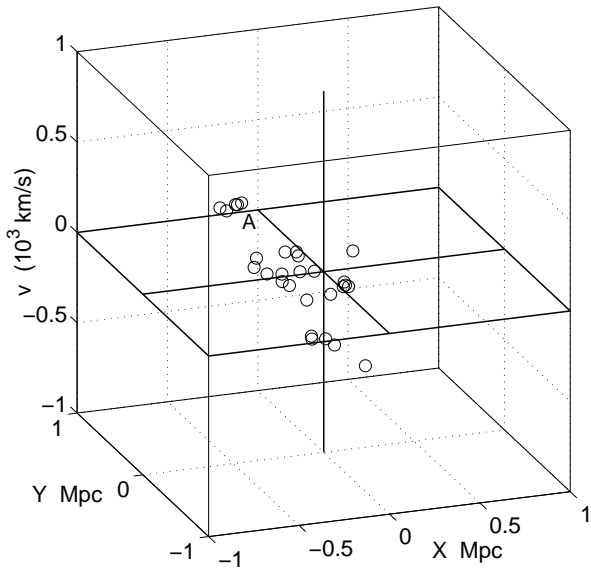,width=0.45\textwidth}
\caption{Local velocity plot of the nearest neighbor positions for Abell 4012 shown in Figure~\ref{a4012}(e).}
\label{a4012_nnlv}
\end{figure}

\subsection{Abell 4013}

Inspection of Figures \ref{a4013}(a),(b), and (d) shows a cluster with a well-defined, 
dense core.  Approximately 0.5 $h^{-1}$ Mpc south of the cluster centre is an additional high 
density group of galaxies having a peak density of greater than 60 gal/$(h^{-1} \mbox{ Mpc})^2$. 
Two small groupings appear west and southwest of 
the centre.  
%Both of these groups have peak densities of 
%greater than 40 gal/$(h^{-1} \mbox{ Mpc})^2$. 
For this cluster, the luminosity density 
is very similar to the number density (compare Figures \ref{a4013}(b) and (c)).

Within the $0.5 \, h^{-1} \mbox{ Mpc}$ dispersion ellipse, the values of $\epsilon$ = 0.29 and $PA$ = 163$^\circ$
are biased by the high-density group of galaxies to the south. The ellipticity 
decreases at larger radii as the influence of this high-density subcluster
decreases.  At a mean radius of 1.0 $h^{-1}$ Mpc, $\epsilon$ =
0.20 with a $PA$ = 116$^\circ$. 
The core fit yilds a core radius of $R_c = 0.17 \pm 0.01$
and a maximum density of $\sigma_0 = 224 \pm 2 \, \mathrm{gal}/(h^{-1} \, \mathrm{Mpc})^2$. The resulting fit 
accurately reproduces the density profile for most of the cluster.
The $\beta$ test indicates
substructure at high significance with $P_{\beta} = 2$\% 
where the high
level of asymmetry is caused by the southern subgroup coupled with the relative
absence of galaxies to the north.

Inspection of the velocity histogram (Figure \ref{a4013}(i)) 
shows a group of fast galaxies well-separated from the rest of the distribution:  
the majority of the cluster members are grouped around a 
velocity of 15,600 km/s, but a small group of 14 are found 
near 17,600 km/s.  This feature
explains the $>$ 99.9\% CL to reject 
the hypothesis that the distribution is normal.  These faster galaxies 
are in three spatial locations, and can clearly be identified in  
Figure \ref{a4013}(a) (marked by `+' symbols).  
Five galaxies are approximately 1.0 
$h^{-1}$ Mpc east of the centre, six 
galaxies appear near the centre of the cluster, and 
three are approximately 1.0 $h^{-1}$ Mpc southwest of 
the centre.

The $\kappa$ test yields an indication of substructure at slightly greater than
the 95\% CL, but inspection of Figure \ref{a4013}(f) shows that the large southern 
subgroup has relatively small bubbles, possibly indicating any dynamical interaction
between core and subcluster is perpendicular to the line of sight. The high velocity galaxies 
to the east with the large $\kappa$ bubbles are also relatively fainter than the majority of
cluster members, and may be background galaxies. But there are also significant $\kappa$
bubbles in the core region that may indicate the merger or dynamical interaction of small 
subgroup(s)(here, spatially unresolved) 
as indicated by the presence of slow and fast galaxies in the core.

\begin{figure*}
\begin{minipage}{115mm}
\centering
\epsfig{figure=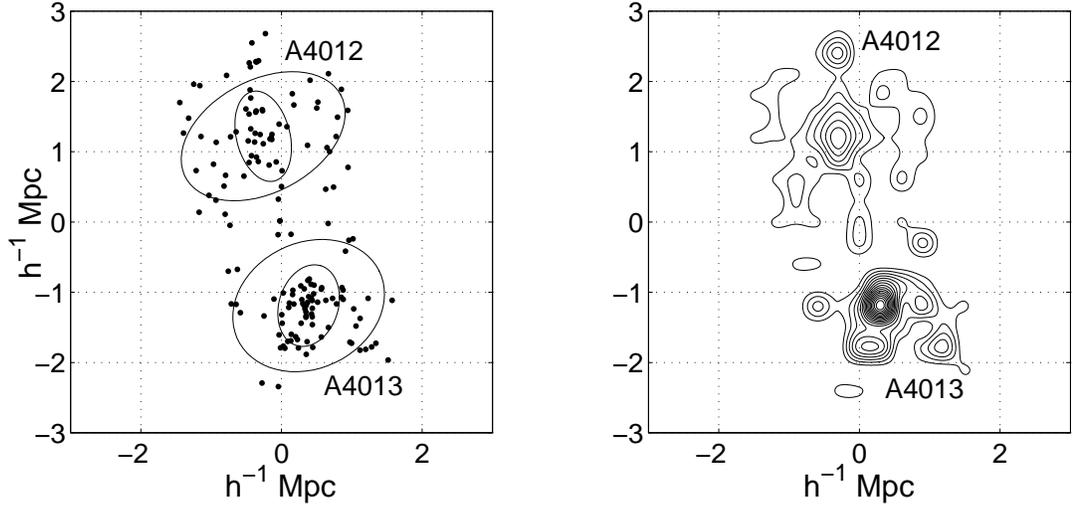,width=0.6\textwidth,angle=-90}
\caption{Relative galaxy positions and number density contours for Abell 4012 and Abell 4013.}
\label{a4012_4013}
\end{minipage}
\end{figure*}

\begin{figure*}
\begin{minipage}{115mm}
\centering
\epsfig{figure=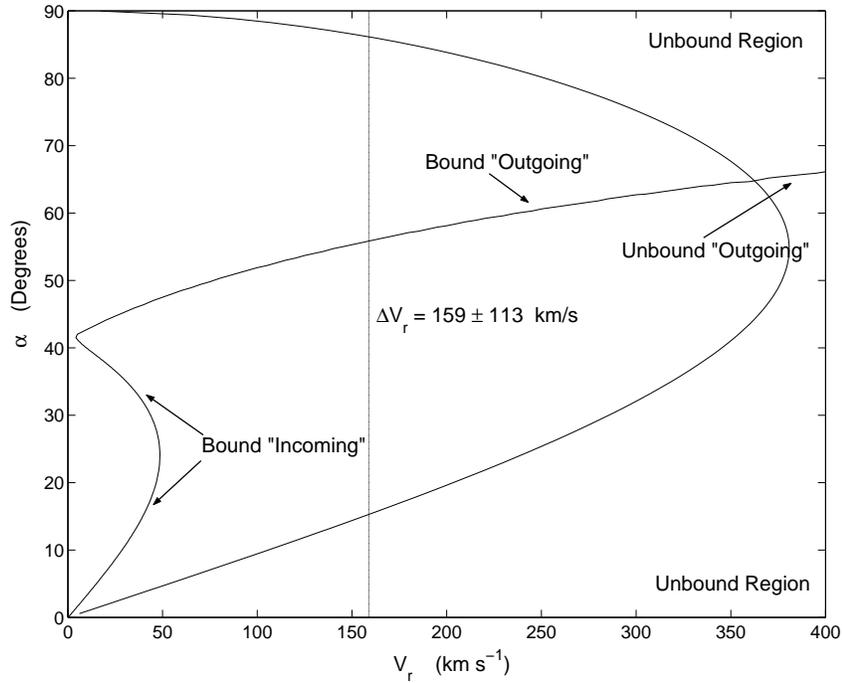,width=0.8\textwidth,angle=-90}
\caption{Bound and unbound solution space for Abell 4012 and 4013. A value of $H_0 = 70$ has been assumed 
for the solution of the parametric equations of motion. The angle $\alpha$ is a projection angle between the
line-of-sight and the vector joining the two cluster centroids.}
\label{a4012_4013_bind}
\end{minipage}
\end{figure*}

\subsection{Abell 4012 and Abell 4013 as a possible bound system}

It is worth
noting the spatial proximity of these two clusters. The projected surface (2-dim) separation
between the two centroids is $\approx 2.5 \, h^{-1}$ Mpc, while their mean redshifts
are $c\overline{z} = 16,230 \pm 498$ km/s and $c\overline{z} = 16,410 \pm 904$ km/s, respectively.
Figure \ref{a4012_4013} shows the galaxy positions and number density contours from which
it can be seen that the southern boundary of A4012 is relatively close to the northern boundary
of A4013. Superimposed
on the plot of galaxy positions are the $0.5 \mbox{ and } 1.0 h^{-1}$ Mpc dispersion ellipses
showing the approximate alignment of the inner regions of the two clusters. 

Given their proximity,
it is worthwhile assessing the likelihood that the two clusters are
gravitationally bound to one another. The cluster masses are estimated using the virial relation
\begin{equation}
M_{vir} = \frac{N}{G} \sum_{i} v_{i}^2 \left( \sum_{i} \sum_{j<i} \frac{1}{r_{ij}} \right)^{-1} \; \mbox{,}
\end{equation}
where $N$ is the number of cluster members, $v_i$ is the peculiar velocity of the $i^{th}$ member, and
$r_{ij}$ is the projected separation between the $i^{th}$ and $j^{th}$ galaxies. Inspection of the velocity
distribution for Abell 4013 reveals 14 galaxies with velocities significantly faster than the other 71
members, and these are removed in the following analysis in order to minimize their effect on the velocity
dispersion (see Figure~\ref{a4013}(i)). Thus, we estimate the mass of Abell 4012 to be $\approx 7 \times
10^{13} \mathrm{M}_{\bigodot}$ with a velocity dispersion of $\sigma_{vel} = 471$ km/s, and a mass of $\approx 4 \times
10^{13} \mathrm{M}_{\bigodot}$ for Abell 4013 with a velocity dispersion of $\sigma_{vel} = 455$ km/s based
on eliminating the ``fast'' galaxies.
Using the Newtonian binding criterion 
\begin{equation}
V_r^2R_P \le 2 G M_{tot}\, \sin^2 \alpha \, \cos \alpha \; \; \mbox{,}
\end{equation}
and the parametric
analysis described in \cite{beers_1} for these masses, projection angle $\alpha$, projected 
centroid separation, and a
radial velocity difference of $V_r = 159$ km/s, we find a $\sim 95\%$ probability that A4012 and A4013 form
a bound system
with a less than $10\%$ probability for the bound `incoming' solutions.
The probability calculation is based primarily on the possible deviation of
the actual radial velocity separation being different from the $V_r = 159$ km/s value 
assuming no worse than a $113 \, \mathrm{km/s}$ RSS measurement error.
From Figure~\ref{a4012_4013_bind}, the system is bound as long as $V_r$ is accurate to within 200 km/s where
the parametric solutions indicated in the figure follow the terminology in \citet{beers_1}.

\subsection{Abell 4038}

Abell 4038 is a fairly rich cluster of richness class $\textrm{R} = 2$, and we
present analysis based on 154 galaxies in this system (the most of any
cluster in this study).  The core of this
cluster is well defined with a central density of $\sigma_0 = 261 \pm 15$ and fitted
optical core radius of $R_c =  0.19$ Mpc.  To the east of the core, a dearth of
galaxies can be seen in the galaxy distribution (Figure \ref{a4038}(a)) at
($-0.5$, $-0.1$) Mpc. This is also evident in Figures~\ref{a4038}(b),(d), and
(e). This apparent gap in the distribution is caused by the bright
star (SAO 192167, $m_v$=4.6) which limits spectroscopy in the
vicinity. The location of the star and a very approximate bounding area
inside which galaxy identification is probably unobtainable are indicated by the (circle + cross)
in Figures~\ref{a4038}(a),(b),(d) and (e). From these figures it can be seen that while
the blanking area covered by the star represents a substantial fraction of the gap area, there
are still portions to the north and south that are clearly devoid of galaxies.
If the gap is not real, one expects to see galaxies just outside the blanking area, so
that it appears that at least part of the galaxy-free region is a true feature of the distribution.
But it should be noted that the inability to detect cluster members
in the vicinity of SAO 192167 could significantly bias the results of any substructure tests
using the global cluster galaxy distribution on the sky if, in fact, cluster members are not
being included. However, substructure tests
that evaluate local parameters or the velocity distribution should be
less affected.

Not surprisingly, the gap in the galaxy distribution strongly 
affects the analysis of this cluster including the ellipticity computation. 
Looking at the smallest radius of 0.5
h$^{-1}$ Mpc, the void has a strong influence on the ellipticity
by cutting off the distribution on the east side. This results in 
a relatively large ellipticity of 0.43 and a position angle 
of $21^\circ$. At the largest radius the ellipticity has risen to 
0.50 and the postion angle has moved to $96^\circ$. This is due to 
the apparently disjoint galaxy clump at ($-1.5$, 0) Mpc and the transition
would possibly be more gradual without the data gap. 

The $\alpha$ and $\beta$ tests indicate the presence
of strong substructure. Since these two tests are more heavily influenced by the global
distribution of galaxies on the sky, the gap in the galaxy distribution
biases the results of these tests. The $\kappa$ test also
indicates the presence of substructure with a very high
significance but since this is a local test, it should be less influenced
by the gap in the data. However, the absolute significance of the test needs
to be investigated in more detail for this cluster.  

The bubble plot in Figure~\ref{a4038}(f) shows at
least four areas with evidence of substructure.  The most obvious are two groupings
$\sim -1.2$ Mpc to the east, and separated from the main cluster
by the gap discussed above. However, these features show up in the
velocity plots (Figures~\ref{a4038}(g) and (h)), and are clearly at 
discrepant velocities compared to the main cluster.  The next most
discrepant grouping is to the south of the main cluster at ($-0.4$, $-1.0$)
Mpc, but this grouping is only weakly apparent in the number or luminosity 
density plots. This subclump appears to be well separated from the main
concentration, but may be affected by the gap near the bright star.  One other
subclump is located west of the centre. The clump
at (1.0,0.0) Mpc shows up as a low-level contour in the number
density map and also can be seen in the local velocity plot. There may be
additional substructure just northeast of the core
(see Figures~\ref{a4038}(b)) and (d)).

Although the $\alpha$, $\beta$, and $\kappa$ tests yield strong signals
for substructure, the $\chi_{vel}^2$ test rejects the normality hypothesis
at only the 96\% CL. Yet the velocity distribution as shown in Figure~\ref{a4038}(i)
does show hints of being a superposition of two (or more) distributions.

This cluster has been previously studied for substructure in the galaxy distribution
by \citet{girardi}, \citet{stein}, and \citet{kriessler},
and using ROSAT X-ray data by \citet{schuecker}. 
However, these studies considered the central region
of the cluster, and did not include the eastern extension that may or may not be associated. 
The velocity statistics and $\Delta$ test value are consistent with those found here for
that region. The Kriessler \& Beers study presents a contour plot of the
central region significantly different in appearance from ours, and their KMM analysis
identifies three modes that we are unable to resolve.
However, a new
study of this cluster in X-ray with longer exposure times should further clarify
the true substructure and resolve the problematic constraints
imposed by SAO 192167.

\subsection{Abell S141}

\begin{figure}
\centering
\epsfig{figure=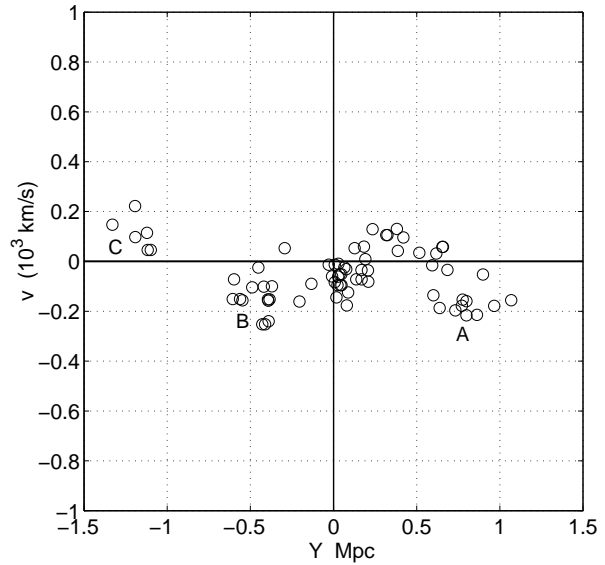,width = 0.45\textwidth}
\caption{Nearest neighbor local velocities versus the y-coordinate for Abell S141. The apparent
clumping in 2D also appears in 3D for at least groupings A, B, and C.}
\label{s141_nn_lv}
\end{figure}

At a distance of $d_p(t_e) = 57 \, h^{-1} \, \textrm{Mpc}$, Abell S141 is the nearest cluster 
in the study ($cz = 5793 \, \textrm{km/s}$).
Inspection of Abell S141 2D plots in Figure~\ref{s141} shows a highly
elongated structure with two apparent clumpings in the south and possibly one in
the north. The core is relatively dense, but, 
together with S1043, S141 has the lowest luminosity of the clusters studied here.

For the four mean radii from 0.5 to 1.25 $h^{-1}$ Mpc,
the ellipticity and position angle have a narrow range
from $\epsilon$ = 0.67 -- 0.58 and $PA = 0^\circ \, (\mathrm{or} 180^\circ)$ --
$168^\circ$ (see Table \ref{ellipt_table}).  We use the
average ellipticity of $\epsilon = 0.64$ and position
angle of $PA = 173^\circ$ to calculate a core radius
$R_c = 0.26 \pm 0.02$ and a central density $\sigma_0 = 158 \pm 18$.

The velocity dispersion of $\sigma_{vel} = 403 \, \mathrm{km/s}$
for the distribution shown in Figure~\ref{s141}(i) is the lowest in
this study. Also from this figure, it is evident that the
distribution is skewed toward higher velocities 
due to the ten relatively fast members labelled with `+' symbols in
Figure~\ref{s141}(a).
The $\chi_{vel}^2$ rejects the normality hypothesis at the 98\% CL,
but it is difficult to correlate the results of this
test as due to any of the groupings except for possibly the one farthest south.

The three apparent subclumpings in the segmentation and 
nearest neighbor plots are labelled as A, B, and C in Figure~\ref{s141}(e).
The local velocities of the nearest neighbors shown in Figure~\ref{s141_nn_lv}
appear to support the hypothesis that groupings B and C form subclusters with  
possibly also a fraction of grouping A forming a subcluster.
However, the statistical tests yield no clear signal for substructure with
the $\beta$ and $\alpha$ tests yielding $P_\beta$ = 23\% and
$P_\alpha$ = 34.6\%.
The lack of a signal in the $\beta$ test is easily explained from
the obvious symmetry in the projected directions. The low velocity dispersion
combined with the 2D symmetry is probably responsible for the lack of a
signal from the $\alpha$ test.

Although the $\kappa$ test result of $P_\kappa = 19.8\%$
using ten nearest neighbor velocities ($\approx \sqrt{N}$) is not statistically
significant, there are some indications of subclustering in the bubble
skyplot in Figure~\ref{s141}(f) corresponding to the groupings described above.
Performing the $\kappa$ test with numbers of nearest
neighbor velocities other than $\sqrt{N}$ did not appreciably change the significance
level. However, for this cluster, the small dispersion of velocities naturally reduces the
discriminating power of the test. This is probably due to the geometrical effects introduced
from analyzing a cluster with one relatively long and two short symmetry axes
along a direction perpendicular to the long axis. Thus, we conclude that in addition
to the possible subclustering, the global shape of this cluster is triaxial with
one long and two short axes (or possibly prolate). This geometry
may be linked with the supercluster environment in which S141 formed.

\subsection{Abell S258}

Abell S258 is highly elongated, as is evident upon
viewing the 2D plots (Figures
\ref{s258}(a) - (e)).  Its length stretches over
3 $h^{-1}$ Mpc in an east to west direction
while its width is typically 1 $h^{-1}$ Mpc.  It has
a moderately dense central region that also appears highly
elliptical in an east to west direction.

%\noindent
The ellipticity and position angle are very
consistent for all mean radii with ranges of
$\epsilon$ = 0.75 -- 0.58 and $PA$ =
$114^\circ$ -- $104^\circ$ (see Table
\ref{ellipt_table}).  This cluster has the highest
ellipticity in the study at a mean radius of 0.5 $h^{-1}$ Mpc
and the highest or second highest ellipticity at mean radii from 0.75 -- 1.25
$h^{-1}$ Mpc.  Using the ellipticity and position angle
for the 1.0 $h^{-1}$ Mpc test radius ($\epsilon = 0.71$,
$PA = 104^\circ$), we find a mean core radius, $R_c = 0.28 \pm 0.04$,
and a maximum density, $\sigma_0 = 108 \pm 1$ (Table
\ref{core}).

%\noindent
The velocity distribution appears bimodal (see Figure
\ref{s258}(i)).  This characteristic, however, is not
sufficient to cause a negative test of normality as 
the $\chi^2_{vel}$ test rejects the normal
hypothesis at an insignificant 70\% CL (Table \ref{velstats}).

%\noindent
Only the $\kappa$ test indicates significant
substructure (Table \ref{sumresults}), but that
test results in a very high significance level with a
$P_\kappa < 0.1\%$.  Inspection of the $\kappa$ bubble
skyplot (Figure \ref{s258}(f)) shows a large number of
members with a high level of $\kappa$ in the far eastern
part of the cluster.  Of the
14 members east of $x = -1.0 \, h^{-1}$ Mpc, 12 are
grouped very close in position and velocity.  These
12 galaxies are found to have a much slower mean velocity 
($\overline{v}$ = 16,863 km/s), are more narrowly dispersed
($\sigma_{vel}$ = 272 km/s) compared to the other members of the cluster
($\overline{v}$ = 17,584 km/s and $\sigma_{vel}$ = 526 km/s), and 
are responsible for the bimodal appearance of the
velocity distribution. They can be seen clearly in the 3D velocity
plots, Figures \ref{s258}(g) and (h). 

\begin{table}
\caption{Abell S258 Comparison of East and West Group.}
\label{s258groups}
\begin{center}
\begin{tabular}{cccc}
\hline\\
 & East Group & West Group & Cluster\\
\hline\hline
 N               &    12       &     75       &     87       \\
$\overline{v}$   & 16,863 km/s & 17,584 km/s  & 17,485 km/s  \\
$\sigma_{vel}$   &    272 km/s &    526 km/s  &    557 km/s  \\
$\overline{b_J}$ &   18.0      &   18.0       &   18.0       \\
$\chi^2_{vel}$   &   -         & CL = 67\% & CL = 70\% \\
\hline
\end{tabular}
\end{center}
\end{table}

It is unclear if this eastern sublcluster is, in fact, part of
Abell S258.  It is possible that this group is a small foreground
cluster.  The eastern group has an average redshift of 767 km/s
less than the remainder of the cluster which could equate to a foreground
distance of 11 Mpc (using $h$ = .70).  However, the velocity difference of 721 km/s is also
within the range of possible orbital velocities estimated for
a subcluster orbiting the main body.  A binding probability analysis will be presented in a
future paper. Comparing velocity versus magnitude for
each member in each group is inconclusive in determining
spatial separation.  Moreover, removing the eastern group members did not
significantly improve the likelihood for normality of the velocity distribution.
Table \ref{s258groups} compares key statistics of each group and the cluster.

There are also possible 
subclusters to the west and north that are more readily apparent from inspection of 
the nearest neighbor plot,
Figure \ref{s258}(e), in which they are labeled `A' and `B'.  Subcluster B  
appears to contain eight galaxies grouped in both position and velocity space with
other galaxies in the surface plot probably not being subcluster members.
The northern A grouping appears to be a result of chance projection in the surface plots.
Finally, the segmentation, nearest neighbor, and 3D local velocity plots
suggest the presence of a possible subcluster just southeast of the core that also
contains the brightest cluster member. It is interesting to note that all of
the apparently real groupings/subclusters lie (roughly) along the position
angle of the major axis of the dispersion ellipse.

\subsection{Abell S301}

Abell S301 (DC-0247-31) is one of the nearest clusters in the
study with an estimated distance of $d_p(t_e) = 65 h^{-1}$
Mpc ($c\overline{z} =  6669$ km/s). The surface plots in Figures~\ref{s301}(a)-(e) 
show the central regions to contain a compact high density core and a smaller second component
to the southeast. Comparison of the luminosity and number densities in the central region
shows them with similar structure (Figures~\ref{s301}(b) and (c)). 
In addition, one other subgrouping having density greater than
$40 \, \mathrm{gal}/(h^{-1} \, \mathrm{Mpc})^2$ appears to the southwest, approximately
1.2 $h^{-1}$ Mpc from the centre, and a lower density grouping appears to
the northeast at approximately $1.0 h^{-1} \, \mathrm{Mpc}$.

Biased by the southeast component, the ellipticity at a mean radius of 0.5 Mpc is
$\epsilon = 0.36$ at a position angle of $PA = 161^\circ$. As the mean radius
increases, the southwestern and northeastern groupings influence the calculation to shift the
position angle with, for example, $\epsilon = 0.61$ and $PA = 48^\circ$ at a
mean radius of 1.0 Mpc (Table~\ref{ellipt_table}). The variation in position
angle is problematic in achieving an elliptical core fit accurate at all radii,
but using the average ellipticity
parameters from the 0.75-1.0 Mpc mean radii is somewhat better than using a
circular fit and yields core parameters of 
$R_c = 0.24 \, h^{-1}$ Mpc and 
$\sigma_0 = 133 \, \mathrm{gal}/(h^{-1} \, Mpc)^{2}$.

Inspection of Figure~\ref{s301}(i) reveals a distribution with most of the
galaxies broadly distributed between $\sim 6300 - 7000 \, \mathrm{km/s}$
and with relatively small tails resulting in a kurtosis of $K = 1.9 = 2.0\sigma_K$.
The $\chi_{vel}^2$ test rejects this as a gaussian distribution at the
99.9\% CL. Two of the slowest galaxies ($cz \approx 5000 \, \mathrm{km/s}$)
are relatively bright compared to the other cluster members and are possibly in the
foreground.

While the $\alpha$ test gives no indication for substructure ($P_\alpha = 24.2\%$),
both the $\beta$ and $\kappa$ tests yield statistically significant signals
($P_\beta = 2\%$ and $P_\kappa = 0.4\%$).
Inspection of the bubble and local velocity plots (Figures
\ref{s301}(f) and (h)) reveals a subcluster in the main
component of the core not evident in the surface plots.
In particular, Figure~\ref{s301}(h) shows that the local velocities of these galaxies 
are significantly faster compared to the cluster mean velocity.
Overall, the
core galaxies have an average velocity of  6968 km/s compared to the mean value for the
cluster of 6594 km/s, but 13 
galaxies in the core have local velocities exceeding 7200 km/s.
These galaxies reside in a very close
group near the centre of the main component ($x = -0.10 \; \mathrm{to} \;  0.10 \,
h^{-1} \, \mathrm{Mpc}$, $y = -0.02 \; \mathrm{to} \; 0.11 \, h^{-1} \, \mathrm{Mpc}$).
The northeastern grouping also appears in
the bubble skyplot, but the farthest southwestern grouping is virtually absent.
Assessing whether the central velocity structure is truly in the core or is
a result of projection effects is required to further determine this cluster's dynamical
configuration and state.

This cluster has been investigated previously for substructure
several times, but with significantly fewer redshifts than the 95 used here.
Using 26 redshifts, \citet{dressler_2} found no substructure ($P_\Delta$
= 34.5\%). Using 29 redshifts, \citet{escalera} found no substructure
from applying the $\Delta$ and Lee tests
($P_\Delta$ = 52.2\% and $P_{Lee}$ = 80.0\%). \citet{stein} 
performed the $\Delta$, Lee, and velocity normality tests
using 25 members and found no substructure ($P_\Delta$
= 29\%, $P_{Lee}$ = 63\%, and $P_{Norm}$ = 43\%).
\citet{salvador_1} and \citet{salvador_2} using autocorrelation techniques found signs of
substructure.  Using a
KMM method, \citet{kriessler} found the central region to
be bimodal ($P$ = $<$ 0.1\%) with the same mode centre locations identified here.

\subsection{Abell S333}

Abell S333 is one of the most diffuse clusters in the sample
with a low surface density core region (see Figures \ref{s333}(a)
and \ref{s333}(b)).  The luminosity-weighted contour
plot (Figure \ref{s333}(c)) shows the brightest area
in the cluster to be separated from the core by approximately 1.0
$h^{-1}$Mpc (projected) northwest of the centre.
The high luminosity is due to
the brightest galaxy in the cluster ($b_J$ = 15.4), but with its velocity  
being several hundred km/s faster than the cluster mean, it seems unlikely this galaxy is
in the foreground.

\begin{figure}
\centering
\epsfig{figure=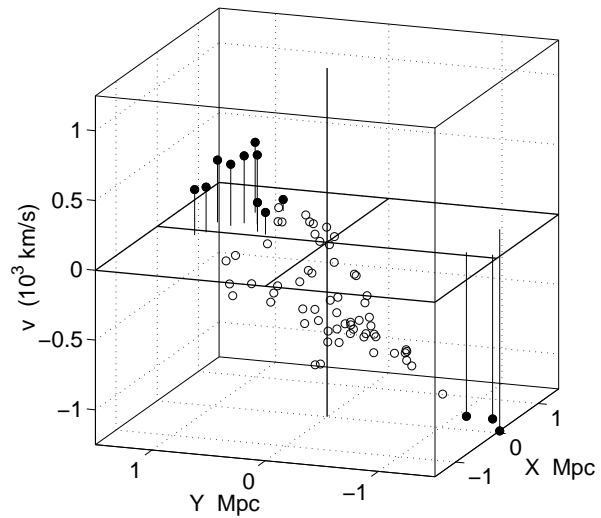, width=0.45\textwidth}
\caption{3D Local Velocity Plot for Abell S333.  This
is Figure \ref{s333}(h) rotated counter-clockwise
50$^\circ$ about the z-axis.  This view clearly shows the spatial asymmetry
in the local velocity with the
northern half of the cluster (the leftmost two quadrants in the
figure) moving away from the line of sight
and with the southern half (the rightmost two quadrants)
moving toward the line of sight.}
\label{s333_lvrotated}
\end{figure}

\begin{figure*}
\begin{minipage}{115mm}
%\centering
\raggedright
\epsfig{figure=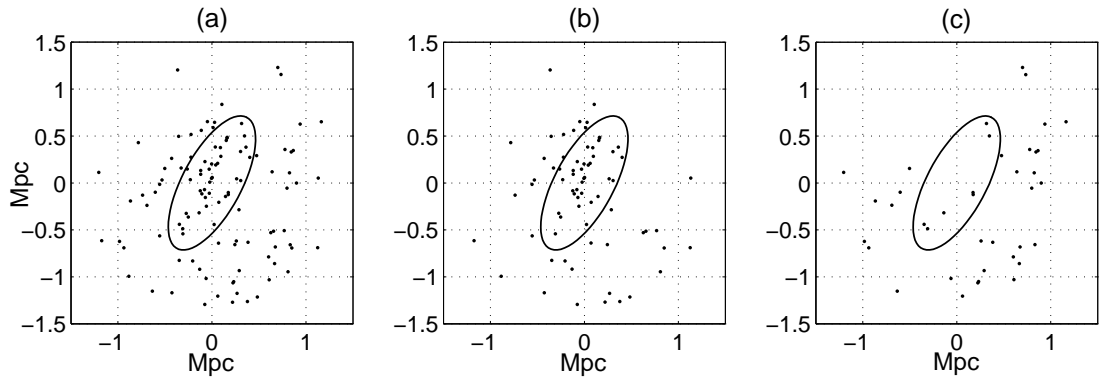,width=0.45\textwidth,angle=-90}
\caption{Galaxy positions for Abell S1043 showing the locations of (a) all galaxies,
(b) galaxies with $\mathrm{v} < 11,000 \, \mathrm{km/s}$,
and (c) galaxies with $\mathrm{v} > 11,000 \, \mathrm{km/s}$.}
\label{s1043_3pos}
\end{minipage}
\end{figure*}

The cluster shows consistent ellipticities and position angles at
all radii tested (see Table \ref{ellipt_table}).  The ellipticity
ranges from $\epsilon$ = 0.59 - 0.41, and the position angle
range of $PA$ = $131^\circ$ -- $156^\circ$ is consistently
aligned with the high luminosity region (Figures \ref{s333}(a) and (c)).
Given the consistent $\epsilon$ and PA at all distances from the centre, we use the
average values ($\epsilon$ = 0.5 and $PA$ = 135$^\circ$) to calculate
the core parameters.  The
resulting values are $R_c = 0.35 \pm 0.01$ 
and $\sigma_0 = 62 \pm 2$.

Inspection of the velocity distribution in Figure \ref{s333}(i)
shows a group of relatively slow galaxies in addition to the unusually-distributed
main group.  The
slow group consists of eight galaxies that have velocities
grouped near 17,500 km/s compared to the cluster mean of
19,372 km/s with their positions being indicated by the `$\circ$' symbol 
in Figure \ref{s333}(a).
Although moving significantly slower than the cluster mean velocity, these galaxies are 
not obvious foreground contamination: six have
magnitudes greater than the cluster mean of 18.1.
Two of the galaxies are
both brighter than average ($b_J$ = 16.9 and 17.8) and
significantly slower than average (v = 17,361 and
17,277 km/s).   While these two galaxies are possible non-members, their local motions 
are consistent with neighboring galaxies that are members (see below).

The histogram of the velocities of the majority of the cluster members 
shows them to be unusually distributed about a narrow peak near the mean velocity.
This is reflected by skewness and kurtosis values that are $2.5\sigma$ 
and $0.9\sigma$ away from normal values, respectively (see Table~\ref{velstats}).
The $\chi^2_{vel}$
velocity test rejects the normal hypothesis with a 99.9\% CL.
If the eight slowest galaxies are removed from the computation, the $S$ and $K$ statistics
move closer to normal; however, the $\chi^2_{vel}$ velocity
test still rejects the normal hyposthesis at a 
93\% CL.

The 3D plots and tests show signs of substructure with
the $\kappa$
test having the strongest indication with a probability
of random occcurence of $P_{\kappa} < 0.1\%$~(Table~\ref{sumresults}).
The substructure detected by the $\kappa$ test is found in
the high luminosity region to the northwest of the core as seen
in the $\kappa$ bubble skyplot (Figure \ref{s333}(f)).  The 3D plots
also highlight this area (Figures \ref{s333}(g) and (h)) where
ten galaxies in the northwest quadrant show local velocities above the 95\% level
of the K-S statistic (represented by the filled circles). 
The spatial asymmetry of the local velocity plot is seen more clearly when Figure \ref{s333}(h) is
rotated counter-clockwise 50$^\circ$ about the z-axis as
shown in Figure \ref{s333_lvrotated}. 

To summarize, Abell S333 has a consistently elongated shape
with two relatively luminous regions (the core and the region to
the northwest).  The luminous region outside the core 
is also where most of the velocity substructure is found. Together with the
(approximate) north-south asymmetry in the local velocity, the cluster shows evidence for dynamics
possibly involving rotation, shear, or infall within its supercluster environment.

%\begin{figure*}
%\begin{minipage}{115mm}
%\raggedright
%\epsfig{figure=fig_32.eps,width=0.45\textwidth,angle=-90}
%\caption{Galaxy positions for Abell S1043 showing the locations of (a) all galaxies,
%(b) galaxies with $\mathrm{v} < 11,000 \, \mathrm{km/s}$,
%and (c) galaxies with $\mathrm{v} > 11,000 \, \mathrm{km/s}$.}
%\label{s1043_3pos}
%\end{minipage}
%\end{figure*}

\begin{figure}
\centering
\epsfig{figure=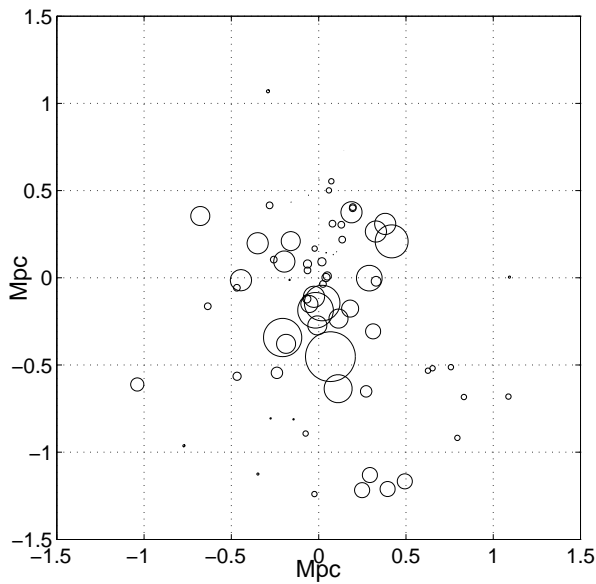,width=0.45\textwidth}
\caption{Abell S1043 $\kappa$ bubble skyplot with v $>$ 11,000 km/s members removed.}
\label{s1043_kappa}
\end{figure}

\subsection{Abell S1043}

Inspection of the scatter and number density
contour plots of Abell S1043 shows a cluster with a well-defined,
relatively dense, elongated central region (Figures \ref{s1043}
(a) and (b)).  A large number of galaxies at distances greater
than $1.0 \; h^{-1}$ Mpc from the centroid are located
to the south and southwest of the
cluster central regions.  As most galaxies to the north are less than
0.5 $h^{-1}$ Mpc from the east-west centre line,
this gives the cluster an appearance of asymmetry.
The luminosity contour plot (Figure \ref{s1043}
(c)) shows two areas outside the core with luminosity
the same or greater than the core. The two
areas of greatest luminosity appear in the 
galaxies to the south and southwest mentioned above where
the four brightest galaxies are located, 
but with velocities showing no evidence for obvious non-membership.

At mean radii of  0.5 to 0.75  $h^{-1}$ Mpc, the
ellipticity ranges from $\epsilon = 0.60$ to $0.43$.
with a position angle from $PA = 152^\circ$  to
$169^\circ$  (Table \ref{ellipt_table}).  At a radius
of 1.0 $h^{-1}$ Mpc, the galaxies to the east and
west of the core tend to reduce the ellipticity ($\epsilon
= 0.03$) with no shift to the position angle ($PA = 162^\circ$).
When the mean radius is increased to 1.25 $h^{-1}$ Mpc,
the galaxies to the south and southwest influence the
calculations such that the ellipticity increases to $\epsilon
= 0.32$ and the position angle shifts to $PA = 16^\circ$.
To fit the central regions as accurately as possible,
the ellipticity and position angles for the $0.5 \; \mathrm{and} \; 0.75
\; h^{-1} \, \mathrm{Mpc}$ dispersion ellipses are used to find
a core radius of $R_c = 0.31 \pm 0.01$ and a maximum density of
$\sigma_0 = 107 \pm 3$.

From Figure~\ref{s1043}(i), the velocity distribution is highly skewed 
(the value of $S = 0.86$ is $2.3\sigma$ away from the expected value of zero) 
with a ``slow'' group of galaxies between 9,000 and 11,000 km/s having a 
normal-shaped distribution, and with an
almost uniform distribution of ``fast'' galaxies from 11,000
to 13,500 km/s. For the entire distribution, the $\chi_{vel}^2$ test rejects the normal 
hypothesis with a CL $>$ 99.9\%.  
However, when the fast galaxies
are removed, the remaining galaxies have a velocity
distribution indistinguishable from a gaussian (CL = 29\% to reject). Further analysis
reveals that relatively few members of
this fast group are located within the 0.5 $h^{-1}$ Mpc
mean radius ellipse (see Figure \ref{s1043_3pos}). 

The two-dimensional plots and tests are
dominated by the galaxies to the south and southwest as shown by the
segmentation plot (Figure~\ref{s1043}(d)), the nearest neighbor plot
(Figure~\ref{s1043}(e)), and the $\beta$ test result of $P_{\beta} = 0.7\%$.
Comparison of Figure~\ref{s1043_3pos} with Figures~\ref{s1043}(b), (d), and (e)
show that ten of the faster
galaxies are located nearly due west of the central regions, but have 
only marginal density or connectedness relative to the other concentrations.

Before partitioning the galaxies into the two velocity groups,
the 3D tests plots indicate a marginal to high level of substructure:
$P_{\alpha} = 10.2\%$ and $P_{\kappa} < 0.1\%$ (Table~\ref{sumresults}).
However, when the fast group of galaxies mentioned above is removed,
the significance is substantially reduced to $P_{\kappa}^{slow} = 8.9\%$. 
Comparison of the $\kappa$ bubble skyplot for the slower group in Figure \ref{s1043_kappa}
with Figures~\ref{s1043}(f), (g), and (h)
reveals that it is the fast galaxies to the west that heavily influence
the $\kappa$ test results for all the galaxies. However, even with the fast
galaxies removed, the central regions still show indications of substructure
dynamics. In particular, there appears to be a possible subcluster near the core that
is the source of the relatively large remaining $\kappa$ bubbles in Figure~\ref{s1043_kappa}.

It is interesting to note that the dispersed galaxies to the south and southwest are not
inconsistent with post-merger asymmetries found in the simulations of \citet{pinkney}.
However other explanations are possible, and a more detailed analysis 
of the cluster is required.

%\begin{figure}
%\centering
%\epsfig{figure=fig_33.eps,width=0.45\textwidth}
%\caption{Abell S1043 $\kappa$ bubble skyplot with v $>$ 11,000 km/s members removed.}
%\label{s1043_kappa}
%\end{figure}

\subsection{APM 917}

APM 917 appears to have a high-density, well-defined
core (Figures \ref{apm917}(a) -- (e)), and does not
appear to have regions of high-density galaxy
groupings outside the central region with the luminosity being strongly
concentrated in the central region (see Figure~\ref{apm917}(c)).
The velocity distribution appears normal.
This appearance of normality is further supported
by the skewness and kurtosis statistics ($S = 0.18$ and
$K = 0.03$) as well as the $\chi_{vel}^2$ test yielding
only a 73\% CL to reject the normal hypothesis (see Table
\ref{velstats}).

\begin{figure}
\centering
\epsfig{figure=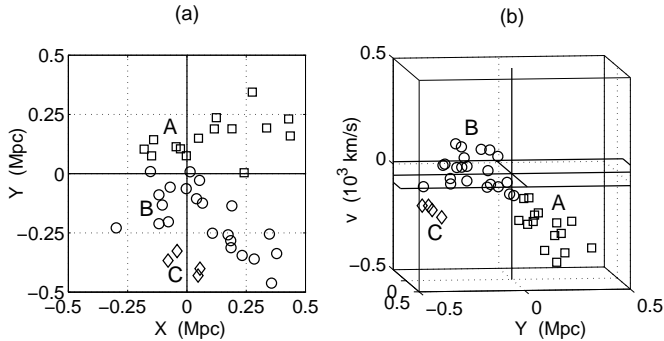,width=0.5\textwidth}
\caption{The central region of APM 917 in 2D and 3D Space.
Plot (a) shows the central region in 2D space while plot
(b) shows the central region in 3D as viewed from the west,
using local velocities for the $z$ direction.
Members of component A are denoted
by squares, component B by circles,
and component C by diamonds.}
\label{apm917_3D}
\end{figure}

The central region, while well-defined, has an unusual
triangular or almost crescent shape.  It has a relatively
moderate level of ellipticity (Table \ref{ellipt_table}) at all
mean radii with varying position angles ($\epsilon$ =
0.45 - 0.27 and $PA = 17^\circ$ -- $56^\circ$
for $R$ = 0.5 - 1.25 $h^{-1}$ Mpc).  The average
of the ellipticity and position angles for the 0.75 and
1.0 $h^{-1}$ Mpc mean radii are used to calculate the
King profile core radius and maximum density (Table \ref{core}).
The cluster core radius at $R_c = 0.14 \pm 0.01$
$h^{-1}$ Mpc is the smallest in the sample and the
maximum density, $\sigma_0 = 270 \pm 1$, is
second highest.

The results of the $\kappa$ test suggest a possibility of
substructure with $P_\kappa = 9.0\%$ (Table \ref{sumresults}).
Inspection of the $\kappa$ bubble skyplot in Figure
\ref{apm917}(f) shows four members to the east of the
centre with a high level of $\kappa$ with some lower
$\kappa$ values in the central region.
The $\alpha$ test indicates a high probability
of substructure with $P_\alpha$ = 0.6\%.  Since the
$\alpha$ test measures the velocity dispersion-weighted
centroid shift when the cluster velocities are randomly
shuffled, and there are no obvious subgroupings in the surface distribution,
it is expected that the results of this test are due to the
presence of significant velocity substructure in the core.
A more detailed examination of the central
region in the 3D surface-velocity space supports this conjecture.

The central region appears to be composed of three components
as seen in Figures \ref{apm917_3D}(a) and (b).
We classify galaxies into one of the three components (A, B, or C) based on
their 3D proximity to the component.  A certain amount of
arbitrary judgment is used, and the 
component boundaries are not meant to be necessarily absolute. When
viewed from the west, it is clear that the A and
C components are moving slower than the central
component, B. All three components have a rather small
dispersion among themselves.  A random reshuffling of member velocities
would result in higher dispersions resulting in lower
weights for the component members and, in turn, a higher
centroid shift in the $\alpha$ test.

\subsection{APM 933}

Inspection of APM 933 galaxy position and number density
contour plots, Figures \ref{apm933}(a) and (b),
shows a high-density, somewhat elongated core, the centre of which 
is offset from the cluster mean centroid by $\approx 0.1 \, h^{-1} \, \mathrm{Mpc}$.
A grouping of galaxies with a density greater than 
$40 \, \mathrm{galaxies}/(h^{-1} \, \mathrm{Mpc})^2$ is located approximately
$1.3 \, h^{-1} \, \mathrm{Mpc}$ to the
north-northeast of the centre. This grouping is also readily apparent in
the segmentation and nearest neighbors plots, Figures \ref{apm933}(d) and (e).
Comparison of the number density and luminosity density contour plots,
Figures \ref{apm933}(b)and (c), shows that the light and mass do not exactly
overlap for this cluster. Including the grouping to the northeast mentioned
above, there are three areas to the north
with luminosities equaling or exceeding that of the cluster central region.
However, correlation of
the velocity-magnitude plot with these high luminosity regions 
reveals no obvious foreground candidates. In fact, the brightest galaxy
in this cluster catalogue (at the position farthest west in Figure~\ref{apm933}(c))
has a measured velocity of more than $1.0\sigma$ faster than the cluster mean indicating that
it is either an exceptionally bright background galaxy or possesses unusual dynamics.

In general, the ellipticity algorithm converges around moderate to low ellipticity
solutions. The elongated appearance of the core is quantified by the ellipticity
computation for the $0.5$ and $0.75 \, h^{-1} \, \mathrm{Mpc}$ mean radius ellipses 
yielding yielding values of $\epsilon = 0.53$ -- $0.40$.
The fitted core parameters are $R_c = 0.36 \pm 0.01$ and $\sigma_0 = 91 \pm 2$.

From Figure~\ref{apm933}(i), the velocity distribution contains 
a peak at $\approx 14,600 \, \mathrm{km/s}$
and an extended tail of galaxies with velocities greater than 16,000 km/s.
These two features result in skewness and kurtosis values of $S = 0.77$ and
$K = 0.81$ that are $2.2\sigma_S$ and $0.9\sigma_K$, respectively, away from the
values expected for a normal distribution.
The skewness
and kurtosis combine such that the $\chi_{vel}^2$ test rejects 
the normal hypothesis with a CL of 97\%.

With a value of $P_\beta <0.1\%$, the $\beta$ test indicates a high probability
for nonrandom asymmetry. Although the $\alpha$ test does not indicate a high probability
for nonrandom substructure ($P_\alpha = 18.5\%$), both the $\kappa$ test ($P_\kappa = 2.6\%$)
and the local velocity plot (Figure~\ref{apm933}(h)) reveal several potentially interesting 
subgroupings and possible bound subclusters labeled A through D in Figure~\ref{apm933}(e).
The 3D plots
again show at least eight of the galaxies in the grouping to the north-northeast 
identified as D probably reside in a bound subcluster (Figures \ref{apm933}(f)--(h)). 
In the 3D local velocity plot, seven of the galaxies in
this subcluster have a $\leq$ 5\% or less probability that the
velocity distribution of their nearest neighbors belong to the cluster
velocity distribution.  The $\kappa$ bubble skyplot and the local
velocity plot also show
substructure in the central regions highlighted by grouping B that is suggestive of two or three
subclusters having motions consistent with merger dynamics or infall, but 
a more detailed analysis is required for unambiguous identification.
A smaller group of six galaxies is found outside the central
regions, and is denoted in the nearest neigbors plot as C.  Four of these
galaxies have velocities within a narrow range of 500 km/s, and likely comprise a small
subcluster while the other two galaxies probably do not belong to this grouping.
Grouping A does not appear to be a dynamically significant subcluster, and may be a binary
with one or two additional galaxies seen in projection in the 2D plots.

\subsection{EDCC 365}

Inspection of EDCC 365 scatter, isodensity, segmentation, and nearest neighbor plots 
(Figures \ref{e365}(a), (b), (d), and (e)) 
shows a relatively diffuse cluster with a 
well-defined core, a large clump of galaxies just northwest
of the core (A), and some structure south and southeast of the
core that does not appear to be subclustering.  The luminosity contour plot, 
Figure \ref{e365}(c), highlights the luminosity of the core although
the dominant luminosity signal is in the south due to the
presence of the brightest cluster galaxy ($b_J = 15.1$). The velocity of
the brightest galaxy is consistent
with cluster membership ($cz = 17,868 \, \mathrm{km/s}$ compared to the
cluster mean redshift of $c\overline{z} = 17,773 \, \mathrm{km/s}$). 
The approximate north-south alignment of the groupings biases the
ellipticity algorithm to yield relatively high ellipticities of
$\epsilon \sim 0.5-0.7$ at
position angles $PA \sim 150-160^\circ$~(see Table~\ref{ellipt_table}).
We compute the core parameters by averaging the ellipticities and position 
angles from the 0.5 and 0.75 $h^{-1}$ Mpc results to find
a core radius and maximum density of
$R_c = 0.37 \pm 0.01 \, h^{-1}$ Mpc and 
$\sigma_0 = 70 \pm 2$.

The velocity histogram (Figure \ref{e365}(i))
appears normal.  The skew and kurtotsis do
show some signs of  anormality ($S = 0.72$
and $K = 1.00$) however, the $\chi^2_{vel}$
test has only a 83\% confidence level to reject
the normal hypothesis.

The cluster asymmetry  measured by the $\beta$ test yields a 
marginally significant $P_\beta = 4\%$ probability of random
occurrence, while the $\alpha$ test gives no indication of substructure
($P_\alpha = 39.1\%$).
However, with $P_\kappa = 0.2\%$, the $\kappa$ test indicates the presence of 
substructure at a high significance level.
From the bubble plot of Figure~\ref{e365}(f), the substructure appears in 
two distinct regions corresponding to the core and the galaxies to the southeast. 
These galaxies are also readily apparent in the peculiar and local velocity plots
in Figures~\ref{e365}(g) and (h),
but do not form a well-defined clump. Correlation of the galaxies in the core with the larger bubbles
and the x-y-local velocity plot reveals no obvious subclustering in this region. The velocities
of the galaxies in grouping A are consistent with this grouping being a subcluster.

\subsection{EDCC 442}

From Figures~\ref{e442}(a)-(e), EDCC 442 is an elongated cluster possessing a relatively  
dense core. Outside the central regions, there are five groupings with number densities
greater than $40 \; \mathrm{gal}/(h^{-1} \, \mathrm{Mpc})^2$ (labelled in Figure~\ref{e442}(d)).
Comparison of Figures~\ref{e442}(b) and (c) shows that the cluster centre of luminosity is
closely correlated with the mean arithmetic centre due to the presence of several of the brightest
galaxies apparently residing near or at the bottom of the cluster potential well.

The ellipticity computation yields higher ellipticities at the larger mean radii due
to the influence of the subgroupings, and smaller ellipticities in the central region,
e.g., $\epsilon = 0.34$ at $R = 0.5 \, h^{-1} \, \mathrm{Mpc}$ while $\epsilon = 0.58$ at
$R = 1.0 \, h^{-1} \, \mathrm{Mpc}$. Note, however, that the position angle of the core
galaxies is aligned with the subgrouping-influenced global cluster position angle
of $PA \approx 35^\circ$. Using the average ellipticities and position angles for the
$0.5 \; \mathrm{and} \; 0.75 \, h^{-1} \, \mathrm{Mpc}$ dispersion ellipses to compute
the core parameters of $R_c = 0.33 \, h^{-1} \, \mathrm{Mpc}$ and 
$\sigma_0 = 113 \pm 4 \; \mathrm{gal}/(h^{-1} \, \mathrm{Mpc})^2$ yields a
significantly better fit to the cluster density profile.   
\begin{figure}
\centering
\epsfig{figure=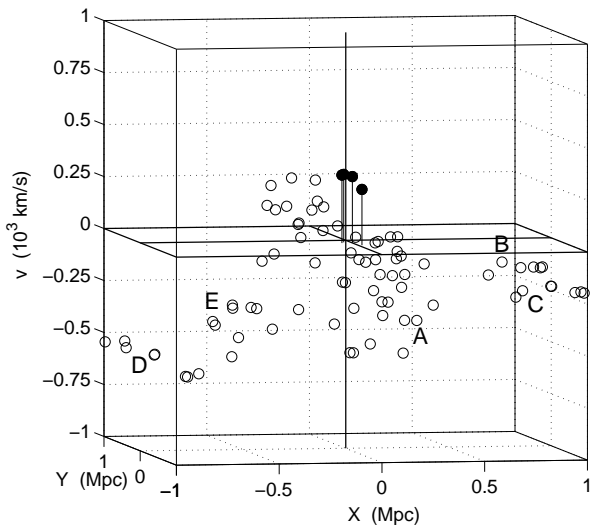,width=0.45\textwidth}
\caption{EDCC442 local velocities of the nearest neighbors shown in
Figure~\ref{e442}(e). Groupings B, C, D, and E are possible
subclusters, and there is velocity structure in the core region.}
\label{e442_nnlv}
\end{figure}

The velocity distribution as shown in Figure~\ref{e442}(i) is 
symmetric ($S = < 0.01$) but somewhat flattened compared to a normal distribution
($K = 1.7 = 2.1\sigma_K$) with a slight indication of bimodality. Nevertheless,
the $\chi_{vel}^2$ test does not reject this distribution as normal (CL to reject
the normal hypothesis = 51\%). The three slowest galaxies with $v < 12,500 \, \mathrm{km/s}$
are probably in the foreground.

Consistent with the appearance of the 2D visualization plots, the $\beta$ test yields
a statistically significant asymmetry of $P_\beta = 3\%$ probability of random occurrence.
The two-dimensional substructure reflects apparently real three-dimensional substructure
in at least four of the subgroupings noted in Figure~\ref{e442}(d). The nearest neighbor
local velocity plot in Figure~\ref{e442_nnlv}  clearly shows that the galaxies in subgroupings B, C,
D, and E have local velocities consistent with their being members of bound subclusters.

The kinematically-weighted centroid shifts tested by the $\alpha$ algorithm do not indicate
significant substructure ($P_\alpha = 24.9\%$). However, with $P_\kappa = 3.5\%$, the $\kappa$ test
does indicate the probable presence of dynamically significant substructure. Correlating
the bubble plot of Figure~\ref{e442}(f) with the segmentation/nearest neighbor plots of
Figures~\ref{e442}(d)-(e) indicates again that the galaxies in groupings C and D 
are probable subclusters, and with the transition between D and E being gradual rather than sharp.
Consistent with the ``stringy'' appearance of the local
velocities of the nearest neighbors in Figure~\ref{e442_nnlv}, the bubble plot also gives 
some indication of 
core kinematics that may indicate non-virialized core dynamics. 
The lack of relatively large bubbles at the
locations of groupings A, B, and E requires further investigation to distinguish whether
these are bound subclusters with velocity distributions indistinguishable from the global
distribution, the local grouping motion is transverse to the line-of-sight, or the
groupings are only a result of projection effects (as appears likely for grouping A although
this grouping does appear to contain a likely bound binary). 

\section{Summary of Results and Discussion}

To begin evaluating the physical significance of the results obtained here, consider that,
in principle, any single test can detect the type of substructure to which it is sensitive
regardless of the results of any other test. 
We believe that the methods used here to detect substructure in the galaxy distribution
are generally accurate to the confidence levels stated due to the discriminating power of the 
3-dimensional tests in substantially reducing (or eliminating) apparent substructure due to projection 
effects. In particular, the $\kappa$ test results combined with the bubble skyplot and 
plot of the local velocities/relative positions of the nearest neighbors
appears to be a particularly useful combination for identifying true substructure. 

However, it should be remembered that it
has been demonstrated
through Monte Carlo simulation that substructure
tests have both ``failure to detect'' and ``false positive'' rates (see, e.g., \citealt{pinkney}).
Thus, before summarizing the
results of the statistical tests applied to the present sample, it is
appropriate to review estimates for the numerical reliability of the $\alpha$ and $\beta$ tests.

Pinkney et al. found that the $\alpha$ test is generally a
robust diagnostic of substructure, sensitive to dispersed subclusters, and that its
effectiveness is not always diminished by a superposition of the core and a subcluster.
In that study, the test identified substructure in 24
out of 36 cases of simulated mergers at a 10\% probability of random
occurrence, second only to the $\Delta$ test.  At a 5\% probability
level the test detected 16 out of 36 subclusters, and at 1\%, 12 out
of 36.  On the other hand, the test is also found to have relatively high
false positive rates at the 10\% and 5\% significance levels.  For example, in a cluster
with 100 members, the test incorrectly detected substructure in 23 out of 100 cases at
a 10\% confidence level.  At a 1\% confidence level, however, its false
positive rate was closer to the other estimators tested~(\citealt{pinkney}).

For the $\beta$ test, the Pinkney et al. study found it to be the second best
two-dimensional test out of the four two-dimensional tests that were evaluated;
their most robust 2-dimensional test was the Fourier elongation test that was essentially replaced
in this study by the ellipticity algorithm.
In that work, The $\beta$ test identified substructure in 16 out of 36 cases with
a 10\% probability of random occurence, 15 cases with a 5\% probability, and nine
cases with a 1\% chance.  The false positive rate was found to be 12\%, 9\%,
and 4\% for significance levels of 10\%, 5\%, and 1\% using a sample cluster size
of 100 galaxies.
Thus, it appears the test has a low detection rate and a high
false positive rate.  This was true for most of the one-dimensional and two-dimensional
tests evaluated.  As one would expect, the $\beta$ test was
effective for detecting deviations from \itshape mirror \upshape symmetry, but not
necessarily \itshape circular \upshape symmetry. 
In terms of merger scenarios,
the test works best at early stages of merger
viewed at right angles but not as well at later stages viewed nearly straight on.
Perhaps most importantly, and not explicitly discussed in previous work, is the sensitivity
of the results to the choice of cluster centroid.

%\noindent
Given the above comments, it is obviously not possible to make an absolute statement regarding the confidence
level at which a particular test yields a true result. In many cases, a final judgment can be rendered
by incorporating high-resolution, long-exposure X-ray data. In other cases, there may be no
absolute determination possible. However, we believe that the approach used here to combine
visualization plots with statistical tests and then to identify sources of positive signatures
is robust in mitigating the relatively small probabilities for the statistical tests to
yield a false positive or negative recognition.
 
\begin{table}
\caption{Summary of statistical tests for substructure.  The number in the table
is the percent probability that the test statistic is the result of a random occurrence.}
\label{sumresults}
\begin{center}
\begin{tabular}{lcccc}
\hline\\
&\multicolumn{4}{c}{Probability in \%} \\
\cline{2-5}
Cluster     & \hspace*{1.5em}$\alpha$\hspace*{1.5em} & \hspace*{1.5em}$\beta$\hspace*{1.5em} & \hspace*{1.5em}$\kappa$\hspace*{1.5em}
        & \hspace*{1.5em}$\chi^2_{vel}$\hspace*{1.5em}      \\
\hline
Abell 930   & ~~3.0            &  30.3           & ~~5.8            & $<0.1$           \\
Abell 957   &  32.1            & ~~0.5           &  14.4            &  51.5            \\
Abell 1139  & $<0.1$           &  14.8           & $<0.1$           &  23.6            \\
Abell 1238  & ~~6.7            &  49.7           & ~~8.0            &  86.4            \\
Abell 1620  &  26.8            & ~~0.1           & ~~0.9            & ~~2.8            \\
Abell 1663  &  28.9            & ~~3.9           & ~~0.1            &  64.4            \\
Abell 1750  & ~~1.9            & $<0.1$          & ~~2.6            & ~~1.8            \\
Abell 2734  &  44.8            & ~~0.4           &  35.3            &  14.9            \\
Abell 2814  & ~~1.3            & ~~8.0           & ~~6.5            &  32.0            \\
Abell 3027  &  44.0            &  18.8           & ~~0.2            & ~~0.1            \\
Abell 3094  &  47.7            & ~~9.9           & $<0.1$           & ~~7.3            \\
Abell 3880  &  15.9            & ~~7.0           &  22.0            &  64.3            \\
Abell 4012  & ~~1.0            &  25.9           &  20.9            & ~~4.0            \\
Abell 4013  &  26.5            & ~~1.8           & ~~4.3            & $<0.1$           \\
Abell 4038  & $<0.1$           & $<0.1$          & $<0.1$           & ~~4.1            \\
Abell S141  &  34.6            &  23.1           &  19.8            & ~~2.0            \\
Abell S258  &  36.7            & ~~6.6           & $<0.1$           &  29.8            \\
Abell S301  &  24.2            & ~~1.8           & ~~0.4            & ~~0.1            \\
Abell S333  &  29.7            &  24.0           & $<0.1$           & $<0.1$            \\
Abell S1043 &  10.2            & ~~0.7           & $<0.1$           & $<0.1$           \\
APM 268     & ~~2.8            & $<0.1$          & ~~2.6            & ~~2.2            \\
APM 917     & ~~0.6            &  38.0           & ~~9.0            &  27.0            \\
APM 933     &  18.5            & $<0.1$          & ~~2.6            & ~~3.3            \\
EDCC 365    &  39.1            & ~~4.1           & ~~0.2            &  17.4            \\
EDCC 442    &  24.9            & ~~2.5           & ~~3.5            &  49.1            \\
\hline
\end{tabular}
\end{center}
\end{table}

The results for the $\alpha$, $\beta$, $\kappa$, and $\chi_{vel}^2$ tests
are collected in Table \ref{sumresults} where it can be seen that only 4 of the 25 clusters
failed to yield at least one signal at or below a 1\% probability for random occurrence (or, equivalently, at
or greater than a 99\% CL where the results have been rounded to the nearest percentile).
Table \ref{scorecard} tabulates the results in terms of the number of clusters that yielded substructure at
less than the 10\%, 5\%, and 1\% probabilities for random occurrence. 
Multimodal central regions were detected in about half of the sample, and many of these appear consistent
with merger scenarios. In general, the luminosity density in these clusters correlates closely with 
the number density, the few exceptions probably being due to the small level of incompleteness leaving
out a few bright galaxies. However, this should be more closely examined in a future study.
%From the detailed analysis presented
%in Section 4, there are a few cases where substructure appears evident even in the absence of a statistically
%significant signal ($< 10\%$ random probability), e.g., Abell 957 possessing a clearly bimodal structure.

\begin{table}
\caption{The number of clusters showing a substructure signal at less than or equal to a specified probability that the signal
is a random occurrence. The entries include clusters whose Monte Carlo probabilities are rounded to the nearest percentile.}
\label{scorecard}
\begin{center}
\begin{tabular}{crrr}
\hline\\
\hspace*{1em} Test \hspace*{1em} & \hspace*{1.5em} 10\% & \hspace{1.5em} 5\% 
& \hspace*{0.5em}  1\% \\
\hline\hline
$\alpha$       & 10 \hspace*{0.3em}  &  8 \hspace*{0.1em} &  5 \hspace*{0.1em} \\
$\beta$        & 17 \hspace*{0.3em}  & 13 \hspace*{0.1em} &  8 \hspace*{0.1em} \\
$\kappa$       & 20 \hspace*{0.3em}  & 16 \hspace*{0.1em} & 11 \hspace*{0.1em} \\
$\chi^2_{vel}$ & 14 \hspace*{0.3em}  & 13 \hspace*{0.1em} &  6 \hspace*{0.1em} \\
\hline
\end{tabular}
\end{center}
\end{table}

It is well known that giant elliptical galaxies in clusters usually reside in a region
of high galaxy density, either in the central region or in bound subclusters, 
and it is believed that galaxy-environment interactions significantly affect galaxy 
formation and evolution.
The absence of clear morphological data for most of the galaxies in this catalogue
restricts analysis here to correlating brightness with clumpiness (as provided by the
luminosity density contour plots). We find that 16 of the 25 clusters contain probable
bound subclusters with 
luminosity density comparable to that in their cores. Of these, 14 clusters contain one or
more subclusters in which the first, second, or third brightest galaxy is found. Especially
noteworthy are A957, A1663, and A1750 as clusters having subclusters comparable in size to the cluster
core and containing a single extremely bright galaxy. The other 11 clusters are
A1238, A1620, A2814, A3027/APM268, A3094, S258, S301, S1043, APM933, and EDCC442.

As mentioned in the Introduction, the clusters selected for this study from the 2dFGRS catalogue were required
to have at least 70 members in order to have confidence in the statistical results, and to provide 
a small enough set to allow analysis of individual clusters. 
With the catalogue utilized for this study not containing clusters with memberships exceeding $N = 154$, 
this selection criterion then naturally resulted in analysing low richness clusters of galaxies
which are structures not as well studied as rich clusters and groups.
Assessing the robustness of the approach here, 
it appears that smaller-sized groups in the 2dFGRS catalogue could reliably be analysed using the same methods.
Future work with this particular set of clusters includes incorporating data from 
the recently published 2PIGG catalogue of clusters \citep{eke} and comparing against
recent N-body simulations of cosmological concordance models. 
More detailed examinations are also planned of selected clusters by including new X-ray data, 
applying estimators to characterise cluster dynamics, and
using analytical tools such as wavelet analysis or the KMM algorithm to further resolve smaller-scale structure.

To summarize, we have presented a survey-level substructure analysis of 25 low richness clusters
of galaxies contained in the 2dFGRS cluster catalogue. Most of the clusters possess
features such as multi-component core/central regions, bound subclusters
well-removed from the centre, pre- and post-merger indicators, and possible signatures of infall,
rotation, or shear dynamics.
Thus, we find that substructure in these
clusters is the rule rather than the exception indicating that, as conjectured by others and
supported by some N-body simulations, low richness clusters relax to structureless equilibrium states
on very long dynamical time scales (if at all). The results here also show that while doubling
the number of redshifts available per cluster does not drastically alter the velocity statistics
found in previous studies, the additional information affords a more comprehensive reconstruction
of cluster substructure.

\vspace*{0.35in}
\noindent{\textbf{Acknowledgments}}

This research has made use of the NASA/IPAC Extragalactic Database (NED) which is operated by the
Jet Propulsion Laboratory, California Institute of Technology, under contract with the National
Aeronautics and Space Administration.

\appendix
\section{Visualization plots for the clusters}

This appendix presents the set of visualization plots described in Section 3.
For each cluster analyzed, plot (a) shows each galaxy position in $(RA, Dec)$ converted to
cartesian $(x,y)$ coordinates with superimposed
dispersion ellipses for mean radii of 0.5 and 1.0 $h^{-1}$ Mpc.
The projected (2-dimensional) galaxy
separations are inferred from the proper distance at photon time of emission, $d_p(t_e)$.
Following the usual convention, north is along the positive y-axis
and east is along the negative x-axis.
The filled circles represent galaxies with velocities within
1.3$\sigma$ of the mean velocity, whereas galaxies slower than 1.3$\sigma$ are
represented by a open circle, and galaxies faster than 1.3$\sigma$ are shown with a plus sign.  For
a normal velocity distribution, approximately 10\% of the galaxies should fall in
the faster or slower group.
Plots (b)-(e) are the 2-dimensional visualization plots described in Section 3.2.

Plot (f) is the $\kappa$ ``bubble skyplot'' described in Section 3.5 where
the size of the circles centered on galaxy positions
is proportional to the $\kappa$ statistic for each galaxy.
As a complement to the substructure implications provided by the velocity statistics and the $\kappa$ test,
visualization of potential substructure correlated with the cluster
velocity distribution is accomplished through two additional types of plots:
the peculiar velocity of each galaxy, $v_{gal} - \overline{v}_{cl}$, plotted
in a 3-dimensional position-velocity format, and the ``local'' velocity constructed from the average
velocity of each galaxy and its nearest neighbors,
$v_{local}~(=\overline{v}_{NN}) - \overline{v}_{cl}$,
plotted in the same 3-dimensional position-velocity format. The zero point for the peculiar velocities
is the cluster average velocity, whereas the zero point for the local velocity is based on the local
velocity of the galaxies residing within one core radius of the cluster centre.
For these latter two plots, the position of
each galaxy is marked by a filled circle when the
$\kappa$ statistic for that galaxy has a $\leq 5\%$ probability of random occurrence
and an open circle otherwise
(the 5\% probability threshold is arbitrary and can be varied to indicate the
``strength'' of a particular grouping). The vertical lines connecting the
position of each filled circle galaxy to the midplane are for visualization purposes only.
As seen in the individual cluster analyses, the local velocity plot
provides a useful visualization complement to the $\kappa$ statistic and bubble plot.
Finally, plot (i) shows the histogram of the velocity distribution using a bin size of 150 km s$^{-1}$.

\clearpage
\begin{figure*}
\centering
\epsfig{figure=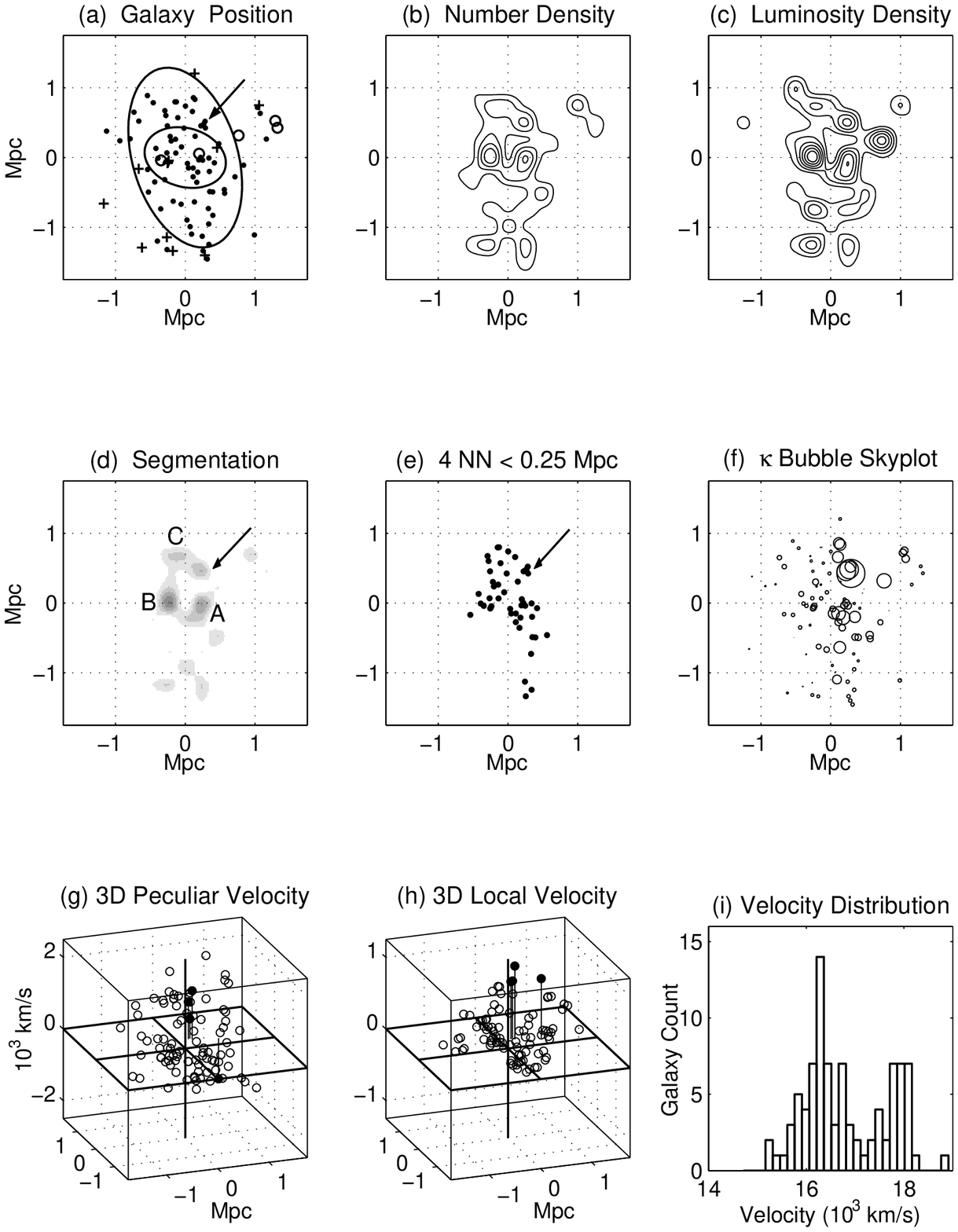}
\caption{Visualization plots for Abell 930.}
\label{a930}
\end{figure*}

\clearpage
\begin{figure*}
\centering
\epsfig{figure=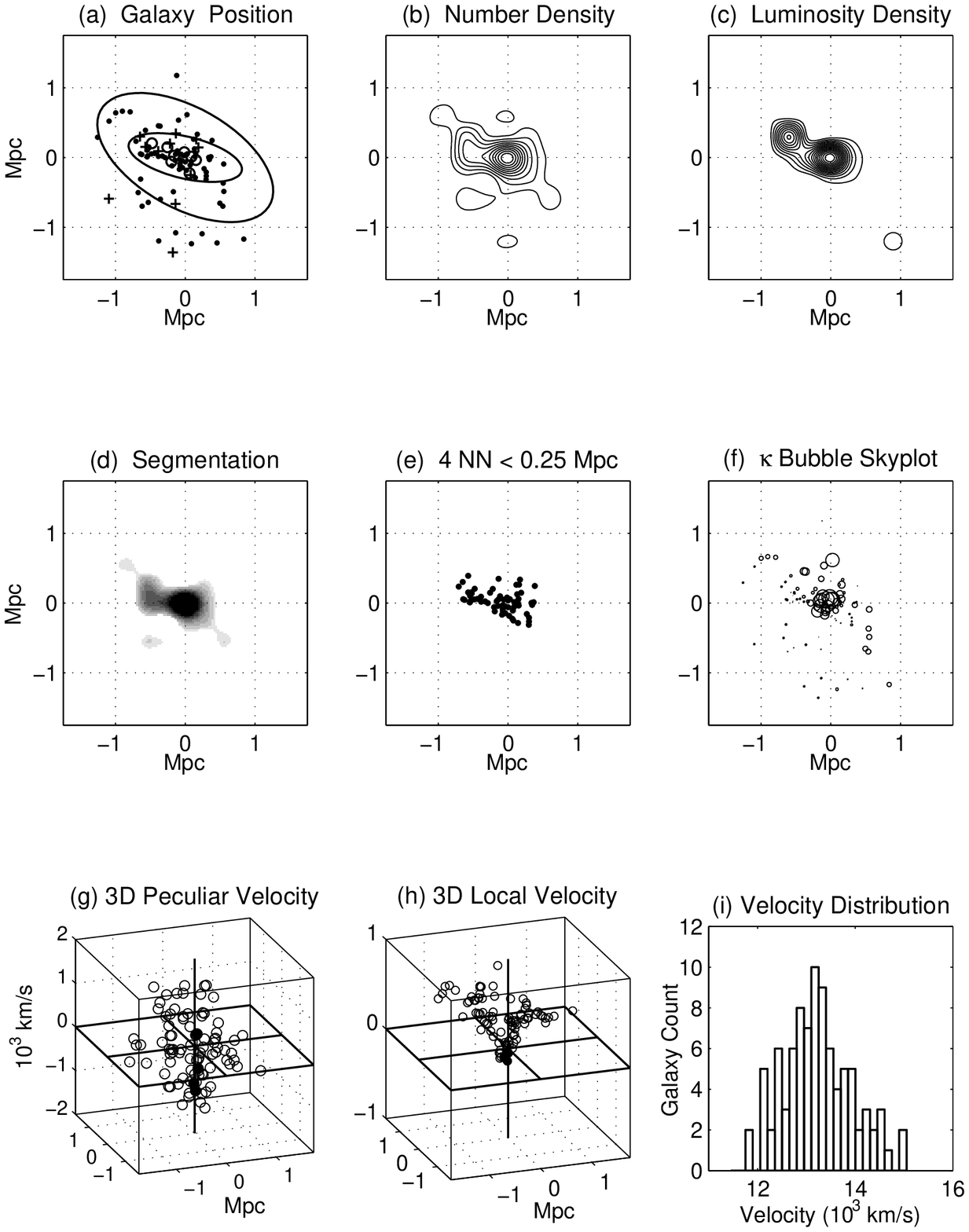}
\caption{Visualization plots for Abell 957.}
\label{a957}
\end{figure*}

\clearpage
\begin{figure*}
\centering
\epsfig{figure=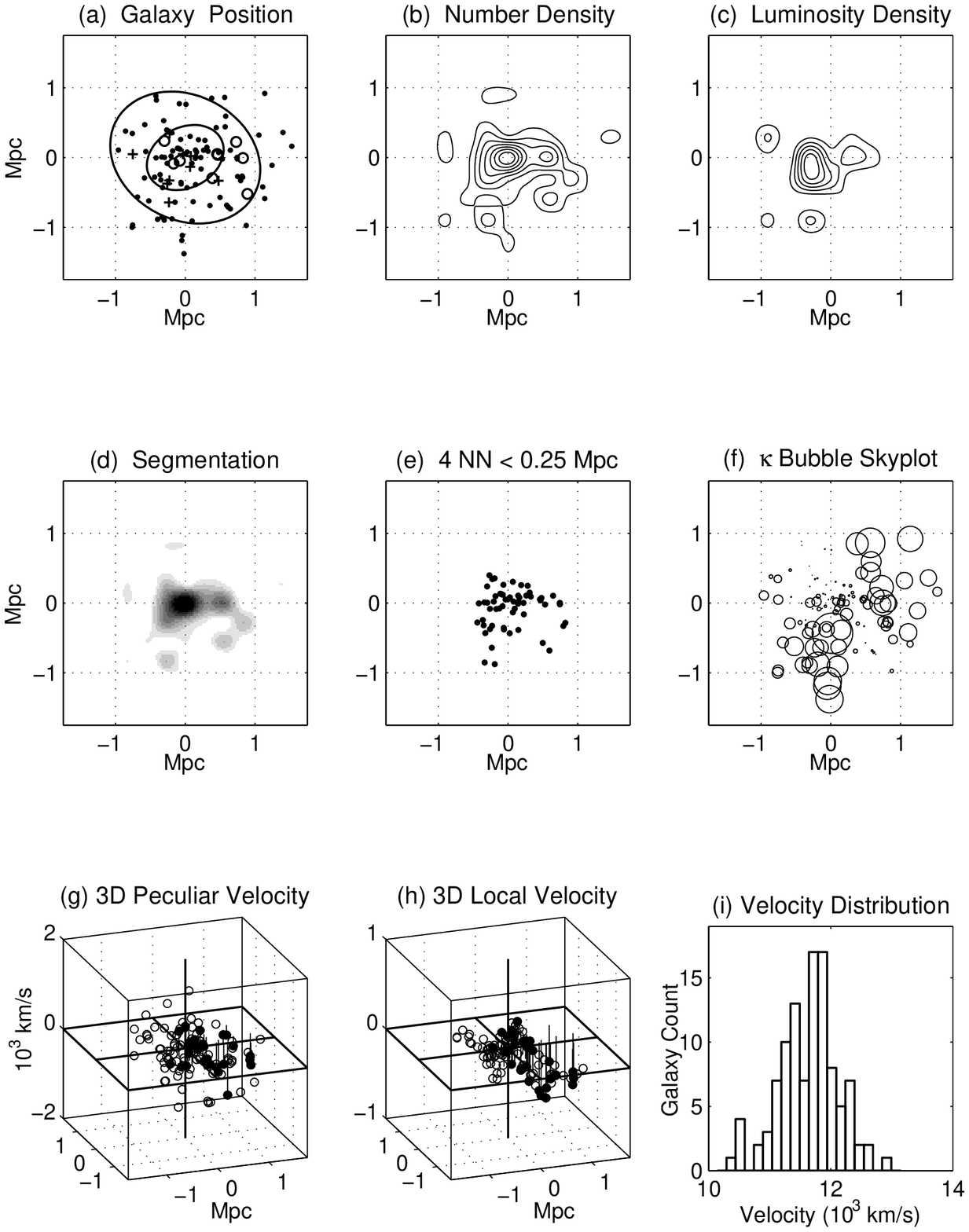}
\caption{Visualization plots for Abell 1139.}
\label{a1139}
\end{figure*}

\clearpage
\begin{figure*}
\centering
\epsfig{figure=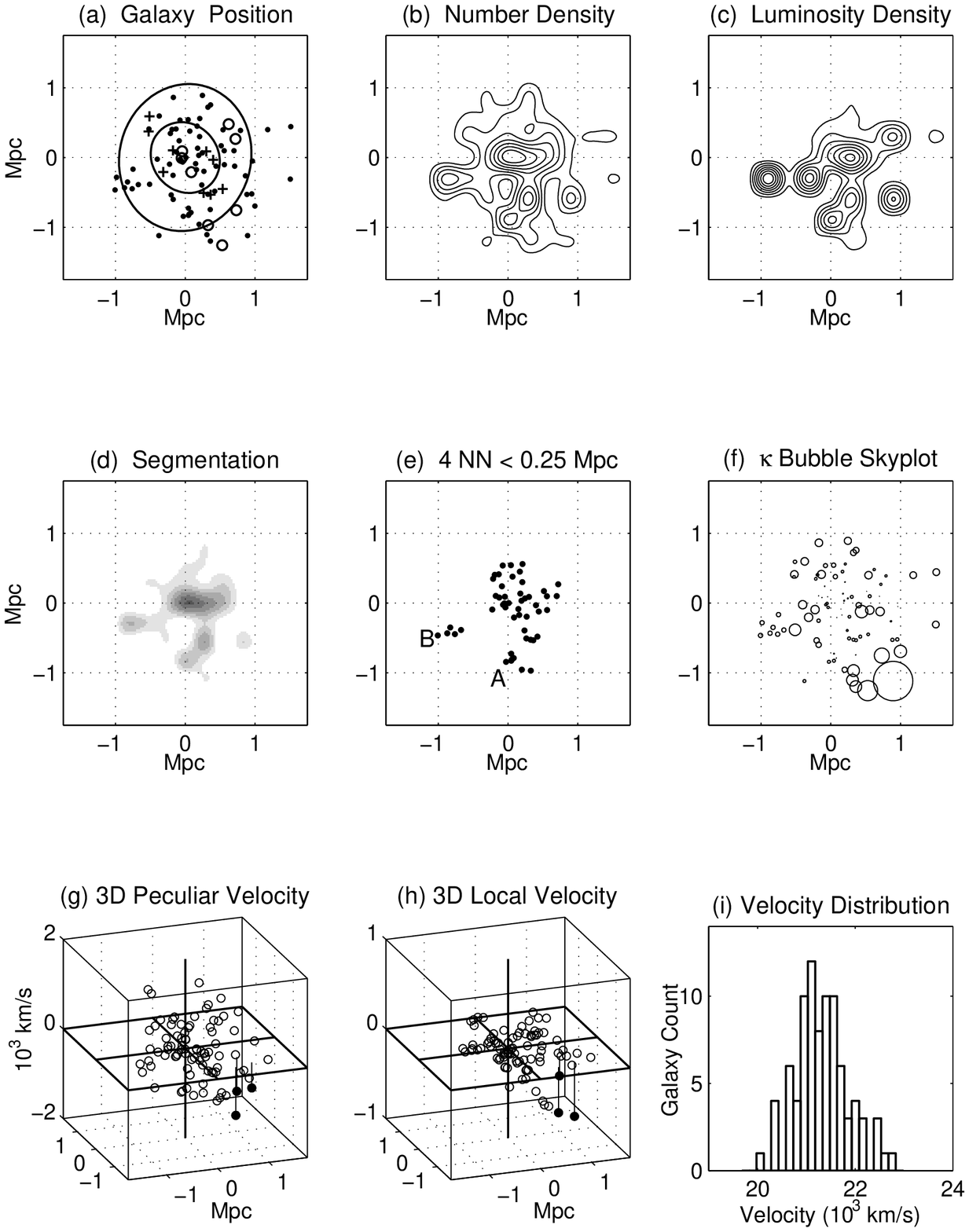}
\caption{Visualization plots for Abell 1238.}
\label{a1238}
\end{figure*}

\clearpage
\begin{figure*}
\centering
\epsfig{figure=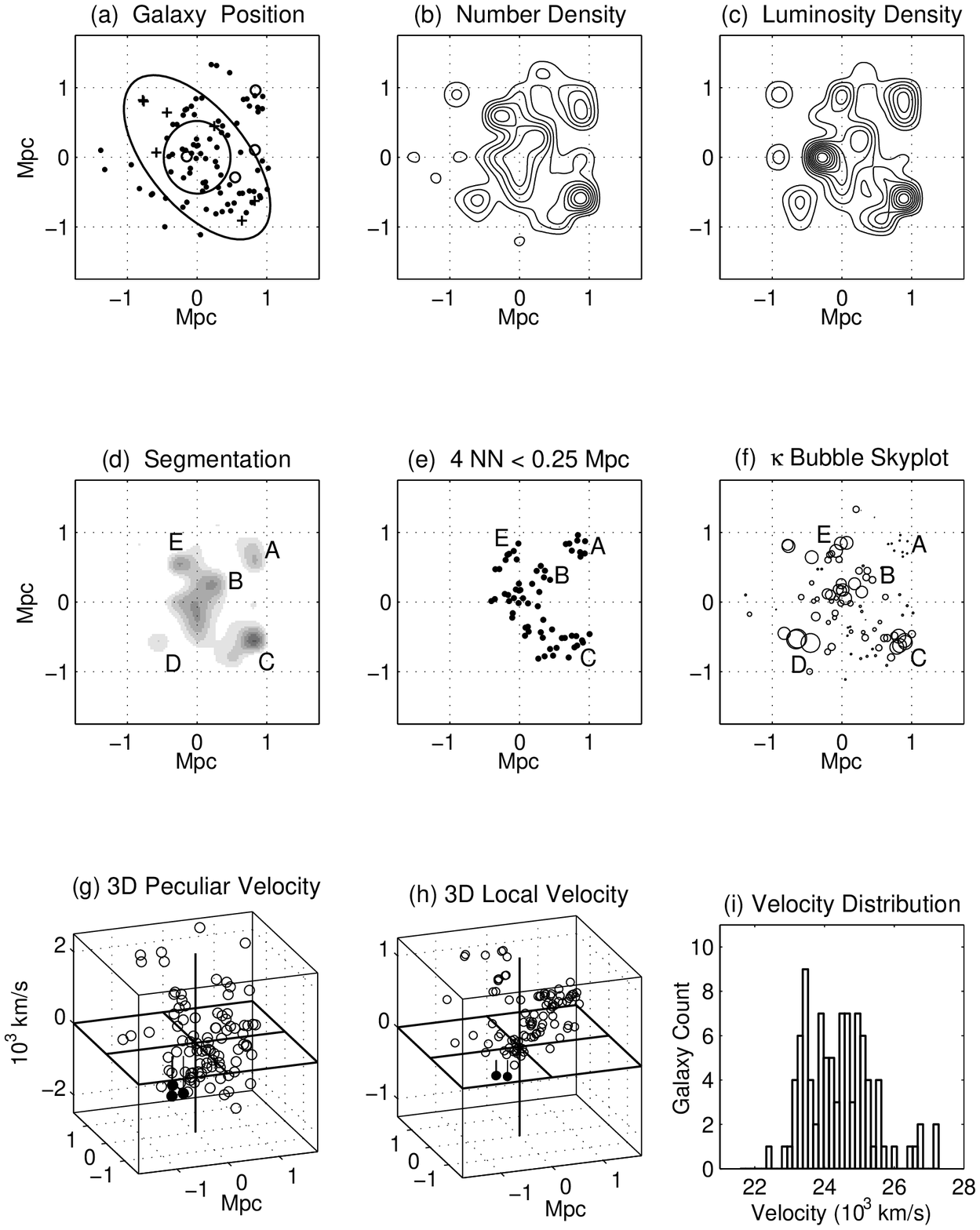}
\caption{Visualization plots for Abell 1620.}
\label{a1620}
\end{figure*}

\clearpage
\begin{figure*}
\centering
\epsfig{figure=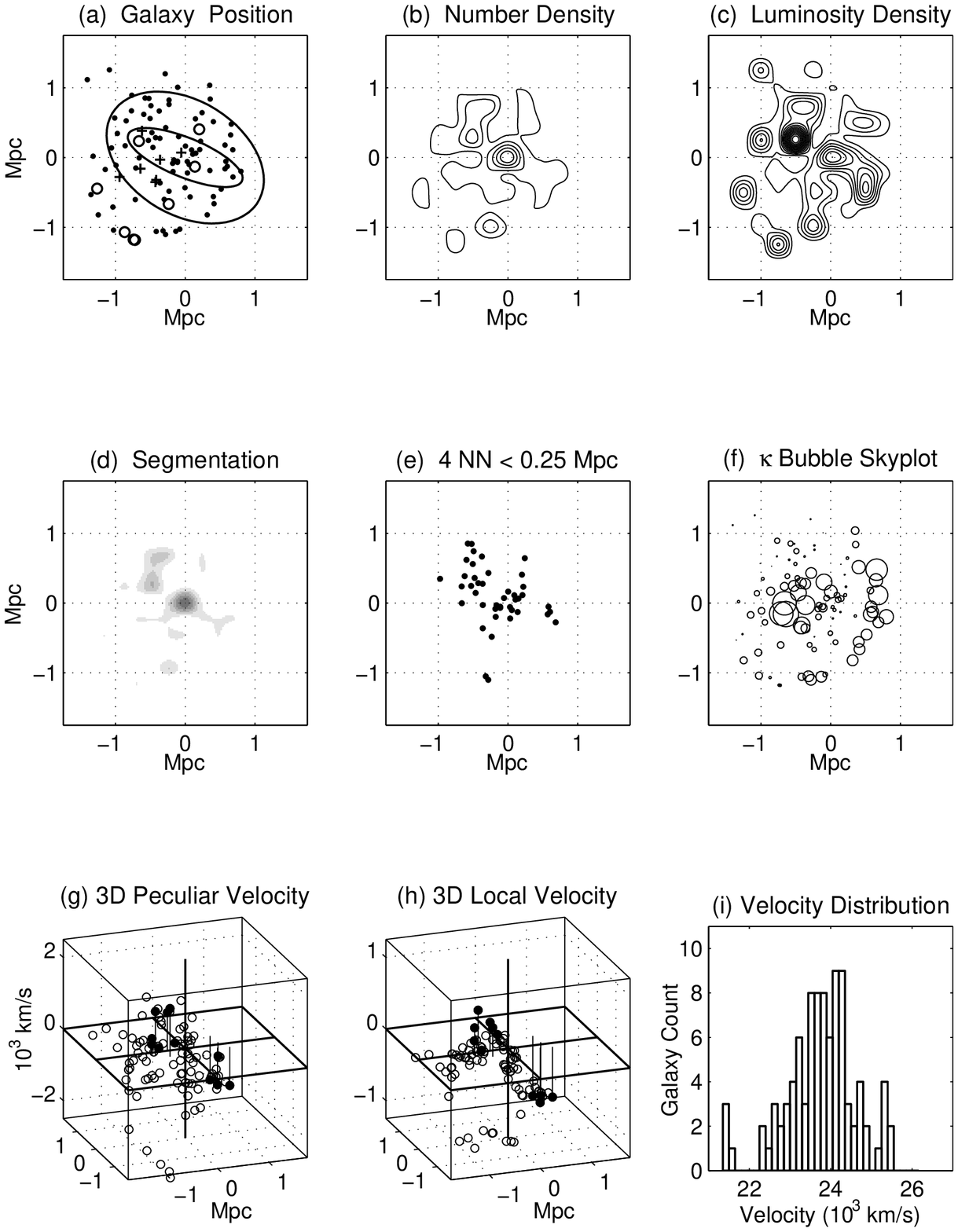}
\caption{Visualization plots for Abell 1663.}
\label{a1663}
\end{figure*}

\clearpage
\begin{figure*}
\centering
\epsfig{figure=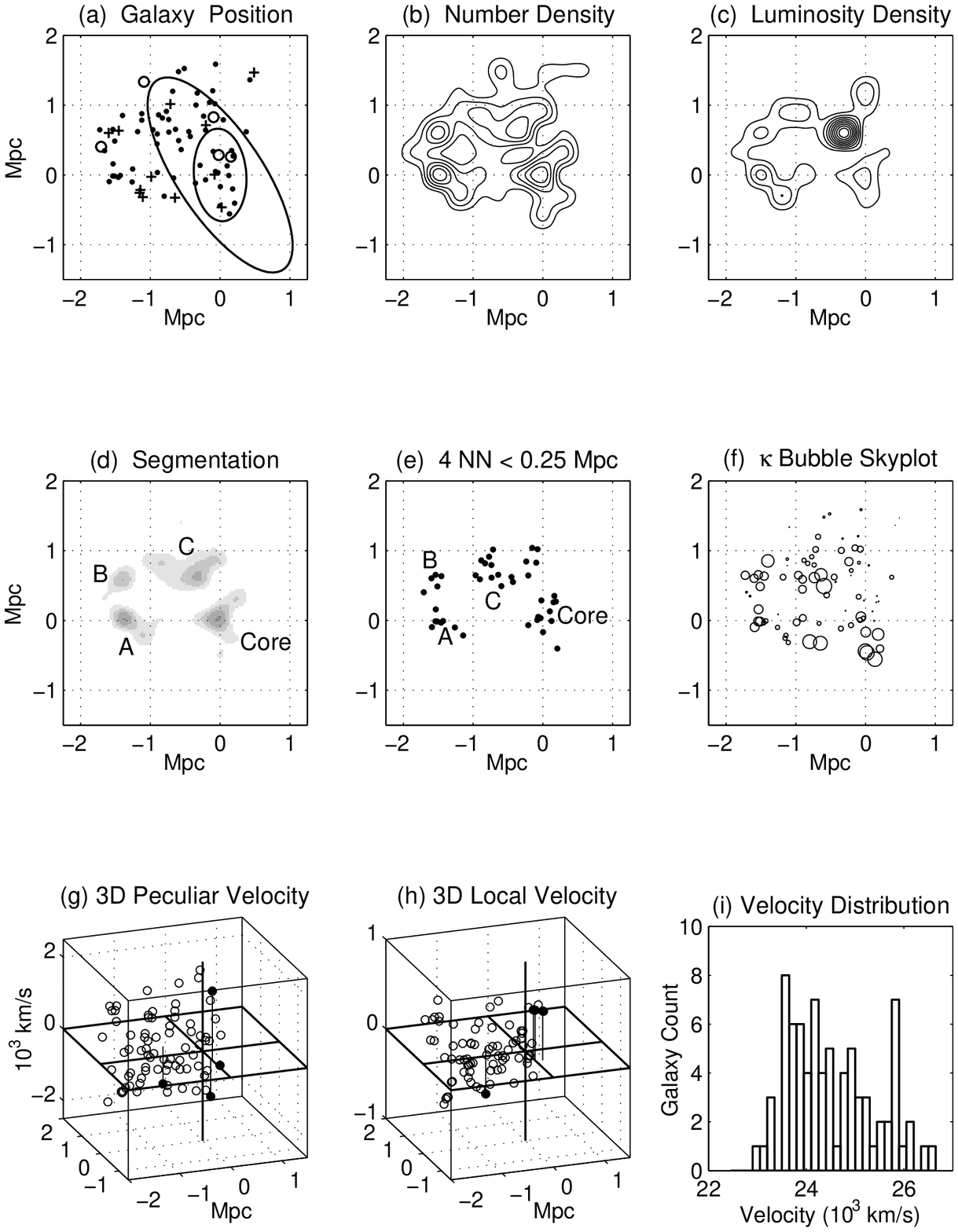}
\caption{Visualization plots for Abell 1750.}
\label{a1750}
\end{figure*}

\clearpage
\begin{figure*}
\centering
\epsfig{figure=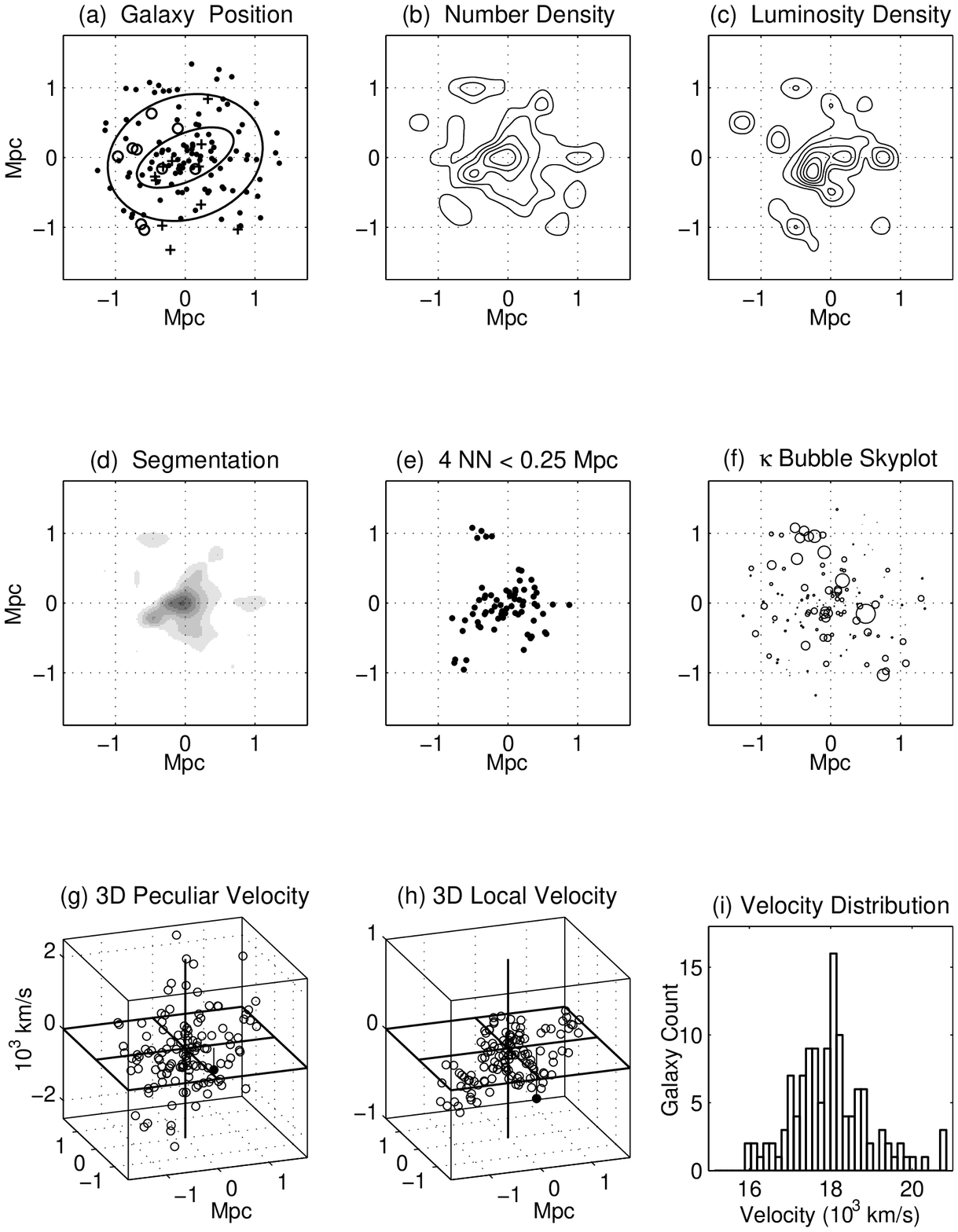}
\caption{Visualization plots for Abell 2734.}
\label{a2734}
\end{figure*}

\clearpage
\begin{figure*}
\centering
\epsfig{figure=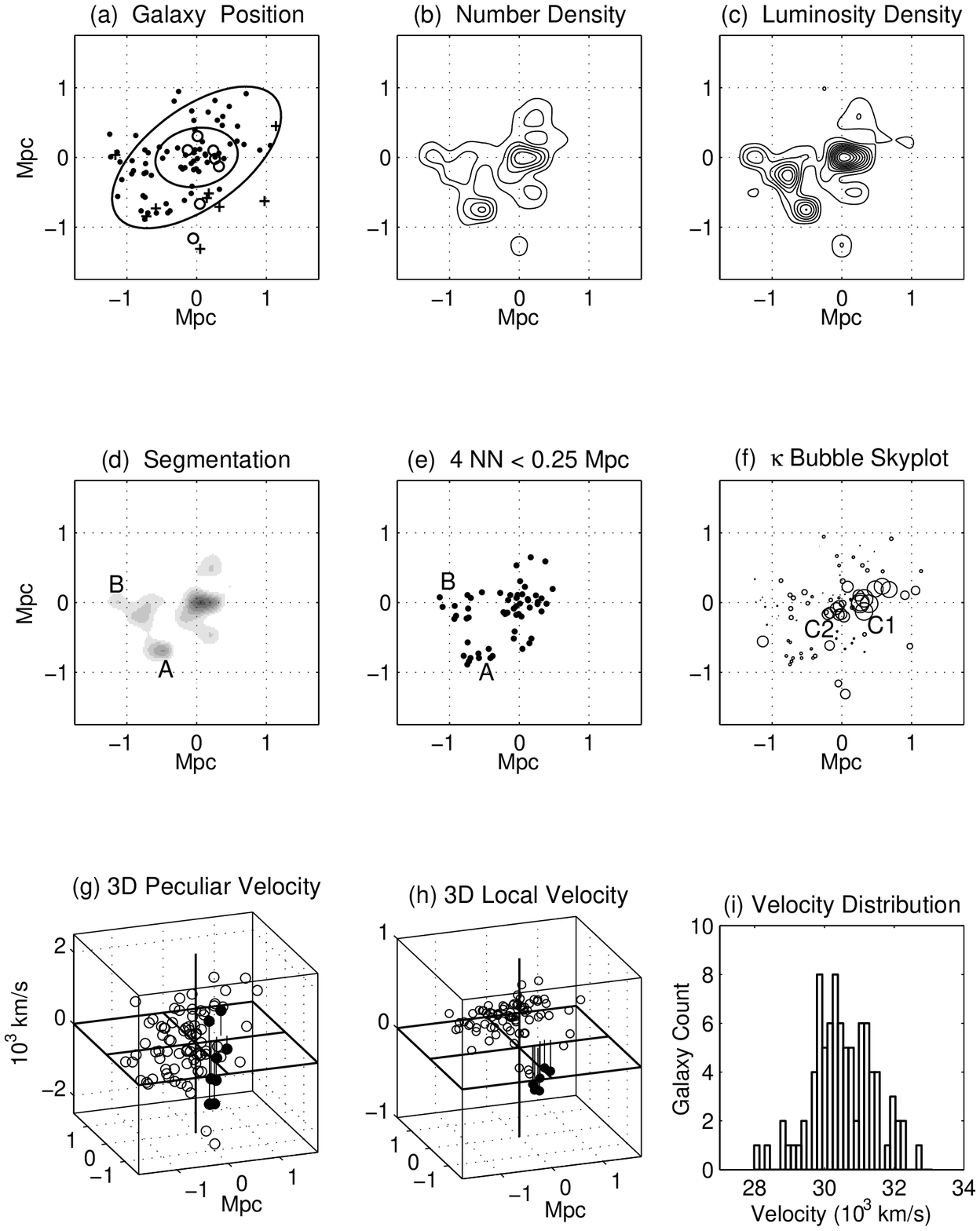}
\caption{Visualization plots for Abell 2814.}
\label{a2814}
\end{figure*}

\clearpage
\begin{figure*}
\centering
\epsfig{figure=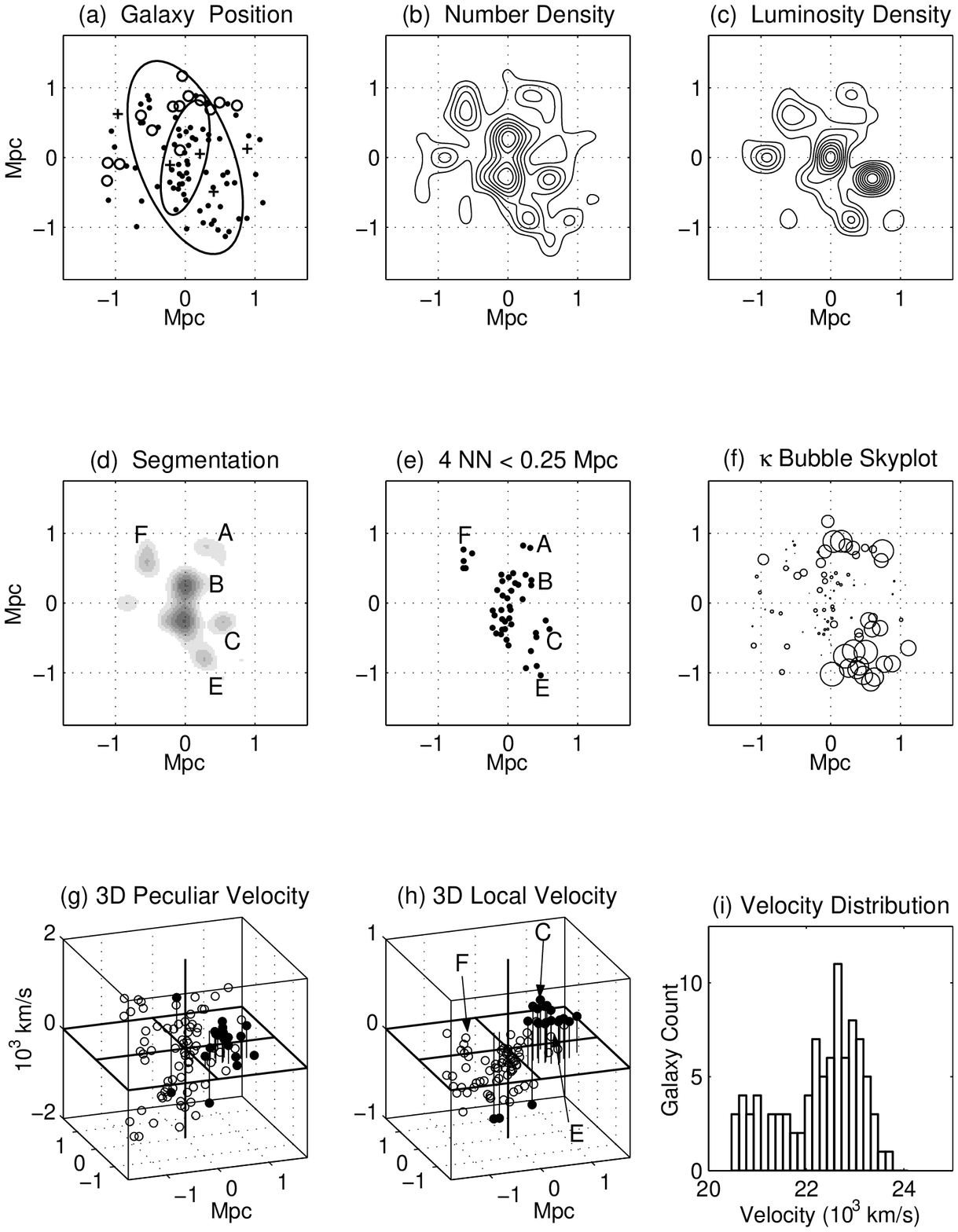}
\caption{Visualization plots for Abell 3027.}
\label{a3027}
\end{figure*}

\clearpage
\begin{figure*}
\centering
\epsfig{figure=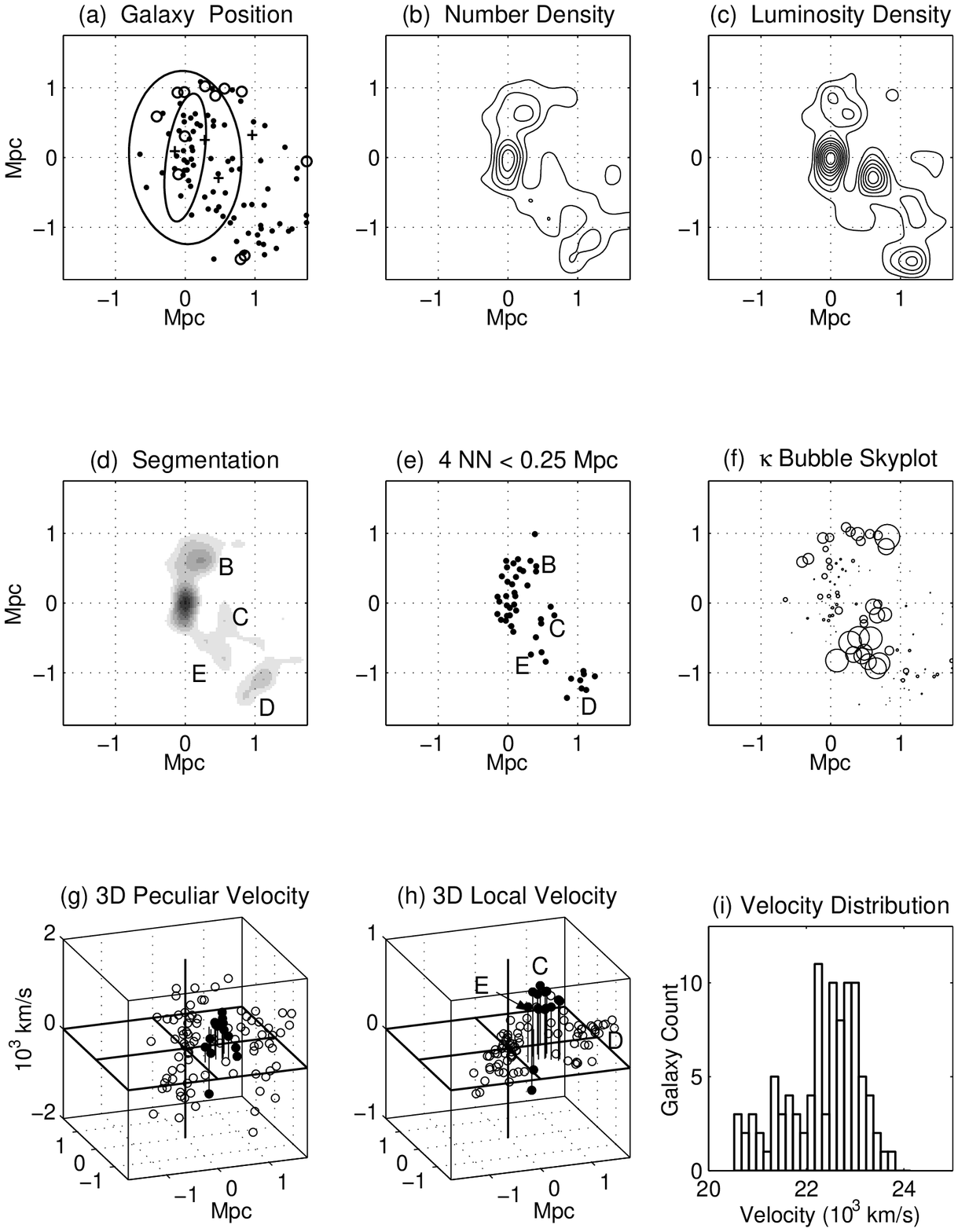}
\caption{Visualization plots for APM 268.}
\label{apm268}
\end{figure*}

\clearpage
\begin{figure*}
\centering
\epsfig{figure=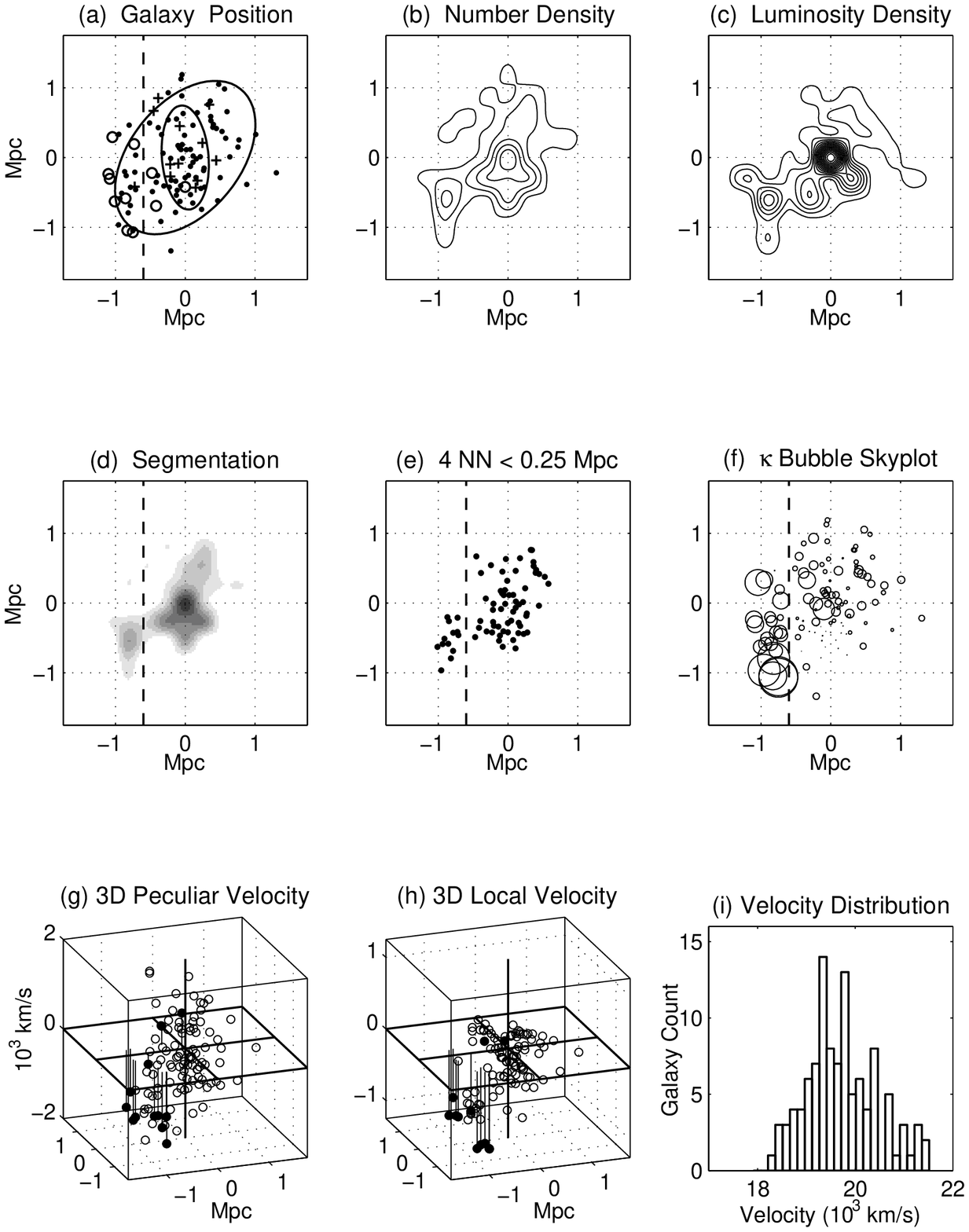}
\caption{Visualization plots for Abell 3094.}
\label{a3094}
\end{figure*}

\clearpage
\begin{figure*}
\centering
\epsfig{figure=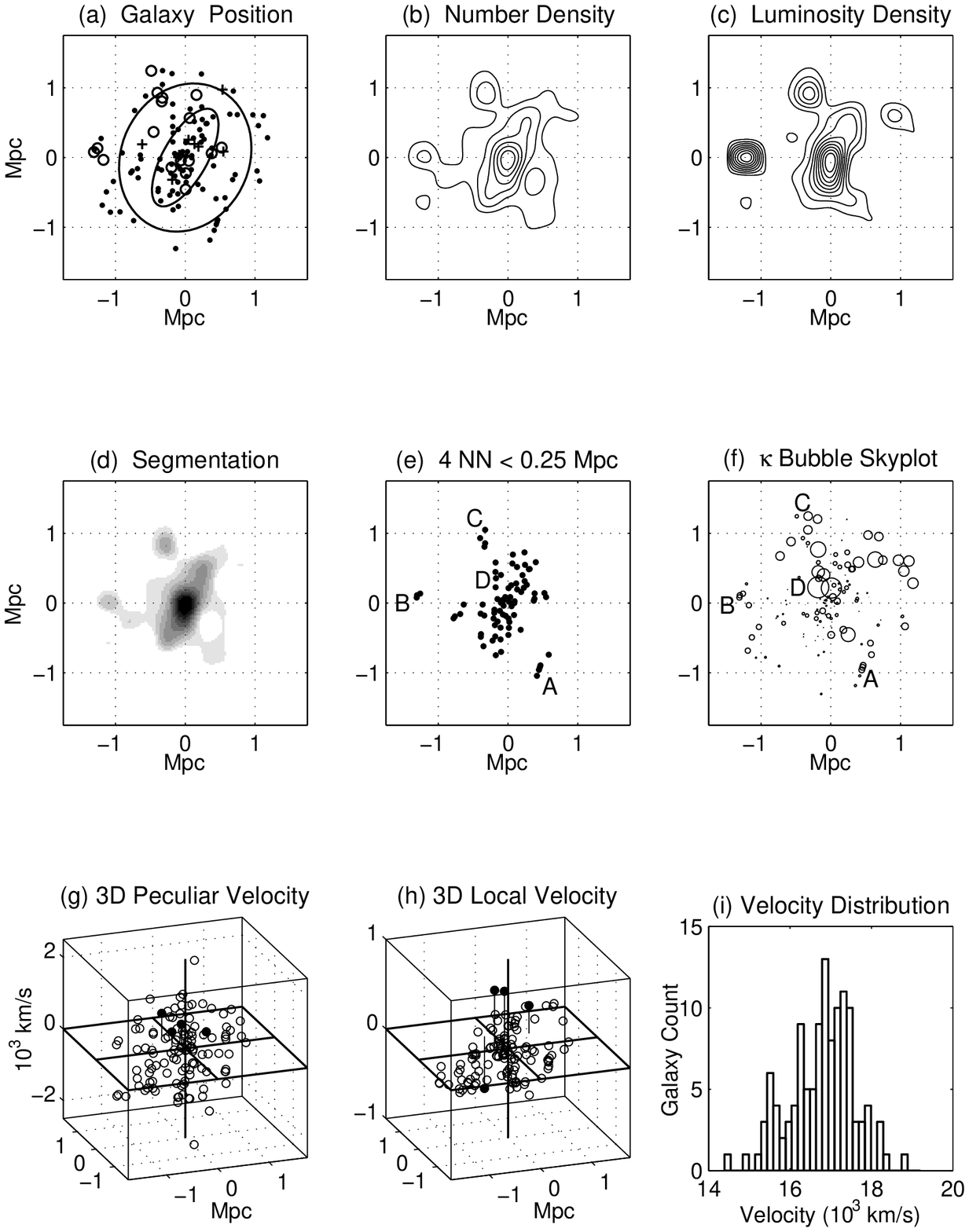}
\caption{Visualization plots for Abell 3880.}
\label{a3880}
\end{figure*}

\clearpage
\begin{figure*}
\centering
\epsfig{figure=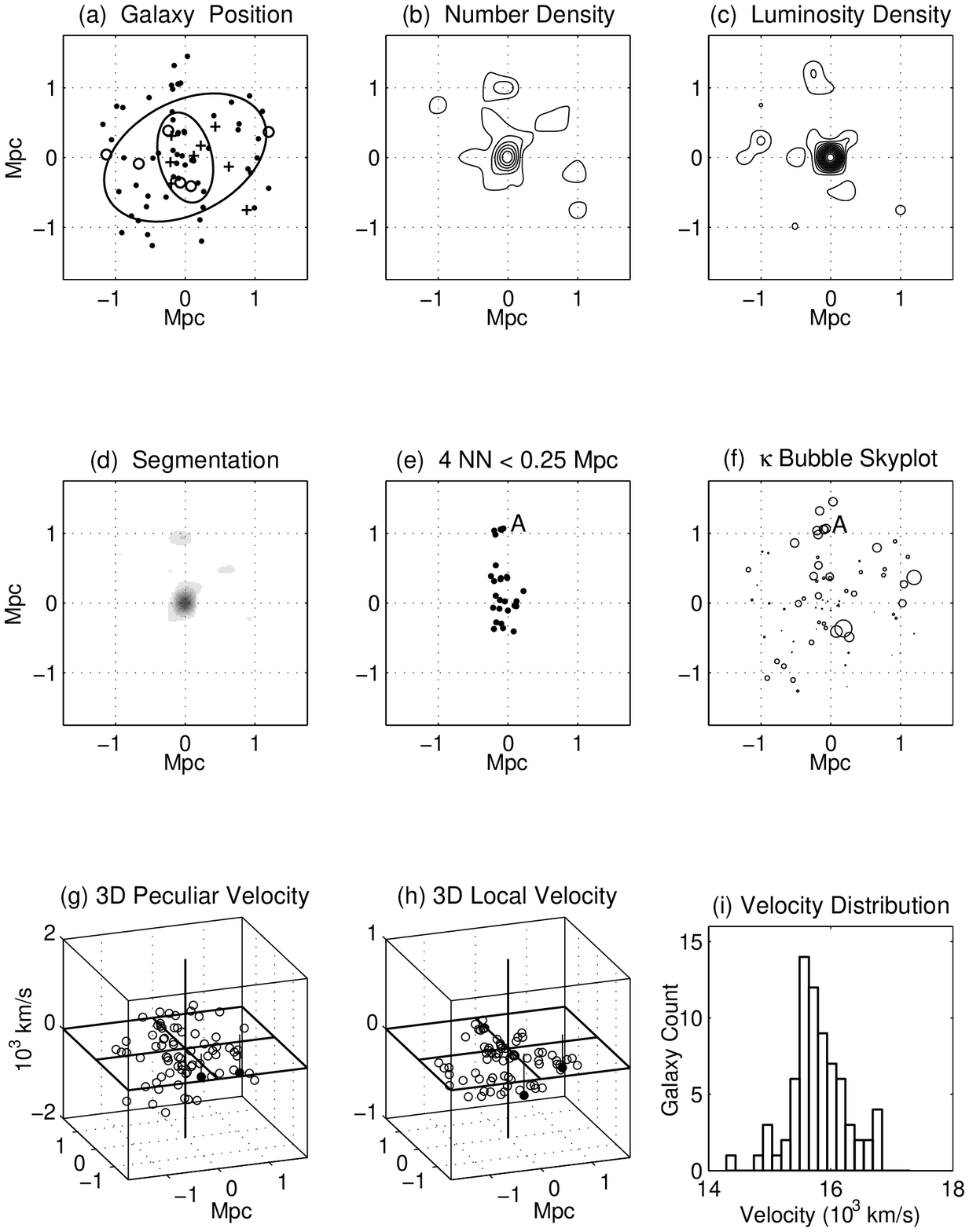}
\caption{Visualization plots for Abell 4012.}
\label{a4012}
\end{figure*}

\clearpage
\begin{figure*}
\centering
\epsfig{figure=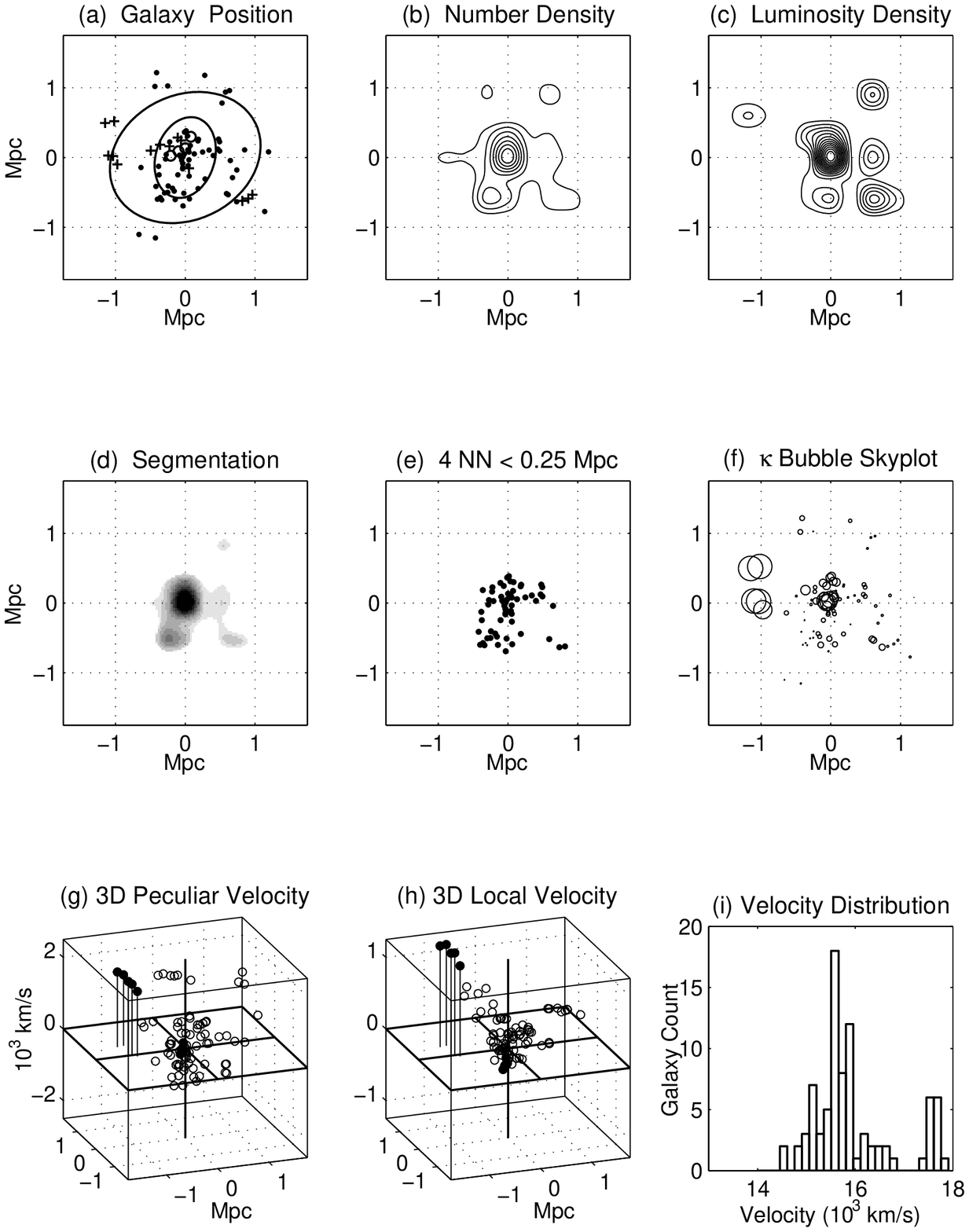}
\caption{Visualization plots for Abell 4013.}
\label{a4013}
\end{figure*}

\clearpage
\begin{figure*}
\centering
\epsfig{figure=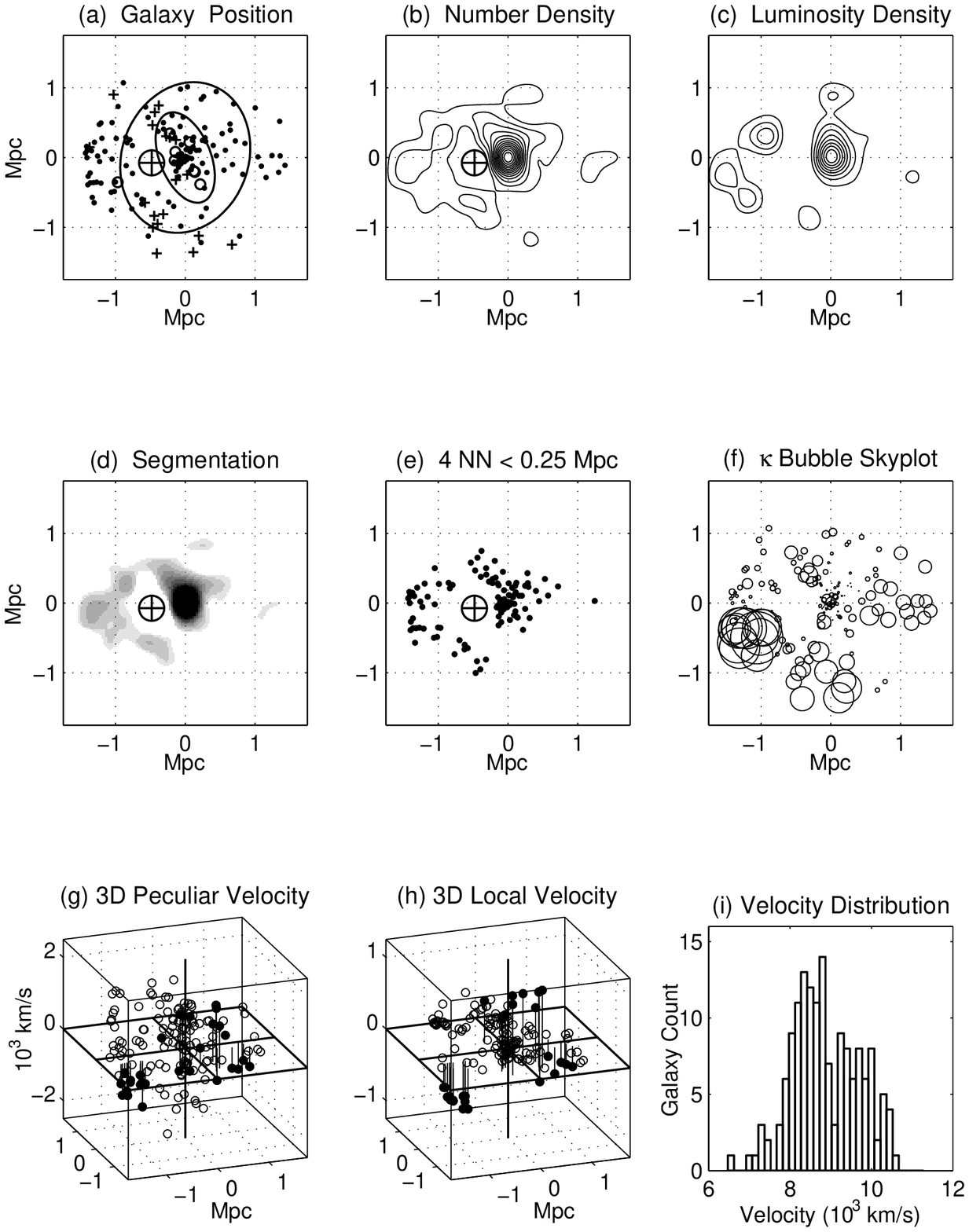}
\caption{Visualization plots for Abell 4038.}
\label{a4038}
\end{figure*}

\clearpage
\begin{figure*}
\centering
\epsfig{figure=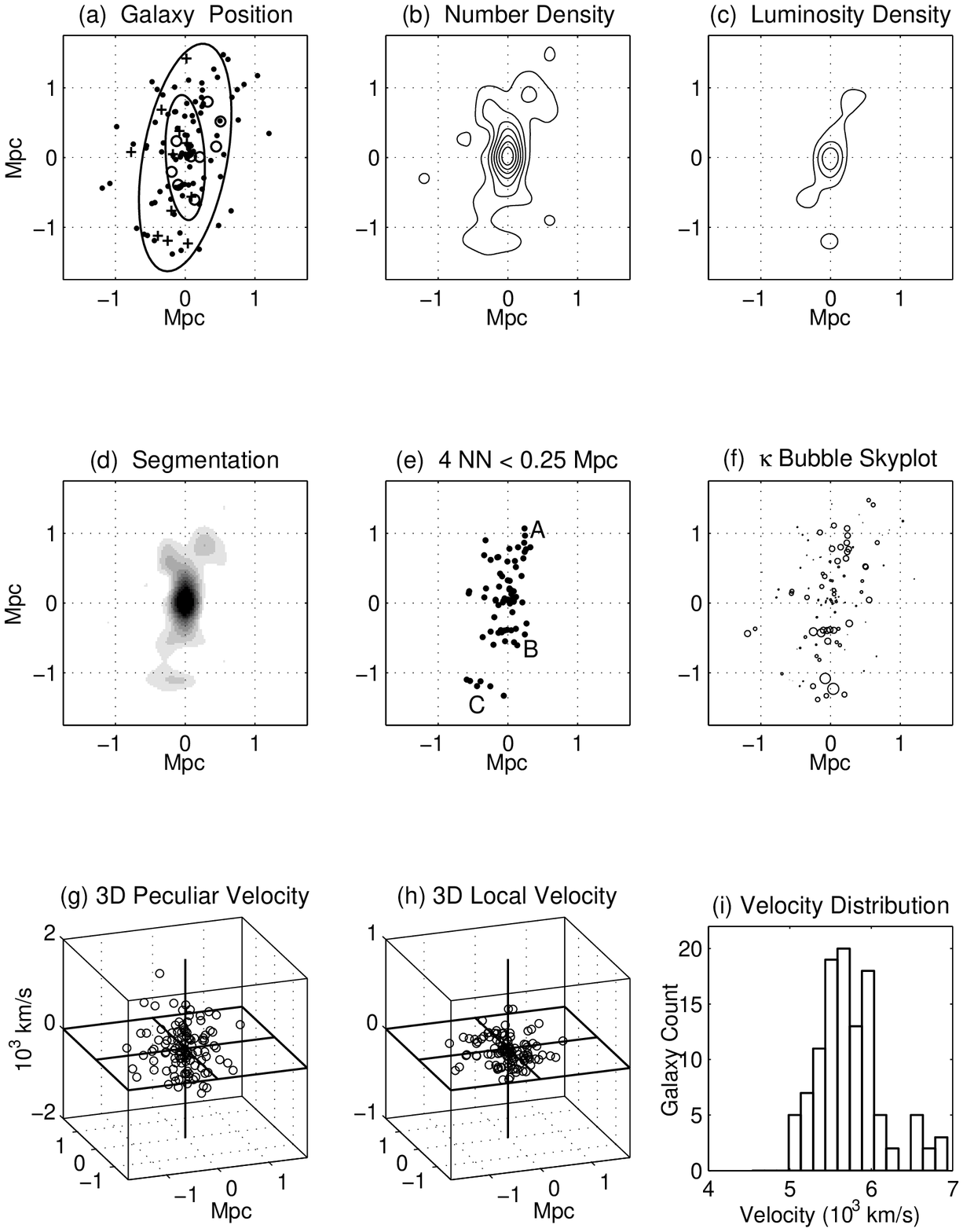}
\caption{Visualization plots for Abell S141.}
\label{s141}
\end{figure*}

\clearpage
\begin{figure*}
\centering
\epsfig{figure=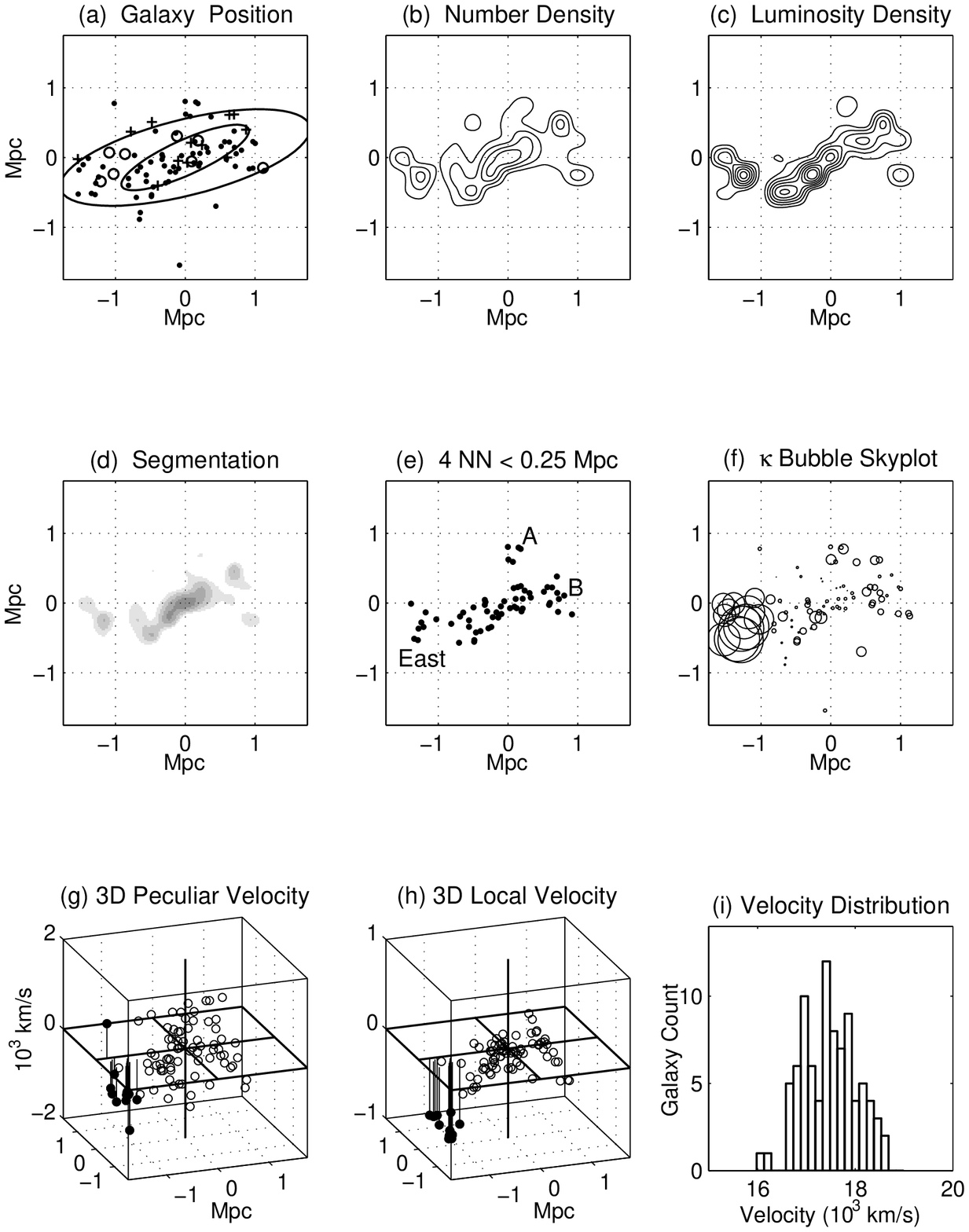}
\caption{Visualization plots for Abell S258.}
\label{s258}
\end{figure*}

\clearpage
\begin{figure*}
\centering
\epsfig{figure=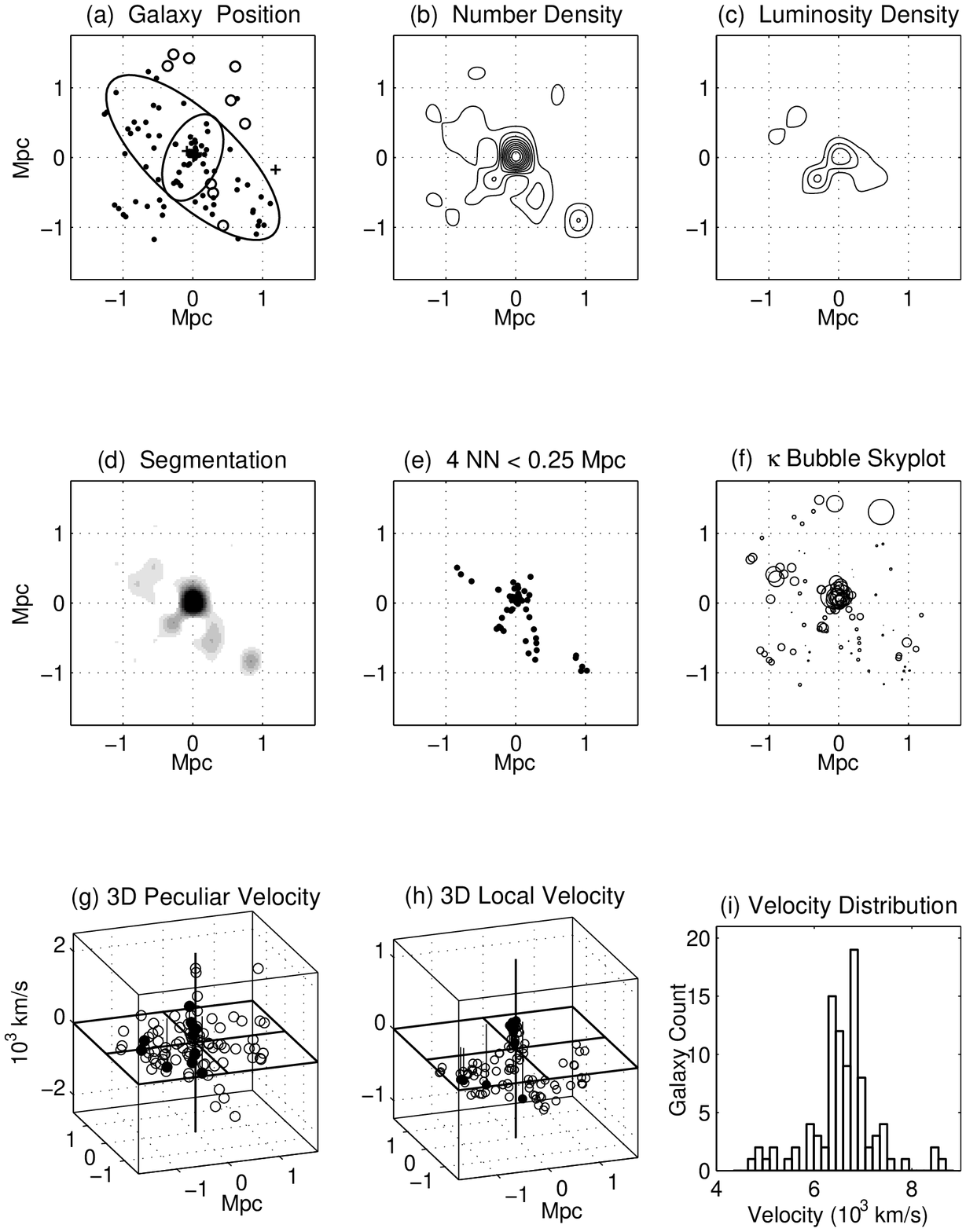}
\caption{Visualization plots for Abell S301.}
\label{s301}
\end{figure*}

\clearpage
\begin{figure*}
\centering
\epsfig{figure=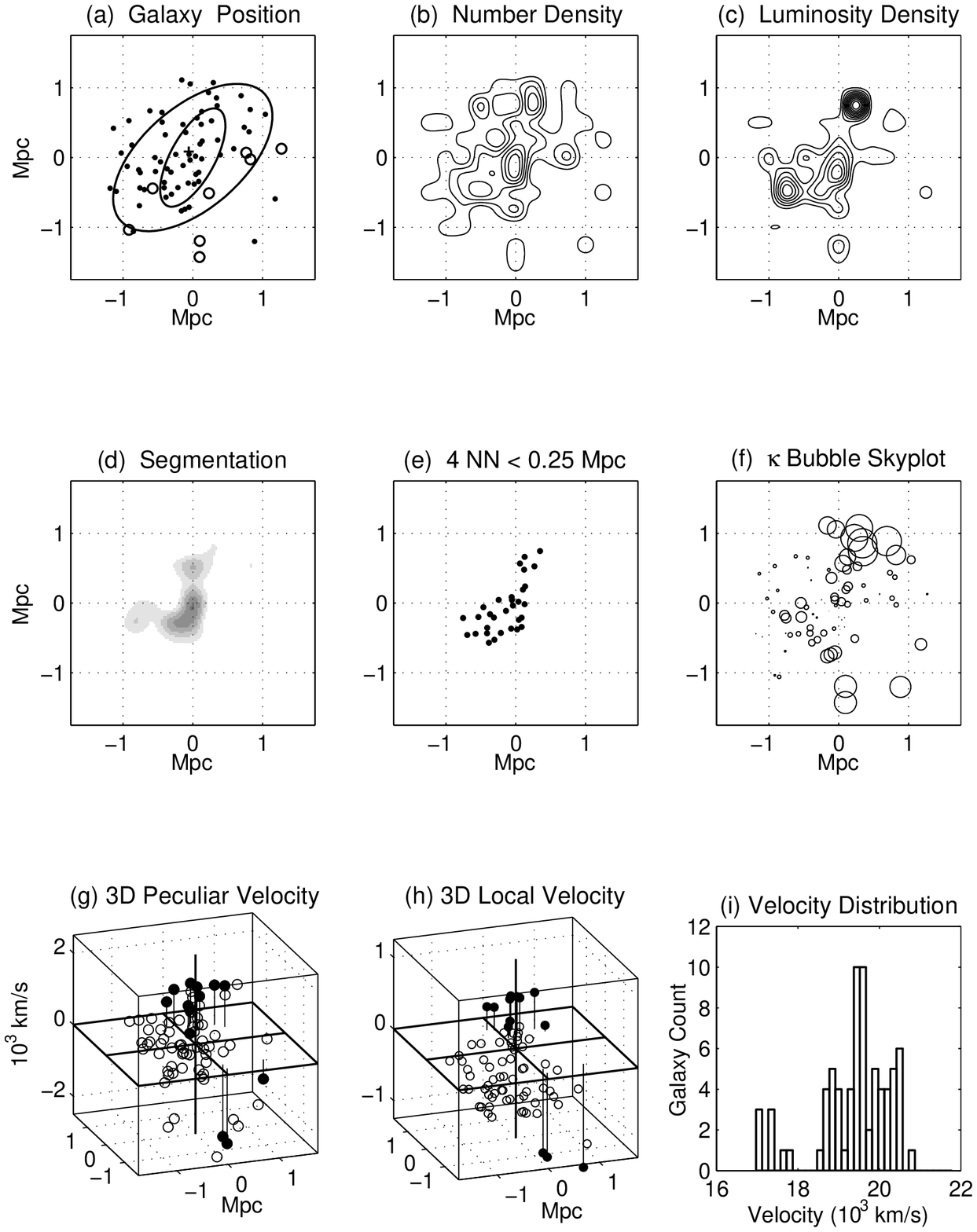}
\caption{Visualization plots for Abell S333.}
\label{s333}
\end{figure*}

\clearpage
\begin{figure*}
\centering
\epsfig{figure=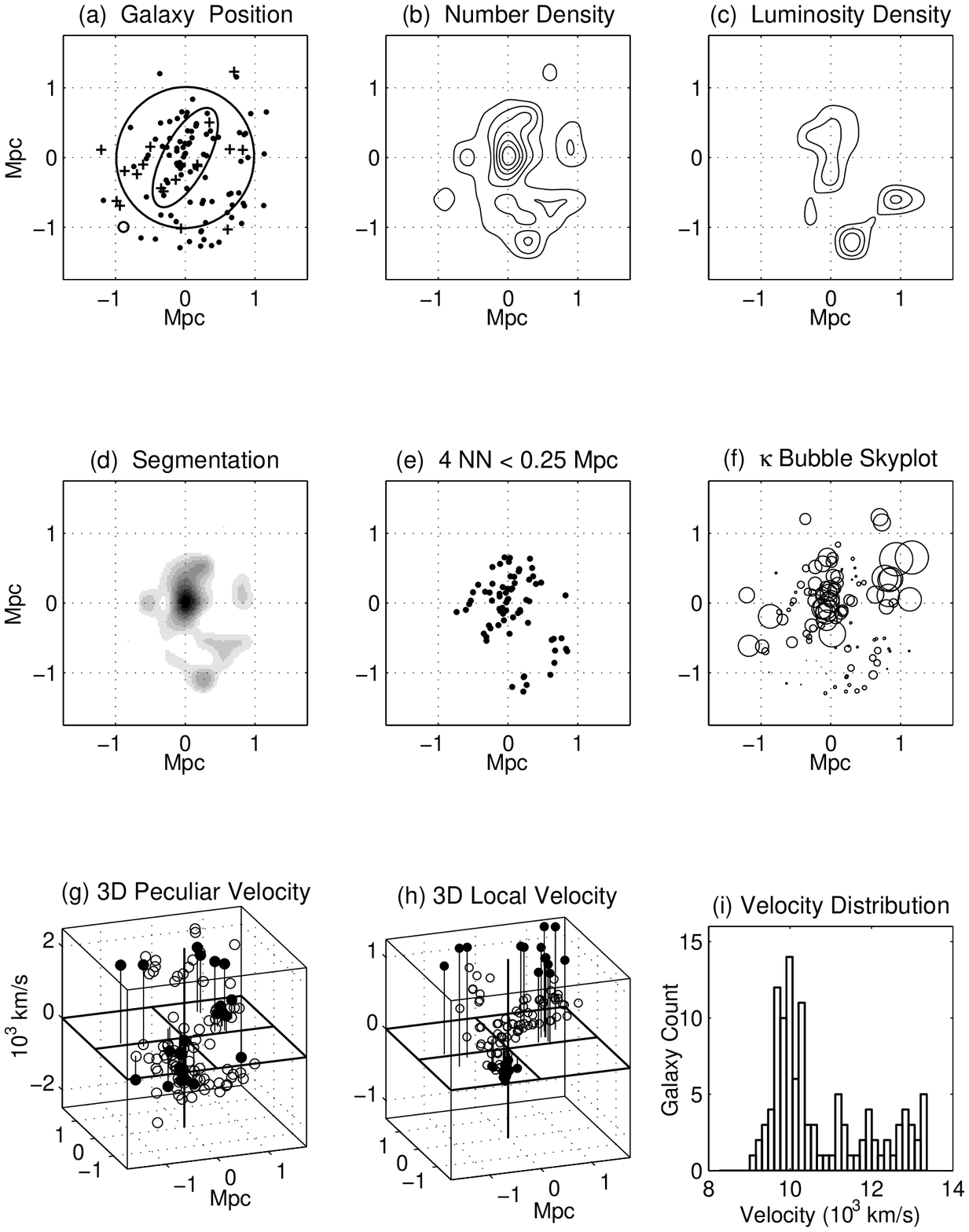}
\caption{Visualization plots for Abell S1043.}
\label{s1043}
\end{figure*}

\clearpage
\begin{figure*}
\centering
\epsfig{figure=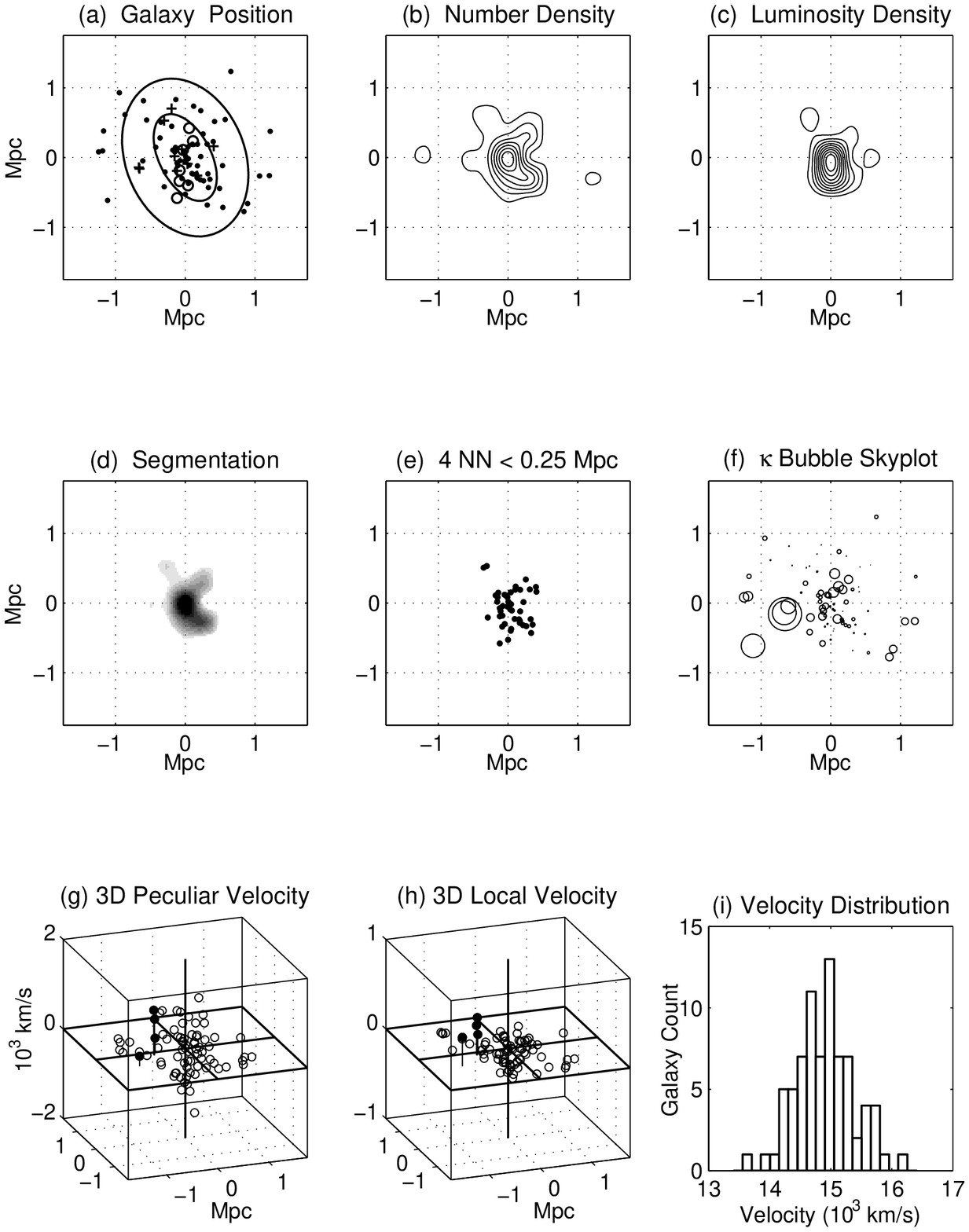}
\caption{Visualization plots for APM 917.}
\label{apm917}
\end{figure*}

\clearpage
\begin{figure*}
\centering
\epsfig{figure=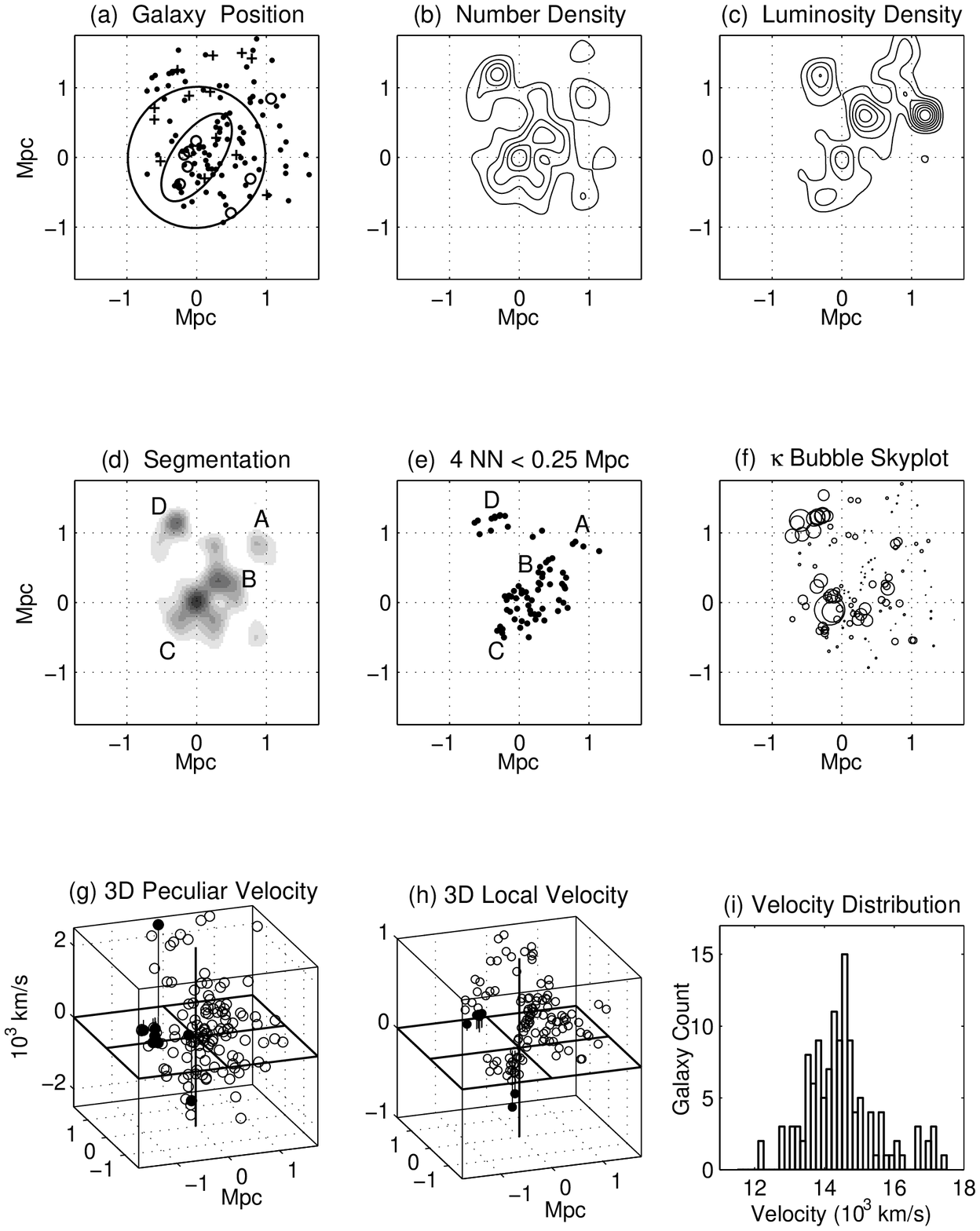}
\caption{Visualization plots for APM 933.}
\label{apm933}
\end{figure*}

\clearpage
\begin{figure*}
\centering
\epsfig{figure=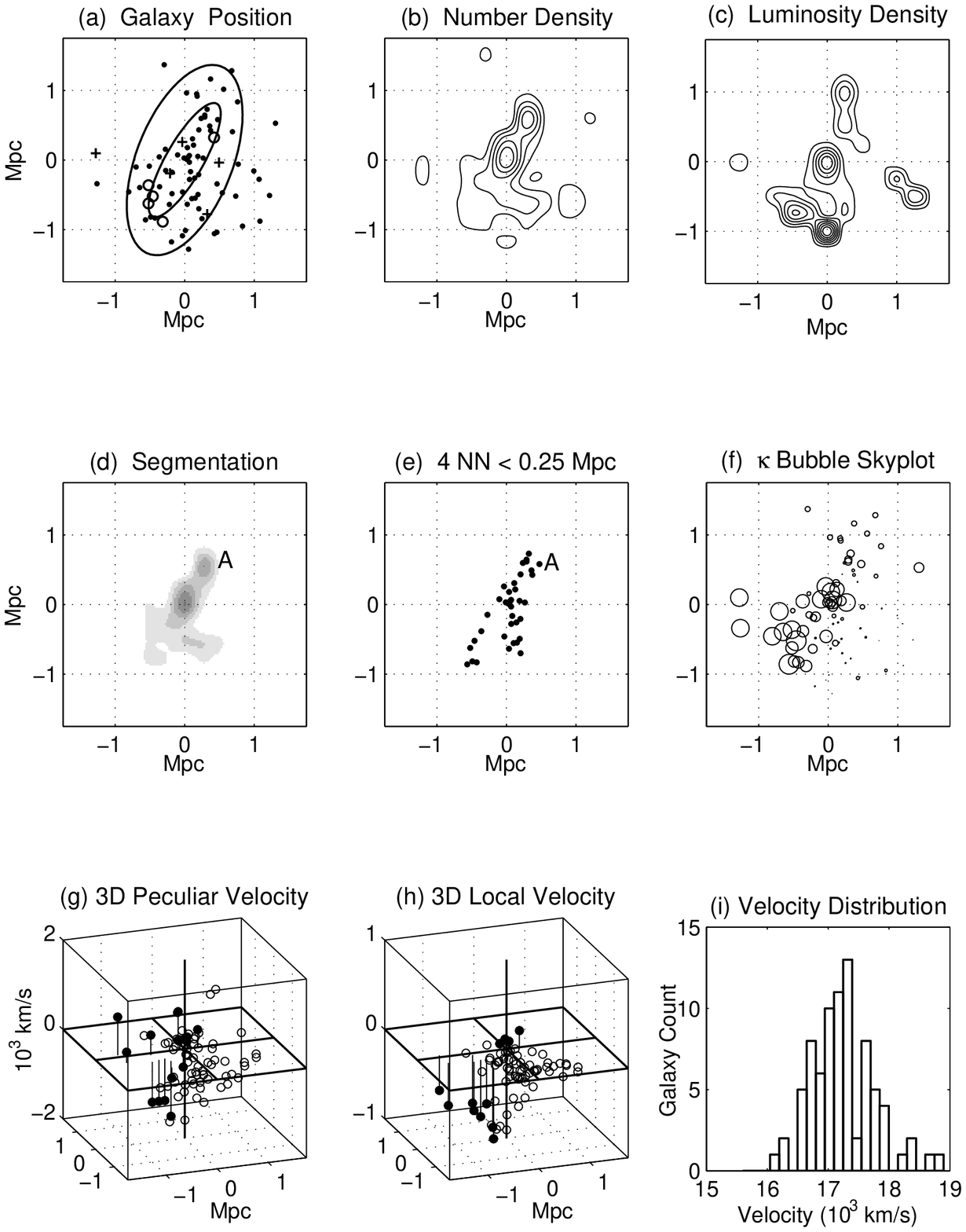}
\caption{Visualization plots for EDCC 365.}
\label{e365}
\end{figure*}

\clearpage
\begin{figure*}
\centering
\epsfig{figure=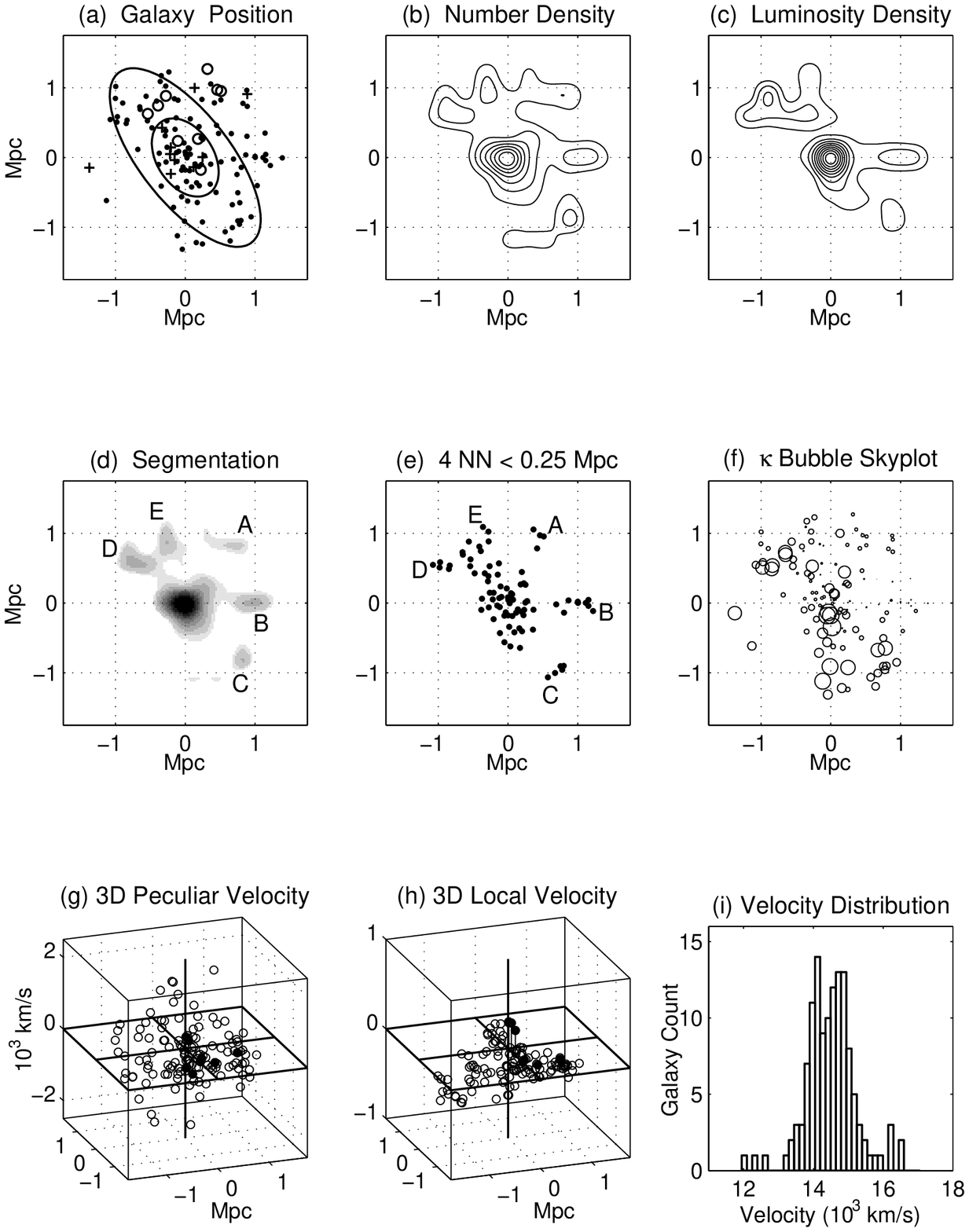}
\caption{Visualization plots for EDCC 442.}
\label{e442}
\end{figure*}

\end{document}